\documentclass [11pt]{article}

\usepackage[utf8x]{inputenc}
\usepackage{xcolor}
\usepackage{amsmath,amssymb,cite,dsfont,mathtools,tensor}
\usepackage{hyperref,verbatim}
\usepackage{float}
\usepackage[font=scriptsize]{caption}
\usepackage{graphicx,lscape,longtable,tabu}
\usepackage{indentfirst}
\usepackage{caption,subcaption}
\usepackage[title]{appendix}

\numberwithin{equation}{section}

\title{{\bf Order and Chaos in the $SU(2)$ Matrix Model \\[1ex] \Large Ergodicity and Classical Phases}}
\author{Chaitanya Bhatt$^a$\footnote{chaitanyab@iisc.ac.in}, Vijay Nenmeli$^b$\footnote{vvn21@cam.ac.uk} \,and Sachindeo Vaidya$^a$\footnote{vaidya@cts.iisc.ernet.in} \\
\begin{small}{\it $^a$Centre for High Energy Physics, Indian Institute of Science, Bangalore, 560012, India}
\end{small}\\
\begin{small}{\it $^b$Cavendish Laboratory, University of Cambridge, Cambridge CB3 0HE, United Kingdom}
\end{small}}

\date{}

\begin{document}

\maketitle
\begin{abstract}
We study the classical non-linear dynamics of the $SU(2)$ Yang-Mills matrix model introduced in \cite{Balachandran2015} as a low-energy approximation to two-color QCD. Restricting to the spin-0 sector of the model, we unearth an unexpected tetrahedral symmetry, which endows the dynamics with an extraordinarily rich structure. Amongst other things, we find that the spin-0 sector contains co-existing chaotic sub-sectors as well as nested chaotic basins, and displays alternation between regular and chaotic dynamics as energy is varied. The symmetries also grant us a considerable amount of analytic control which allows us to make several quantitative observations. Next, by noting that several features of the model have natural thermodynamic interpretations, we switch from our original chaos-theoretic viewpoint to a more statistical perspective. By so doing, we see that the classical spin-0 sector has a rich phase structure, arising from ergodicity breaking, which we investigate in depth. Surprisingly, we find that many of these classical phases display numerous similarities to previously discovered \textit{quantum} phases of the spin-0 sector \cite{Pandey2017}, and we explore these similarities in a heuristic fashion.
\end{abstract}

\section{Introduction}
Quantum chromodynamics is an $SU(3)$ non-Abelian gauge theory that plays an indispensable role in the physics of strong interactions. It is however, subtle and complicated: not only is it nonlinear and possesses an infinite number of degrees of freedom, it also has an infinite-dimensional gauge group. Progress in understanding the theory has been made mostly in the perturbative regime, or by approximating the theory by simpler models. One such model is the $SU(3)$ gauge matrix model (such as those studied in \cite{Berenstein2017}, \cite{Kurkcuoglu2021}, \cite{Kurkcuoglu2003}) obtained as the extreme low-energy limit of the full gauge field theory on $S^3 \times \mathbb{R}$: it has been successful in predicting the masses of light hadrons with surprising accuracy \cite{Pandey2017}.

In this work we will study an even simpler model, the $SU(2)$ gauge matrix model and in particular, its classical dynamics. Although nonlinear, the model has a finite number of degrees of freedom: there are three rotational, three gauge and three non-compact gauge-invariant degrees of freedom. 
Angular momentum conservation naturally allows a decomposition of the full dynamics into non-rotating and rotating sectors. 
Here we will restrict our attention to the former, which we shall henceforth refer to as the `spin-0 sector' of the matrix model.  

Despite this restriction, the spin-0 sector still has a six-dimensional phase space. Coupled with the dearth of quantitative methods inherent to non-linear systems, even this reduced system seems, at first glance, intractable. However, as we shall see, the discovery of a hidden tetrahedral symmetry simplifies matters enormously. In this avatar, the model is a three-dimensional isotropic oscillator perturbed by cubic and quartic non-linearities.

A theorem of Weinstein \cite{Weinstein1973} assures that integrable Hamiltonian systems continue to have periodic orbits even when perturbed by a small nonlinearity. 
The existence of periodic orbits and a use of group theoretic 
methods for classifying them allows us to systematize our study. 
While Hamiltonian chaos is 
typically studied using the apparatus of Kolmogorov-Arnold-Moser (KAM) theory \cite{Kolmogorov:430016,Moser:430015,Arnold1963}, physical insights are sometimes masked by the abstract nature of the necessary computations. A study of periodic orbits from the point of view of their (in)stability will prove to be ideal for our setup and will help us develop a much more intuitive feel for the dynamics. 


Second, the periodic orbits come with symmetries of their own (\cite{Montaldi1988},\cite{Efstathiou2003}), which allows us to further simplify our analysis. In particular, 
we shall see that a good fraction of the orbits live on a four dimensional submanifold of the full phase space and can be 
separately studied using an appropriately reduced \textit{effective} four dimensional system.  This is not unlike the 
Kepler problem, where the rotational symmetry renders generic orbits planar. The effects of the extra dimensions are mostly 
cosmetic, so that we lose no generality by treating these orbits using the effective four-dimensional system and eventually 
reverting back to the full model. 

Lastly, 
the symmetries 
also simplify the expressions governing time evolution along certain periodic orbits, to the point where analytic solutions can 
be obtained, and uncover dynamics that is far more 
intricate than one usually encounters.
It turns out that the set of \textit{all} trajectories over the phase space can be partitioned into  classes, with each class stemming from the destabilization of a specific type of periodic orbit. Although 
the idea of regarding chaotic trajectories as destabilisations of periodic orbits is not new, any `memory' of the parent orbit is usually rapidly erased in the chaotic domain,
and perturbations 
about different periodic orbits quickly cease to be distinguishable from one another. What \textit{is} novel here is that such a 
 `memory loss' does not occur as long as energies belonging to particular bands. As a result, for such energies, the phase space displays the peculiar feature of \textit{multiple 
co-existing chaotic `basins'}.  Systems containing co-existing attractors are quite rare -- the Rabinovich-Fabrikant model \cite{Rabinovich} being the 
prototypical example -- and are normally rather artificial. It is thus 
extremely interesting to see this phenomenon arising naturally in the setting of a gauge matrix model. 


The spin-0 sector thus possesses an extraordinarily rich dynamics, worthy of a study even as a standalone non-linear system. Our eventual aim however, is to work out how such dynamics ties in with the physics of gauge theory. Such a mapping can be carried out by identifying chaotic and regular sectors of the non-linear system with \textit{classical phases} of the underlying gauge-matrix model \cite{Berenstein2017}.  While such themes will indeed feature in our analysis, albeit in a more nuanced manner, they will only form one half of a two stage procedure. This is because we have, in addition to our non-linear analysis, a thorough repository of the \textit{quantum} dynamics of the matrix model \cite{Pandey2017}.  In particular, the quantum matrix model has been shown to admit quantum phases via superselection sectors; phases which, remarkably enough, bear some resemblance to the classical phases associated with certain classes of periodic orbits. The classical spin-0 sector of the full matrix model in some sense retains some `memory' of its innate quantum nature! These links between the classical and quantum regimes can be exploited both ways: in one direction, we can use techniques from the phase study of the quantum theory to better elucidate their classical counterparts. On the other hand, our classical-quantum correspondence is not perfect -- as we shall see, there are classical phases of the spin-0 sector which have no apparent quantum analogs. It is thus natural to use well-established methods, such as the Gutzweiller trace formula \cite{Gutzwiller1990}, to attempt to search for quantum counterparts to these classical phases or, should they not exist, to understand the limits of this correspondence. These questions are by no means trivial, and will constitute the subject of a future work. In this article, we will just provide a heuristic outline of the various connections between classical and quantum phases.

To explain the peculiar features of the spin-0 sector dynamics, we shall use three distinct but interlocking diagnostic tools, designed for similar but not identical purposes. 

Our first tool involves quantifying the growth of fluctuations about individual periodic orbits. The resulting fluctuation equations are identical in form to those describing the eigenstates of a quantum particle in a certain periodic potential. This correspondence allows us to connect well-known results of band theory to novel analogs in the the study of fluctuations. As is well known from solid-state physics, the spectrum of a quantum particle in a periodic potential comprises several energy bands, separated from one another by band gaps \cite{Ashcroft76}. As we will show, such features manifest on the nonlinear side of the correspondence as \textit{alternations between regularity and chaos} as we vary the energy. Although alternations between regularity and chaos (\textit{intermittency}, as it is termed \cite{Ott2002}) have been documented in literature,  such alternations are usually irregular, with no clear-cut methods for identifying regions of stability or instability. In contrast, the analytic control (which we owe to the tetrahedral symmetry) we have over our fluctuation equations allows us to make far more precise statements on the locations of transition points. Specifically, we will, for a particular class of periodic orbits, work out the \textit{exact} energy at which the \textit{first} transition from instability to stability occurs. For this same class of orbits, we will also be able to obtain asymptotically valid expressions for transition points in the high energy limit. This analysis follows from a study of the \textit{monodromy matrix $\mathcal{U}$} \cite{Teschl2012-co}. More precisely, it is the \textit{spectrum} of $\mathcal{U}$ that proves to be a reliable indicator of orbit stability. In our case, it turns out the symmetries of the spin-0 sector and the associated simplification in time evolution allow us to assess orbit stability using just a \textit{single} spectral invariant ($\text{Trace}\,\,\mathcal{U}$) rather than its entire spectrum. As we will later see, this reduction will additionally grant us an unusually strong analytic handle over the chaotic dynamics, and will help us derive a good number of precise quantitative results. 

Next, given that we are dealing with a highly non-linear system, it is natural to consider Lyapunov exponents and Poincar\'e sections -- the standard indicators of chaos. While these constructs do not normally yield analytical information, the latter is an excellent qualitative diagnostic for chaos, while the former reliably quantifies the `degree of chaos' present. Adapted to our system, Poincar\'e sections wonderfully bring out the numerous substructures underlying the full dynamics, particularly the phenomena of ergodicity breaking and co-existing chaotic `basins'. Lyapunov exponents complement the visual aids provided by the Poincar\'e sections and also serve as an excellent independent identifier for ergodicity breaking.

A third set of diagnostic tools is drawn from the thermodynamics of small systems. Statistical constructs such as temperature and entropy, while usually applied to  many-body systems, can also be discussed in the context of chaotic dynamics owing to the common theme of ergodicity which underlies these constructs. We find that Gibbs entropy and temperature \cite{Hilbert2014}, first discussed in a non-linear dynamical context in \cite{Berdichevsky1991}, beautifully illustrate the ergodicity breaking inherent to our model. Additionally, it serves as a useful verification alongside the other diagnostic tools we have mentioned, and naturally lends weight to our interpretation of ergodicity breaking as classical phases.

 This article is organized as follows: In Section \ref{matrixmodelsetup}, we describe the Yang-Mills matrix model, its Lagrangian and Hamiltonian formalisms, and obtain the equations governing the spin-0 sector. In Section \ref{sec3}, we outline the symmetries of the spin-0 sector, investigate their  topological and dynamical effects, and introduce the families of periodic orbits that they generate. Section \ref{sec4} builds on this with a thorough enumeration of the structure and properties of these aforementioned families, adding an extra pair along the way. This is followed, in Section \ref{StabilityAnalysis}, with an extensive study of the stability properties of each family of orbits using monodromy matrix theory. We then pursue a traditional chaos study (Poincar\'e sections and Lyapunov exponents) in Section \ref{CT}, where we also tie these results to those of the monodromy analysis of the previous section. The thermodynamic viewpoint is pursued in Section \ref{TD}, where we examine the relation between ergodicity (and its breaking) and Gibbs temperature. We also compare our observations with the results of our non-linear dynamical analysis of Section \ref{CT}. Section \ref{sec8} then explores the classical phase structure of the spin-0 sector, using the substantial collection of results developed in preceding sections. Section \ref{sec9} suggests evidence for the links between the classical phases and a host of quantum phases uncovered in a previous work of one of the authors (SV). Section \ref{sec10} provides a summary of this work and indicates directions for future work.

\section[Setting up the SU(2) Matrix Model]{Setting up the $SU(2)$ Matrix Model}\label{matrixmodelsetup}

The $SU(2)$ matrix model contains nine degrees of freedom grouped into a single matrix variable $M\in M_3(\mathbb{R})$ \cite{Balachandran2015,Pandey2017}. The dynamics of the system is governed by the Lagrangian 
\begin{align}
\label{MMLagrangian}
	L_{YM}&=\frac{1}{2g^2}\left(E_i^a E_i^a-B_i^a B_i^a\right), \quad i,a=1,2,3.
	\end{align}
Here $g$ is the Yang-Mills coupling, and $E$ and $B$ are the chromoelectric and chromomagnetic fields respectively, defined as
	\begin{align}
	E_i^a=\dot{M}_{ia}+ \epsilon_{abc}M_{0b}M_{ic},\quad
	B_i^a=\frac{1}{2}\epsilon_{ijk}F_{jk}^a,\quad
	F_{ij}^a=-\epsilon_{ijk}M_{ka}+\epsilon_{abc}M_{ib}M_{jc}.\quad
	\end{align}

Since the action possesses an $SU(2)$ gauge symmetry, we may use the associated gauge freedom to fix $M_{0a}$ to zero.

Rewriting the Lagrangian \eqref{MMLagrangian} in terms of the matrix variable $M$, we obtain
	\begin{align}
	L_{YM}=\frac{1}{2g^2}\textrm{tr}(\dot{M}^{\mathrm{T}}\dot{M})-\frac{1}{2g^2}\textrm{tr}(M^{\mathrm{T}}M)+\frac{3}{g^2}\textrm{det}M
	-\frac{1}{4g^2}[\textrm{tr}(M^{\mathrm{T}}M)]^2+\frac{1}{4g^2}\textrm{tr}[(M^{\mathrm{T}}M)^2].\label{Lagrangian}
	\end{align}
This Lagrangian is invariant under a left $O(3)$ action (physical rotations plus parity) and a right $SO(3)$ action (gauge transformations). The left and right actions give rise to two sets of conserved charges -- the physical angular momentum $J=\dot{M}M^T-M\dot{M}^T$, arising from the left $SO(3)$ action, and the gauge angular momentum $\Gamma=\dot{M}^TM-M^T\dot{M}$,  associated with the action of the gauge group. 

A new set of coordinates $(R, A, S)$, similar to the coordinates of singular value decomposition (SVD) \cite{Iwai2010,Iwai20103D} will prove to be very convenient. The matrix $M$ decomposes as  $M=RAS^T$ with $\ R \in O(3),\ S \in SO(3)$ and $A$ a real diagonal matrix $\mathrm{diag}(a_1,a_2,a_3)$. 
Introducing the angular velocities  $\Omega\equiv R^T\dot{R}$ and $\Lambda\equiv S^T\dot{S}$, the Lagrangian naturally separates into a kinetic term $T$ and a potential term $U$, and may thus be expressed as
\begin{align}
	L_{YM}&=\frac{1}{g^2}(T-U), \quad \textrm{where}\\
	T&=\frac{1}{2}\textrm{tr}(\dot{A}^2-A^2(\Omega^2+\Lambda^2)+2\Omega A \Lambda A)\quad \textrm{and}\\
	U&=U(a_1, a_2, a_3)
	=\frac{1}{2}[(a_1-a_2a_3)^2+(a_2-a_3a_1)^2+(a_3-a_1a_2)^2].\label{Uexp}
	\end{align}
The Lagrangian is independent of the `angular' coordinates $R$ and $S$, and in particular, the potential $U$ depends solely on the  variables $(a_1,a_2,a_3)$. With the $(R, A, S)$ coordinates, the angular momentum $J$ and the gauge angular momentum $\Gamma$ take the form 
\begin{align}
J=R(\Omega A^2+A^2 \Omega-2A \Lambda A)R^T,\quad
\Gamma=S(\Lambda A^2+A^2 \Lambda-2A \Omega A)S^T.
\end{align}

For the phase space formulation, we begin by defining the canonical momenta
	\begin{align}
	p_A=\frac{\partial L}{\partial \dot{A}}=\frac{1}{g^2}\dot{A},\quad
	p_{\Omega}=\frac{\partial L}{\partial \Omega}=\frac{1}{g^2}R^TJR,\quad
	p_{\Lambda}=\frac{\partial L}{\partial \Lambda}=\frac{1}{g^2}S^T\Gamma S.
	\end{align}

In terms of the (phase space) coordinates $(R, A, S, p_{\Omega}, p_A, p_{\Lambda})$, the Hamiltonian is
	\begin{align}
	H_{YM}&=\langle p_{\Omega}, \Omega\rangle_{\mathfrak{so}(3)}+\langle p_{\Lambda}, \Lambda\rangle_{\mathfrak{so}(3)}+\langle p_A, \dot{A}\rangle_{\mathfrak{so}(3)}-L, \\
	 &=\frac{g^2}{2}\langle p_A, p_A\rangle_{\mathfrak{so}(3)}+\frac{g^2}{2}\langle p_{\Omega}, \Omega\rangle_{\mathfrak{so}(3)}+\frac{g^2}{2}\langle p_{\Lambda}, \Lambda\rangle_{\mathfrak{so}(3)}+\frac{1}{g^2} U(A),
\label{Hamiltonian} \\
\textrm{where } \langle\xi, \eta\rangle_{\mathfrak{so}(3)} &\equiv \frac{1}{2}\textrm{tr}(\xi^{\textrm{T}}\eta) \nonumber.
	\end{align}
	The Gauss law requires us to fix $\Gamma=0$, i.e. $p_{\Lambda}=0$. We will thus omit any equations involving these variables.
The equations of motion (EOM) are then 
	\begin{align}
	\frac{dA}{dt}=\frac{\partial H}{\partial p_A},\quad
	\frac{dp_A}{dt}=-\frac{\partial H}{\partial A},\label{eom1}\\
	\frac{dp_{\Omega}}{dt}=[p_{\Omega}, \Omega],\quad
	\Omega=\frac{\partial H}{\partial p_{\Omega}}.\label{eom2}
	\end{align}
Since $\Omega$ is an antisymmetric $3\times 3$ matrix, it can be completely specified by a real triplet $(\omega_1, \omega_2, \omega_3)$ via
	\begin{align}
	\Omega=
	\begin{bmatrix}
	0 & -\omega_3 & \omega_2 \\
	\omega_3 & 0 & -\omega_1 \\
	-\omega_2 & \omega_1 & 0
	\end{bmatrix},
	\end{align}
	with the triplet transforming as a three vector $\omega$ under $SO(3)$ rotations. In terms of  $\omega_{i}$ and $a_{i}$, we can explicitly rewrite the Hamiltonian as 
	\begin{align}
	H_{YM}&=\frac{g^2}{2}\left(p_{a_1}^2+p_{a_2}^2+p_{a_3}^2\right)
	+\frac{g^2}{2}\left(\frac{a_2^2+a_3^2}{(a_2^2-a_3^2)^2}p_{\omega_1}^2+\frac{a_3^2+a_1^2}{(a_3^2-a_1^2)^2}p_{\omega_2}^2+\frac{a_1^2+a_2^2}{(a_1^2-a_2^2)^2}p_{\omega_3}^2\right)\nonumber\\
	&+\frac{1}{2g^2}\left((a_1-a_2a_3)^2+(a_2-a_3a_1)^2+(a_3-a_1a_2)^2\right).\label{Hamiltonian9D1}
	\end{align}
On canonically rescaling the coordinates and momenta as $a_{i}\rightarrow g a_{i},p_{a_{i}}\rightarrow \frac{p_{a_{i}}}{g}$, 
we obtain
	\begin{align}
	H_{YM}&=\frac{1}{2}\left(p_{a_1}^2+p_{a_2}^2+p_{a_3}^2\right)
	+\frac{1}{2}\left(\frac{a_2^2+a_3^2}{(a_2^2-a_3^2)^2}p_{\omega_1}^2+\frac{a_3^2+a_1^2}{(a_3^2-a_1^2)^2}p_{\omega_2}^2+\frac{a_1^2+a_2^2}{(a_1^2-a_2^2)^2}p_{\omega_3}^2\right)\nonumber\\
	&+\frac{1}{2}\left((a_1-g a_2a_3)^2+(a_2-g a_3a_1)^2+(a_3-g a_1a_2)^2\right).\label{Hamiltonian9D2}
	\end{align}
With these coordinates, the EOM {\ref{eom1},\ref{eom2}} become
\begin{align}
\dot{a}_1&=p_{a_1},\quad \dot{a}_2=p_{a_2},\quad \dot{a}_3=p_{a_3},\\
\dot{p}_{a_1}&=-\frac{1}{2}  \bigg(\frac{2 a_1 p_{\omega _2}^2}{\left(a_1^2-a_3^2\right){}^2} - 
	\frac{4 a_1 \left(a_1^2+a_3^2\right) p_{\omega _2}^2}{\left(a_1^2-a_3^2\right){}^3} + 
	\frac{2 a_1 p_{\omega _3}^2}{\left(a_1^2-a_2^2\right){}^2} - 
	\frac{4 a_1 \left(a_1^2+a_2^2\right) p_{\omega _3}^2}{\left(a_1^2-a_2^2\right){}^3}\bigg) \nonumber \\
		&-\frac{2g^2 a_1 a_2^2-6 g a_3 a_2+2 g^2  a_1 a_3^2+2 a_1}{2},\\
\dot{p}_{a_2}&=-\frac{1}{2}  \bigg(\frac{2 a_2 p_{\omega _1}^2}{\left(a_2^2-a_3^2\right){}^2} - 
	\frac{4 a_2 \left(a_2^2+a_3^2\right) p_{\omega _1}^2}{\left(a_2^2-a_3^2\right){}^3} + 
	\frac{4 a_2 \left(a_1^2+a_2^2\right) p_{\omega _3}^2}{\left(a_1^2-a_2^2\right){}^3} + 
	\frac{2 a_2 p_{\omega _3}^2}{\left(a_1^2-a_2^2\right){}^2}\bigg)\nonumber \\
	&-\frac{2 g^2 a_2 a_1^2-6 g a_3 a_1+2 g^2 a_2 a_3^2+2 a_2}{2},\\
\dot{p}_{a_3}&=-\frac{1}{2}  \bigg(\frac{4 a_3 \left(a_2^2+a_3^2\right) p_{\omega _1}^2}{\left(a_2^2-a_3^2\right){}^3}+\frac{2 a_3 p_{\omega _1}^2}{\left(a_2^2-a_3^2\right){}^2}+\frac{4 a_3 \left(a_1^2+a_3^2\right) p_{\omega _2}^2}{\left(a_1^2-a_3^2\right){}^3}+\frac{2 a_3 p_{\omega _2}^2}{\left(a_1^2-a_3^2\right){}^2}\bigg)\nonumber \\
	&-\frac{2 g^2 a_3 a_1^2-6 g a_2 a_1+2 g^2 a_2^2 a_3+2 a_3}{2},\\
	\dot{p}_{\omega _1}&=-\frac{\left(a_2^2-a_3^2\right) \left(-3 a_1^4+\left(a_2^2+a_3^2\right) a_1^2+a_2^2 a_3^2\right) g^4 p_{\omega _2} p_{\omega _3}}{\left(a_1^2-a_2^2\right){}^2 \left(a_1^2-a_3^2\right){}^2},\\
	\dot{p}_{\omega _2}&=\frac{\left(a_1^2-a_3^2\right) \left(-3 a_2^4+a_3^2 a_2^2+a_1^2 \left(a_2^2+a_3^2\right)\right) g^4 p_{\omega _1} p_{\omega _3}}{\left(a_1^2-a_2^2\right){}^2 \left(a_2^2-a_3^2\right){}^2},\\
	\dot{p}_{\omega _3}&=-\frac{\left(\left(a_2^2+a_3^2\right) a_1^4-\left(a_2^4+3 a_3^4\right) a_1^2+3 a_2^2 a_3^4-a_2^4 a_3^2\right) g^4 p_{\omega _1} p_{\omega _2}}{\left(a_1^2-a_3^2\right){}^2 \left(a_2^2-a_3^2\right){}^2}.
	\end{align}
	From these equations, it is easy to see that we have a consistent set of solutions with $p_{\omega_{i}}$'s set to zero. Physically, this corresponds to the irrotational sector of the matrix model, and it is these equations that we will study under the name of the spin-0 sector. Explicitly, the equations governing the dynamics of the spin-0 sector are then
		\begin{align}
	\dot{a}_1(t)= p_{a_1}(t),\quad
	\dot{a}_2(t)= p_{a_2}(t),\quad
	\dot{a}_3(t)= p_{a_3}(t), \label{EOM21}
	\end{align}
	\begin{align}
	\dot{p}_{a_1}(t)&=-\frac{2 g^2 a_1(t) a_2(t)^2-6 g a_3(t) a_2(t)+2 g^2 a_1(t) a_3(t)^2+2 a_1(t)}{2}, \label{EOM22} \\
	\dot{p}_{a_2}(t)&=-\frac{2 g^2 a_2(t) a_1(t)^2-6 g a_3(t) a_1(t)+2 g^2 a_2(t) a_3(t)^2+2 a_2(t)}{2},  \label{EOM23} \\
	\dot{p}_{a_3}(t)&=-\frac{2 g^2 a_3(t) a_1(t)^2-6 g a_2(t) a_1(t)+2 g^2 a_3(t) a_2(t)^{2}+2 a_3(t)}{2}. \label{EOM24}
\end{align}
These equations emerge from the variation of the Hamiltonian
\begin{align}
	H_{0}&=\frac{1}{2}\left(p_{a_1}^2+p_{a_2}^2+p_{a_3}^2\right)+\frac{1}{2}\left(a_1^2+a_2^2+a_3^2-6ga_1a_2a_3+g^2(a_1^2a_2^2+a_2^2a_3^2+a_3^2a_1^2)\right), \label{Hamiltonian5}
\end{align}
which is simply the full Hamiltonian \eqref{Hamiltonian9D2}, with the $p_{\omega_{i}}$ fixed to zero. For the remainder of this article, we will always assume zero angular momentum and work exclusively with equations {\eqref{EOM21}--\eqref{Hamiltonian5}} . 
\section{Symmetries of the Spin-0 Sector}
\label{sec3}
\subsection{The Action of the Tetrahedral Group}
Since the Hamiltonian is independent of the `angular' coordinates $R$ and $S$, the non-trivial dynamics is in the evolution of the $a_{i}$'s. 
Remarkably, Hamiltonian \eqref{Hamiltonian5} is \textit{further} invariant under the action of a discrete group. Explicitly, the action of an arbitrary element of this discrete symmetry group on the phase space variables is given by compositions of the following
\begin{enumerate}
\item  $a_{i}\rightarrow a_{P(i)}, p_{a_{i}}\rightarrow p_{a_{P(i)}}$, where $P$ is an element of the permutation group $S_{3}$.
\item $a_{i}\rightarrow {s_{i}}a_{i}, \,p_{a_{i}}\rightarrow {s_{i}}p_{a_{i}}$, where $s_{i}$ is $-1$ for two values of $i$ and $1$ for the remaining $i$. For example, $(a_{1},a_{2},a_{3})\rightarrow(a_{1},-a_{2},-a_{3}),\,\,(p_{a_{1}},p_{a_{2}},p_{a_{3}})\rightarrow( p_{a_{1}},-p_{a_{2}},-p_{a_{3}})$.
\end{enumerate}
Transformations of the second kind form a $\mathbb{Z}_{2}\times\mathbb{Z}_{2}$ subgroup of the full symmetry group, while transformations of the first kind constitute an $S_{3}$ subgroup. Both sets of transformations clearly do not commute. The full symmetry group can in fact be shown to be a semi-direct product of these two subgroups and is isomorphic to the tetrahedral group $T_{d}$. 

The Hamiltonian further possesses an additional $\mathbb{Z}_{2}$ time-reversal 
symmetry $T: p_{a_i} \to -p_{a_i}$. Along with the time-reversal group $\mathcal{T}$, the full discrete symmetry group of the spin-0 sector is thus  $T_d \times 
\mathcal{T}$. We emphasize that the $T_{d}$ symmetry of Hamiltonian \eqref{Hamiltonian5} is a non-trivial 
consequence of the SVD and in particular, bears no relation to the continuous rotational symmetries of the 
original Lagrangian $\eqref{MMLagrangian}$. This unexpected symmetry will play a crucial role in 
understanding the dynamics of the spin-0 sector in several ways, and will in particular hand us far more 
analytic control than is usually available in non-linear systems.
\subsection{Equipotential Surfaces of the Spin-0 Sector}
\begin{figure}[H]
	\centering
	\begin{subfigure}[c]{0.24\textwidth}
	\includegraphics[scale=0.24,width=\textwidth]{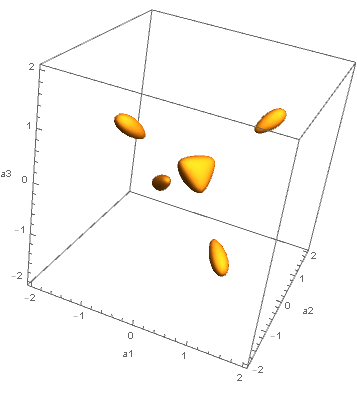}
	\caption{$E=0.05$}
	\label{EqS0.05}
	\end{subfigure}
	\hfill
	\begin{subfigure}[c]{0.24\textwidth}
	\includegraphics[scale=0.24,width=\textwidth]{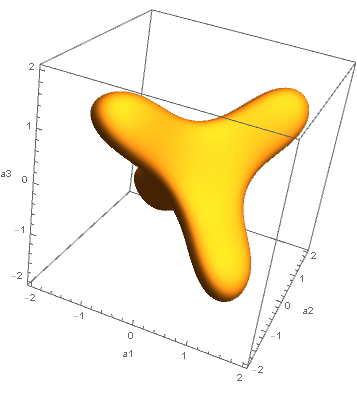}
	\caption{$E=0.5$}
	\label{EqS0.5}
	\end{subfigure}
	\begin{subfigure}[c]{0.24\textwidth}
	\includegraphics[scale=0.24,width=\textwidth]{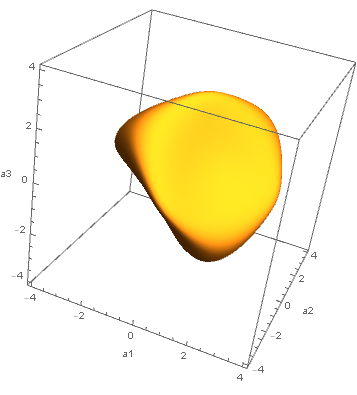}
	\caption{$E=5$}
	\label{EqS5}
	\end{subfigure}
	\hfill
	\begin{subfigure}[c]{0.24\textwidth}
	\includegraphics[scale=0.24,width=\textwidth]{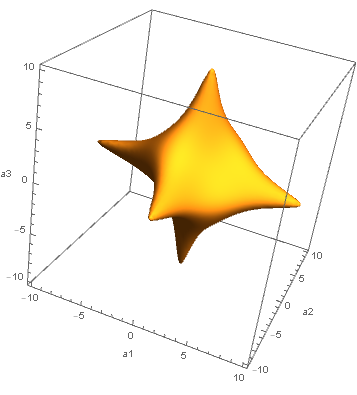}
	\caption{$E=50$}
	\label{EqS50}
	\end{subfigure}
	\caption{Configuration space equipotentials of the spin-0 sector Hamiltonian}
	\label{EqS}
\end{figure}

The tetrahedral symmetry is best seen by looking at the equipotential surfaces of the Hamiltonian of the spin-0 sector. Equipotential surfaces for various energies have been displayed in Figure~\ref{EqS}. There are two points of interest to note:
\begin{enumerate}
\item The tetrahedral symmetry, while present at all energies, is less visible at intermediate energies (Figure~\ref{EqS5}) and apparently transits to an octahedral symmetry at high energies (Figure~\ref{EqS50}). This transition is only approximate, and can be attributed to the decreasing significance of the cubic term in the potential at high energies.
 \item The \textit{topology} of the equipotential surface changes as we cross a certain critical energy $E_{c}$.  Equipotentials at `subcritical' energies (Figure~\ref{EqS0.05}) are disconnected and are composed of a central lobe and a set of four side lobes. `Supercritical equipotentials' (Figures~\ref{EqS0.5}-\ref{EqS50}), in contrast, are connected surfaces in configuration space.  From the geometry of the equipotentials, it is clear that the critical energy $E_{c}$ is simply the energy $E$ at which the number of solutions of the equation $V(a,a,a)=E$ is exactly one. Solving this, we obtain $E_{c}=\frac{3}{32 g^{2}}$.
\end{enumerate}
Given this topological feature of the potential, it is natural to partition the dynamics into `subcritical' and `supercritical' regimes, and study each one separately. Indeed, we will later find that the dynamics of the two regimes are quite different, with each zone displaying peculiarities of different kinds. 

\subsection{Symmetries and Periodic Orbits}
The EOMs \eqref{EOM21}-\eqref{EOM24} are highly nonlinear and, as we shall see, lead to chaotic dynamics. Chaotic Hamiltonian systems are frequently studied using techniques closely associated with the Kolmogorov-Arnold-Moser (KAM) theorem \cite{Kolmogorov:430016,Moser:430015,Arnold1963}.  Such KAM investigations involve splitting the Hamiltonian into an integrable portion and  non-integrable perturbations, and then using perturbative methods to study the dynamical effects of these corrections. 

The Hamiltonian \eqref{Hamiltonian5} governing the dynamics of the spin-0 sector has a natural interpretation as a perturbed system of three decoupled simple harmonic oscillators (SHOs), with $g$ playing the role of a perturbation parameter.  However, it turns out that $g$ is \textit{not} the ideal candidate for the perturbation parameter. To see this, we note that if the 6D phase space vector $(a_{1}(t),a_{2}(t),a_{3}(t),p_{a_{1}}(t),p_{a_{2}}(t),p_{a_{3}}(t))$ is a solution to the EOM with $g=1$ and energy $E$, then $(a_{i},\frac{p_{a_{i}}}{g^2})$ is also a solution to the EOM with  coupling $g$ and energy $\frac{E}{g^{2}}$. As a result, the qualitative features of solutions -- orbit shapes, time averages, measures of chaos/stability, to name a few -- depend not on the specific values of energy and coupling, but a particular combination thereof. The above scaling arguments show that $g^{2}E$ is the correct choice. Thus we may as well set $g$ to $1$ and observe the entire spread of dynamics by varying just the energy. It is worth emphasising that with this convention, we have $E_{c}=\frac{3}{32}$.

   
  Hamiltonian systems possess periodic orbits sufficiently close to an integrable limit \cite{Weinstein1973}. Models with tetrahedral symmetry have been thoroughly studied and their orbits classified in \cite{Montaldi1988,Efstathiou2003}. 
 Similar approaches involving simplification of periodic orbit analysis by discrete group symmetries have been applied to the Henon-Heiles system\cite{Henon1964}. 
 In fact, the spin-0 sector of the full matrix model can itself be regarded as an instance of a specific class of higher dimensional analogs of the Henon-Heiles system, first put forward in \cite{Efstathiou2003} . 

 The $T_{d}\times \mathcal{T}$ symmetry of the Hamiltonian of the spin-0 sector implies the existence of multiple families of periodic orbits. Most of these orbits persist at low energies, but get destroyed on increasing energy and moving away from the integrable limit. We will refer to these as \textit{non-linear normal modes} (NLNMs). The NLNMs of the spin-0 sector can be classified by symmetry properties. More precisely, the NLNMs may be classified according to their stabilizers $G$. They fall into five classes, listed in table~\ref{ESOrbits1}. (Here $\mathcal{T}_2= \lbrace 1,C_2 T \rbrace $ and $\mathcal{T}_s= \lbrace 1,C_s T \rbrace $.)
\begin{table}[H]
    \centering

    \begin{tabular}{|l|l|l|}
    \hline
        Conjugacy class of stabilizer & Shorthand notation & Number of modes \\ \hline
        $D_{2d} \times \mathcal{T}$ & $A_4$ & 3 \\ \hline
        $C_{3v} \times \mathcal{T}$ & $A_3$ & 4 \\ \hline
        $C_{2v} \times \mathcal{T}$ & $A_2$ & 6 \\ \hline
        $S_4 \wedge \mathcal{T}_2$ & $B_4$ & 6 \\ \hline
        $C_3 \wedge \mathcal{T}_s$ & $B_3$ & 8 \\ \hline
    \end{tabular}
        \caption{Periodic Orbits of the Spin-0 Sector}
               \label{ESOrbits1} 
\end{table}
The presence of NLNMs is formally established by considering a `reduced' phase space, obtained by quotienting the full six-dimensional phase space by the orbits of the decoupled SHO limit. Correspondences can then be drawn between properties of objects living in the original phase space and their counterparts residing on the reduced phase space. In particular, the above NLNMs of \eqref{Hamiltonian5} can be mapped to critical points of an appropriate Hamiltonian living in the reduced phase space. Morse theoretic methods can then be used to demonstrate the existence of fixed points of the reduced Hamiltonian, or alternately NLNMs of the full Hamiltonian \eqref{Hamiltonian5}. 
An additional family of twelve orbits corresponding to non-critical points of the Hamiltonian, with stabilizer $C_s \wedge \mathcal{T}_2$, can also be shown to exist for the spin-0 sector. The full details of this procedure can be found in \cite{Efstathiou2003}.

\begin{figure}[H]
	\begin{subfigure}[b]{0.32\textwidth}
	\includegraphics[scale=0.3]{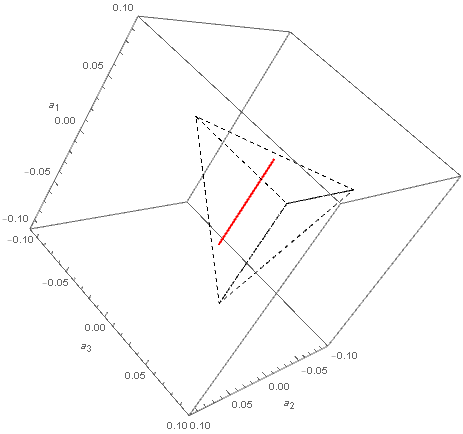}
	\caption{$A_4$}
	\label{ESA4}
	\end{subfigure}
	\begin{subfigure}[b]{0.32\textwidth}
	\includegraphics[scale=0.3]{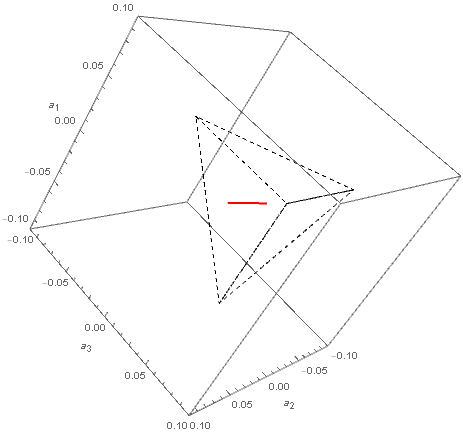}
	\caption{$A_3$}
	\label{ESA3}
	\end{subfigure}
	\begin{subfigure}[b]{0.32\textwidth}
	\includegraphics[scale=0.3]{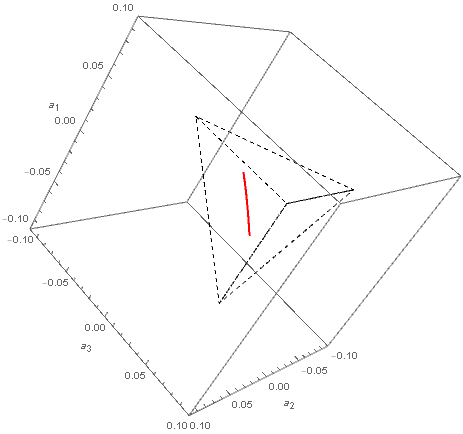}
	\caption{$A_2$}
	\label{ESA2}
	\end{subfigure}
	\\
	\\
	\begin{subfigure}[b]{0.32\textwidth}
	\includegraphics[scale=0.3]{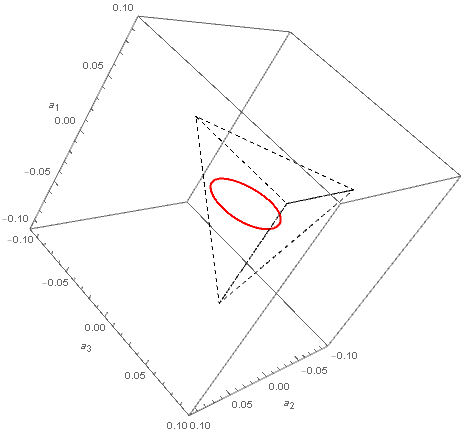}
	\caption{$B_4$}
	\label{ESB4}
	\end{subfigure}
	\begin{subfigure}[b]{0.32\textwidth}
	\includegraphics[scale=0.3]{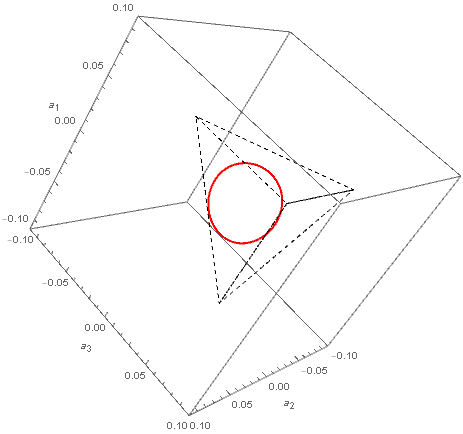}
	\caption{$B_3$}
	\label{ESB3}
	\end{subfigure}
	\begin{subfigure}[b]{0.32\textwidth}
	\includegraphics[scale=0.3]{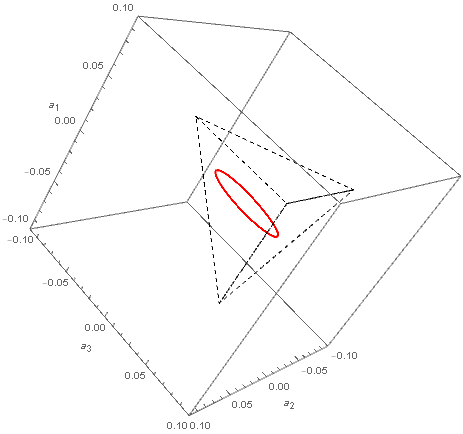}
	\caption{Non-critical Orbit}
	\label{ESnc}
	\end{subfigure}
\caption{Configuration Space Plots of NLNMs}
\label{OrbitsPlots}
\end{figure}

Representative plots for each family of orbits have been shown in Figure~\ref{OrbitsPlots}. 
\subsection{Nested Non-Linearity and Reduced Dynamical Systems}\label{RMMDefn}
One would definitely expect the larger dimensionality of the phase space to present difficulties. 
 Once again, the symmetries of the spin-0 sector come to our aid. They do so by essentially constraining trajectories to lower dimensional subsets of the full phase space. Trajectories constrained in such a manner can then be described by the dynamics of a \textit{reduced} system living on a lower dimensional subset of the full phase space. Happily, it turns out that a thorough study of relevant reduced dynamics is, with some modifications, enough to reproduce several salient features of the \textit{full} six-dimensional model. 
 
  As an example of reduced dynamics, let us consider trajectories with all $a_{i}$'s initially set to a common value $a_{0}$ and all $p_{a_{i}}$'s initially equal to a common value $p_{a_{0}}$. The tetrahedral symmetry of the EOM ensures that these relations will be undisturbed by time evolution. Such trajectories form a subclass of all the possible orbits and are solutions of a reduced system nested in the full model. This reduced system is governed by the dynamical equations
\begin{align}
    \label{RMM2D}
    \dot{a}(t) = p_{a}(t), \quad
    \dot{p}_{a}(t) = -a(t)+3a(t)^{2}-2a(t)^{3},
\end{align}
where, $a/p_{a}$ denotes the common value of the coordinates/momenta. This is simply the dynamics of a particle in the one dimensional double well $V_{DW}(a)=\frac{1}{2}(a(a-1))^{2}$.  Formally, subsets of the phase space which are mapped to (subsets of) themselves by time evolution are referred to as \textit{invariant sets}. We have thus simply identified a two dimensional invariant subset of our model - the set of phase space points with all coordinates equal and all momenta equal. Note that the dynamics in this invariant set is governed by a Hamiltonian, in fact the Hamiltonian obtained by setting coordinates and momenta in \eqref{Hamiltonian5} to a common pair $a,p_{a}$.

In this case, the resulting reduced dynamics is regular, as it should be - the reduced Hamiltonian is two-dimensional and therefore integrable. A far more interesting invariant set is obtained by setting just \textit{two} of the coordinates and their corresponding momenta to common values. Once again, the $T_{d}$ symmetry of the EOM {\eqref{EOM21}-\eqref{EOM24}} render these relations time invariant. Assuming, without loss of generality, that $a_{1}$ serves as the `lone' coordinate, so that $a_{2}=a_{3}=a$ and $p_{a_{2}}=p_{a_{3}}=p_{a}$, the equations governing the reduced dynamics are then
\begin{align}
    \label{4DRMM}
\dot{a}_{1}(t)&=p_{a_{1}}(t),\quad
 \dot{p}_{a_{1}}(t)=-a_{1}(t)(1+2a(t)^{2})-3a(t)^{2}, \\
 \dot{a}(t)&=p_{a}(t),\quad
 \dot{p}_{a}(t)=-a(t)(1+a_{1}(t)^{2}+a(t)^{2})-3a_{1}(t)a(t).
\end{align}

The reduced dynamics in this case resides on a four dimensional subset of the phase space, specifically the subset defined by the relations $a_{2}=a_{3}$ and $p_{a_{2}}=p_{a_{3}}$. 

We shall henceforth distinguish the full six dimensional dynamics from these reduced four dimensional subsystems by referring to the latter as `Reduced Dynamical Systems' (RDSs). In particular, we can choose to fix any two coordinates (and their corresponding momenta) equal to one another and the resulting reduced dynamics for any choice will qualify as an RDS. Since any two choices are related by a symmetry transform, we will fix the convention  $a_{2}=a_{3}=a$ and $p_{a_{2}}=p_{a_{3}}=p_{a}$ for any explicit computations hereafter.

As it turns out, several of the NLNMs are constrained to lie on RDS subspaces. For this reason, a thorough study of the RDSs suffices to explain a good fraction of the full six-dimensional dynamics. Surprisingly, the RDS dynamics \textit{also} have ties to the quantum phases of the spin-0 sector of the $SU(2)$ matrix model, as we shall later see.

\section{Periodic Orbits and their Classification}
\label{sec4}
Having built up the kinematical aspects of the model, we shall now proceed with our analysis in the following three stage fashion:
\begin{enumerate}
\item Enumerate the periodic orbits and understand their geometry and dynamics. This requires some qualification, which we do below.
\item Individually study their stability and destabilization.
\item Correlate the destabilization of these orbits with the generically observed chaotic dynamics.
\end{enumerate} 
 The lack of analytic control inherent to non-linear systems makes it impossible to identify \textit{all} of the periodic orbits. For our purposes however, it will suffice to confine our attention to those orbits whose destabilization has noticeable imprints on the chaotic dynamics. As it turns out, these sets of orbits are composed of NLNMs \textit{and} two sets of orbits that stem from geometric rather than group-theoretic considerations. These two families of `geometric' orbits, along with the NLNMs, can together provide convincing explanations for all the observed peculiarities of the chaotic dynamics, and will thus be the focus of our study.
 We thus begin with an analysis of the various classes of NLNMs, following which we shall briefly explore the origins and properties of the geometric orbits.

\subsection{NLNMs}\label{NLNMs}
We find numerically that all but two families of NLNMs exist only at low energies and are rapidly destroyed as we move away from the integrable regime.  Only the $A_{3}$ and the $A_{4}$ orbits are present at \textit{all} energies (they are protected by their high symmetry), and their stability properties display surprising subtleties. We will elaborate on this in section \ref{StabilityAnalysis}. We will thus devote individual subsections to each of these classes, and follow this up with an enumeration of the basic properties of the remaining NLNMs.
\subsubsection[A4 Orbits]{$A_4$ Orbits}\label{A4Defn}
While the equations of motion (\ref{EOM21}-\ref{EOM24}) are  highly non-linear, \textit{all} non-linear corrections to a given coordinate's evolution involve \textit{only} the remaining two coordinates - there are no non-linear `self interactions'. As a result, setting two of the coordinates to zero at some point in time renders the \textit{instantaneous evolution} of the last coordinate purely harmonic. In fact, by setting their corresponding momenta to zero as well, we can actually `freeze' these coordinates at zero and render the dynamics of the third `lone' coordinate \textit{completely} harmonic. Such trajectories are classified as $A_4$ orbits, and despite their characterization as NLNMs, evolve harmonically with time. Mathematically, the $A_{4}$ orbits evolve as $(a_{i}(t),p_{a_{i}}(t))=(A\sin(t+\phi),0,0,A\cos(t+\phi),0,0)$, or suitable permutations thereof. Individual orbits of the $A_{4}$ type are thus completely specified by an amplitude $A$ (having energy $E=\frac{A^{2}}{2}$)  and a phase $\phi$. $A_{4}$ orbits clearly exist at all energies and, for subcritical energies, are confined to the central lobe of the allowed configuration space. A representative orbit is shown in Figure~\ref{ESA4}.

\begin{figure}[H]
   \centering
     \begin{subfigure}[b]{0.3\textwidth}
         \centering
         \includegraphics[width=\textwidth]{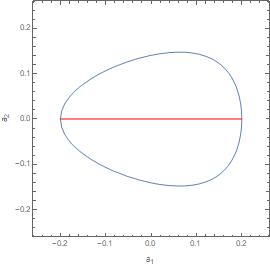}    
     \end{subfigure}
       \caption{RDS Projected $A_{4}$ Orbit}
         \label{A4Red}
\end{figure}

Since the $A_4$ orbits have two coordinates and their corresponding momenta set to zero, they lie on RDS subspaces. More precisely, each RDS possesses harmonic orbits with the common coordinate and the common momentum frozen to zero. The projection of an $A_{4}$ orbit onto the corresponding RDS is shown alongside the relevant constant energy RDS hypersurface in Figure~\ref{A4Red}

\subsubsection[A3 Orbits]{$A_3$ Orbits}\label{A3Defn}
We have already encountered $A_3$ orbits earlier in equation \eqref{RMM2D}. Their dynamics is governed by a double well potential $V_{DW}(a)=\frac{1}{2}(a(a-1))^{2}$. These trajectories and their images under the $T_{d} \times \mathcal{T}$ action are collectively referred to as $A_{3}$ orbits. Initial conditions depicting an $A_{3}$ orbit must thus be of the form $(a_{i},p_{a_{i}})=(a_{0},a_{0},a_{0},p_{a_{0}},p_{a_{0}},p_{a_{0}})$ (or its transform under $T_{d} \times \mathcal{T}$). As with $A_{4}$, individual $A_{3}$ orbits are uniquely specified by the two parameters $a_{0}$ and $p_{0}$ which together fix both the energy of the orbit and a suitable zero reference.  

As solutions to a quartic potential, the $A_3$ orbits are periodic and their time evolution may be expressed in terms of elliptic integrals. Additionally, depending on whether or not the total energy exceeds the `well depth' $\frac{1}{2}$, trajectories either spread across both basins of the double well (the `supercritical' regime) or lie confined to one of the two basins (`the subcritical' case). Correspondingly, the matrix model possesses \textit{subcritical} $A_3$ orbits  for all $E<E_{c}$ that are confined to either the central lobe or one of the side lobes \ref{EqS} and supercritical orbits for all $E>E_{c}$, which live in both central and side lobes. 
 Symmetry considerations tell us that we have eight $A_3$ orbits for any subcritical energy and 4 for any supercritical energy. A representative subcritical orbit is shown in Figure~\ref{ESA3}.

\begin{figure}[H]
   \centering
     \begin{subfigure}[b]{0.3\textwidth}
         \centering
         \includegraphics[width=\textwidth]{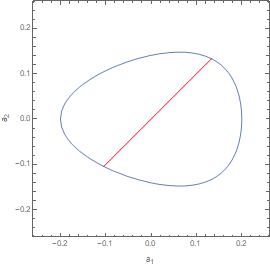}    
     \end{subfigure}
       \caption{RDS Projected $A_{3}$ Orbit}
         \label{A3Red}
\end{figure}

Once again, these orbits can be embedded in RDSs with the lone and common coordinates (and momenta) set equal to one another. The projection of an $A_{3}$ orbit onto the corresponding RDS is shown alongside the relevant constant energy RDS hypersurface in Figure~\ref{A3Red}.
\subsubsection{Other NLNMs}\label{others}
Unlike the $A_{4}$ or $A_{3}$ orbits, the remaining classes of NLNMs exist only for low energies and are rapidly destroyed as we leave the integrable regime. At energies where they \textit{do} exist, initial conditions leading to such orbits can be implicitly specified by relations between the coordinates and momenta derived from \cite{Efstathiou2003}. In the list below, we enumerate the required relations for each class of orbits. We also list the numerically obtained energies at which these orbits cease to exist.
\begin{enumerate}
\item $A_2$:
$a_1=p_{a_1}=0, a_2=a_3, p_{a_2}=p_{a_3}$ and $T_d \times\mathcal{T}$ transformations thereof. These orbits are destroyed at $E\simeq0.001$.
\item $B_4$:
$a_1 = p_{a_1} = 0, a_2 = - p_{a_3}, a_3 = + p_{a_2}$ and $T_d \times\mathcal{T}$ transformations thereof. These orbits are destroyed at $E\simeq0.01$.

\item $B_3$:
$a_2 = \frac{1}{2}\left(-a_1 + \sqrt{3} p_{a_1}\right),
a_3 = \frac{1}{2}\left(-a_1 - \sqrt{3} p_{a_1}\right),
p_{a_2} = \frac{1}{2}\left(- \sqrt{3} a_1 - p_{a_1}\right),
p_{a_3} = \frac{1}{2}\left( \sqrt{3} a_1 - p_{a_1}\right)$ and $T_d \times\mathcal{T}$ transformations thereof. These orbits are destroyed at $E\simeq0.01$.

\item Non-critical NLNMs:
$a_2 = a_1,
a_3 = \sqrt{5} p_{a_1},
p_{a_2} = p_{a_1},
p_{a_3} = - \sqrt{5} a_1$ and $T_d \times \mathcal{T}$ transformations thereof. These orbits are destroyed at $E\simeq0.006$.

\end{enumerate}
Again, individual orbits of each class are uniquely specified by two parameters, which together fix the energy and provide a suitable zero-reference. Representative figures are shown in Figures~\ref{ESA2}-\ref{ESnc}. Amongst these classes of orbits, only the $A_{2}$ and non-critical orbits have two coordinates and their corresponding momenta set to common values and thus possess RDS analogs. The $B_{4}$ and $B_{3}$ orbits, by contrast, are genuinely non-planar NLNMs.
\subsection{Geometric Orbits}
The methods we will utilize for finding geometric orbits was first used in the context of the Henon-Heiles system \cite{Churchill1979ASO}.

As stated earlier, the study of the NLNMs alone is not sufficient for a comprehensive understanding of the dynamics. We also find two families of geometric orbits which do not arise from stabilizer subgroups of the full $T_{d}\times\mathcal{T}$ action. We call them geometric because they emerge from constraints imposed by the requirement of continuity of certain phase space observables over equipotentials of the \textit{RDSs}. In contrast to the NLNMs, the geometric orbits are initially defined over the RDSs and then translated to the full spin-0 sector using a canonical inclusion map. Despite these differences, both the geometric orbits and the NLNMs have their origins in the symmetries of their respective systems. Consequently, we must begin our search for the former by investigating the symmetries of the RDSs.

\begin{figure}[H]
     \centering
         \includegraphics[width=0.4\textwidth]{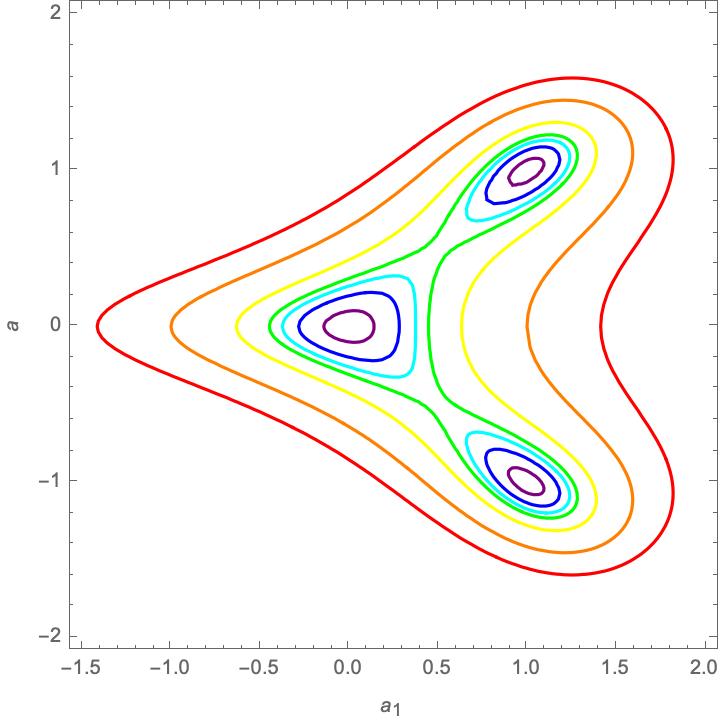}
     \caption{Equipotential Surfaces of an RDS}
     \label{RMMContours}
     \hfill
     \end{figure}

The tetrahedral symmetry of the spin-0 sector reduces to a more modest $\mathbb{Z}_{2}$ symmetry for the RDSs. The sole non-trivial symmetry transformation induced by the action of this reduced symmetry group is, in phase space, simply $a_{1}\rightarrow a_{1}, a\rightarrow -a, p_{a_{1}}\rightarrow p_{a_{1}}, p_{a}\rightarrow -p_{a}$. This abstract action translates to a geometric symmetry of the RDS equipotentials about the $a_{1}$ axis. These equipotentials are described by contours of the form
\begin{equation}
\label{RMMEnergy}
\frac{p_{a_{1}}^{2}}{2}+p_{a}^{2}+\frac{1}{2}(a_{1}^{2}+2a^{2}-6a_{1}a^{2}+2a_{1}^{2}a^{2}+a^{4})=E,
\end{equation}
where the LHS is simply the Hamiltonian \eqref{Hamiltonian5} with the replacements $a_{2},a_{3}\rightarrow a$ and $p_{a_{2}},p_{a_{3}}\rightarrow p_{a}$. The structure of the equipotentials of the spin-0 sector thus directly translate to the equipotentials of the RDSs, which therefore also undergo a topology change at $E_{c}=\frac{3}{32}$. Representative equipotentials are shown in Figure~\ref{RMMContours}. The key to constructing geometric orbits lies in utilizing the symmetries of the equipotentials in conjunction with those of the trajectories. The latter can be neatly formulated in terms of relevant constructs which we term \textit{return maps}. The return maps and the precise algorithms for constructing geometric orbits are outlined in the following subsections.

 Following \cite{Churchill1979ASO}, we will often refer to them as $\Pi_1$ and $\Pi_2$ orbits.
\subsubsection[Pi1 Orbits]{$\Pi_{1}$ Orbits}
The return map required for constructing a $\Pi_{1}$ orbit of energy $E^{0}$ is defined over the surface of the $E^{0}$ equipotential of the RDS. Specifically, given a point $(a_{1}^{0},a^{0})$ on this equipotential, we consider the unique trajectory starting from rest at this point , i.e. $p_{a_{1}}^{0}=p_{a}^{0}=0$. This trajectory, or more accurately its configuration space projection, traces out a curve confined to the interior of the $E^{0}$ equipotential which (in principle) crosses the $a_{1}$ axis, at some time $t_{0}$. The return map $\mathcal{R}$ is defined to output the angle made by the tangent to the curve at $t=t_{0}$ with the $a_{1}$ axis.

\begin{figure}[H]
     \centering
     \begin{subfigure}[b]{0.45\textwidth}
         \centering
         \includegraphics[width=\textwidth]{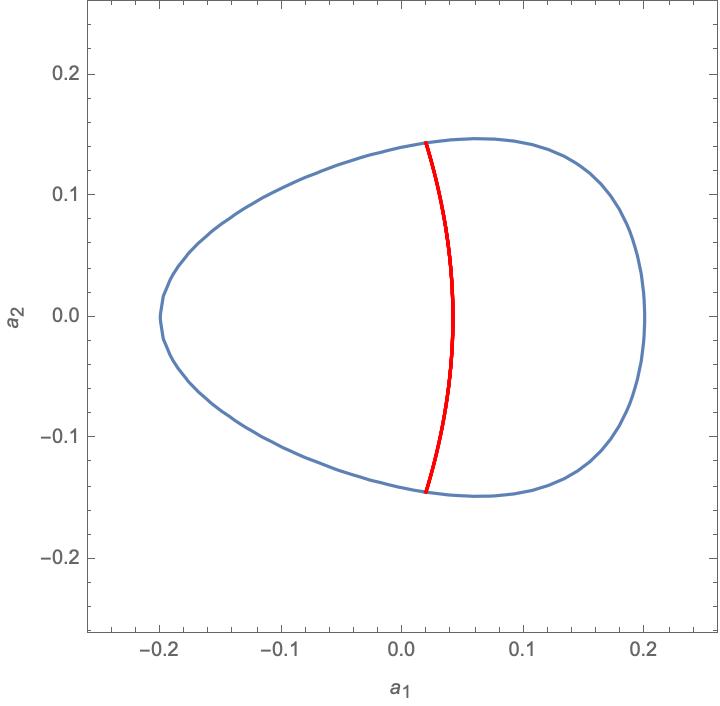}
         \caption{RDS Projection of $\Pi_1$ Orbit}
         \label{Pi1Red}
     \end{subfigure}
     \hfill
     \begin{subfigure}[b]{0.45\textwidth}
         \centering
         \includegraphics[width=\textwidth]{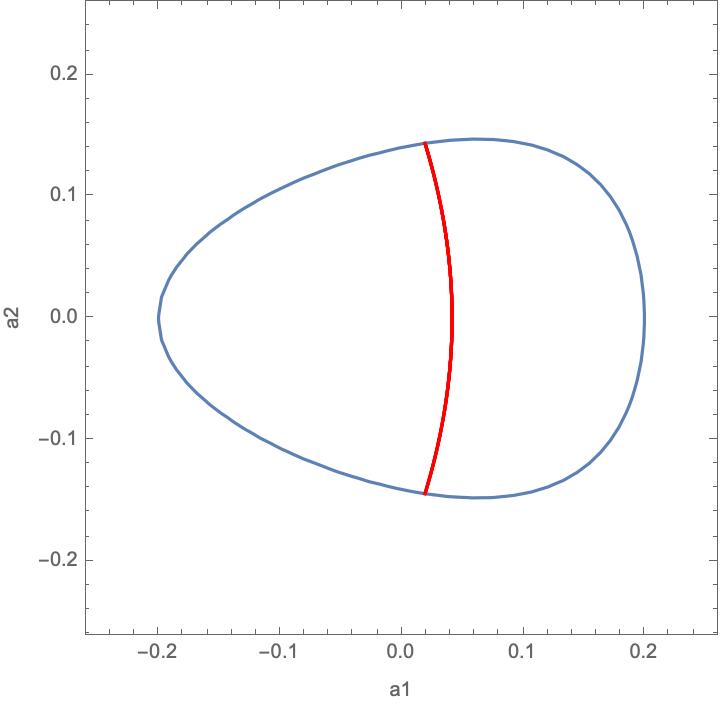}
         \caption{$\Pi_1$ Orbit}
         \label{Pi1}
     \end{subfigure}
     \caption{$\Pi_{1}$ Orbits}
     \label{Pi1Orbits}
     \hfill
     \end{figure}

The crucial observation behind constructing $\Pi_{1}$ orbits can be concisely formulated in terms of the return map. Specifically, \textit{points $Q$ on the equipotential satisfying $\mathcal{R}(Q)=\pm\frac{\pi}{2}$ generate periodic orbits}. This follows from the action of the full symmetry group $\mathbb{Z}_{2}\times\mathcal{T}$. Consequently, the question of generating $\Pi_{1}$ orbits reduces to one of finding solutions to the equation $\mathcal{R}(Q)=\pm\frac{\pi}{2}$. Since we have, for each energy, a pair of $A_{3}$ orbits yielding return map outputs of $\frac{\pi}{4}$ and $\frac{3\pi}{4}$, the intermediate value theorem guarantees at least one solution to the above equation. As it is a trivial task to locate the intersections of the $A_{3}$ orbits with the $E^{0}$ equipotential, we may then use these as reference points to initiate a binary search algorithm to obtain solutions to the above equation. Numerically, we can then establish the existence of a single $\Pi_{1}$ orbit for any energy. These orbits, initially constructed over the RDS phase space, can be trivially extended to the full spin-0 sector. Representative pictures are shown in Figure~\ref{Pi1Orbits}.

\subsubsection[Pi2 Orbits]{$\Pi_{2}$ Orbits}
\begin{figure}[H]
     \centering
     \begin{subfigure}[b]{0.45\textwidth}
         \centering
         \includegraphics[width=\textwidth]{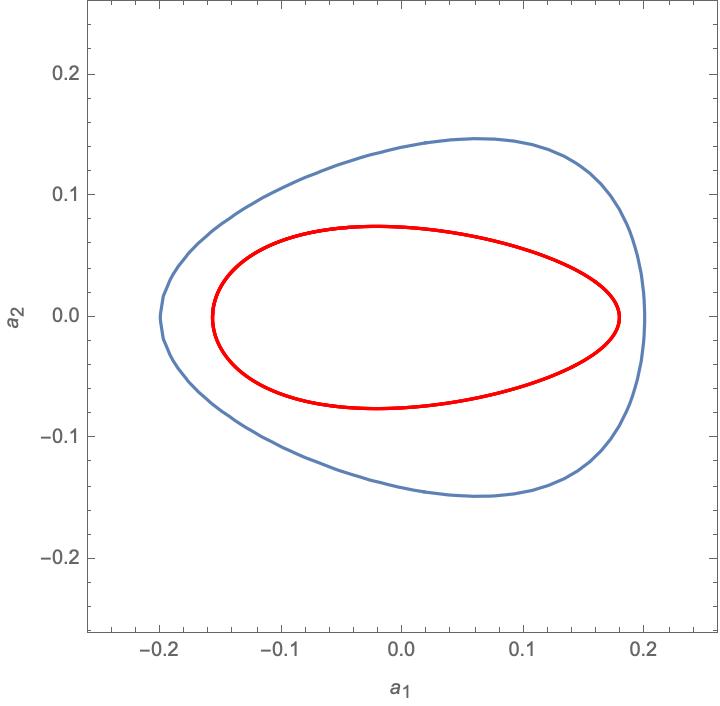}
         \caption{RDS Projection of $\Pi_2$ Orbit}
         \label{Pi2RedPic}
     \end{subfigure}
     \hfill
     \begin{subfigure}[b]{0.45\textwidth}
         \centering
         \includegraphics[width=\textwidth]{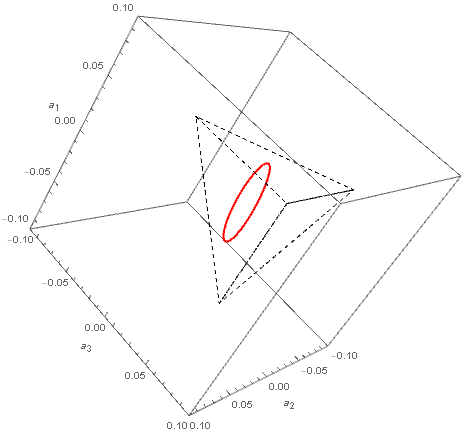}
         \caption{$\Pi_2$ Orbit}
         \label{Pi2Pic}
     \end{subfigure}
     \caption{$\Pi_{2}$ Orbits}
     \label{Pi2}
     \hfill
     \end{figure}

A second set of geometric orbits can be constructed by formulating a different type of return map, essentially the same as our earlier one, but defined over the $a_{1}$ axis rather than over equipotential surfaces. More precisely, given an arbitrary energy $E_{0}$, we consider generic points on the $a_{1}$ axis with $p_{a_{1}}$ set to zero initially and $p_{a}$ fixed by the energy constraint. As before, this trajectory generates a curve whose angle with the $a_{1}$ axis is then captured by this second return map $\mathcal{R}_{2}$. Once again, solutions to the equation $\mathcal{R}_{2}=\pm\frac{\pi}{2}$ yield periodic orbits, this time \textit{closed} orbits in configuration space, which we categorize as $\Pi_{2}$. Again, we can set up binary search methods for numerically solving the generating equation, with reference points being the intersections of the $E_{0}$ equipotential with the $a_{1}$ axis. Unlike the $\Pi_{1}$ orbits however, there are no continuity arguments for justifying the presence of the $\Pi_{2}$ orbits. Indeed, numerical evaluations tell us that the $\Pi_{2}$ orbits cease to exist beyond a threshold energy $E_{\Pi_{2}}\simeq26$, a second unexpected energy scale of the spin-0 sector. That said, $E_{\Pi_{2}}$ lies in the far supercritical regime, so that $\Pi_{2}$ orbits do persist over a good range of energies. Representative orbits are shown in Figure~\ref{Pi2}.

\section{Monodromy Analysis of Periodic Orbits}\label{StabilityAnalysis}
Having enumerated the features of relevant periodic orbits, we will next outline the methods we shall use for assessing their stability. Our strategy rests on the properties of a construct known as the \textit{monodromy matrix} \cite{Teschl2012-co}, which we define below. 

Consider an $n$-dimensional non linear system $\dot{x}(t)=F(x,t)$. Let $x_{p}(t)$ be a periodic solution of this system with time period $T_{p}$. An infinitesimal fluctuation $\delta x(t)$ about $x_{p}(t)$ can be shown to \textit{linearly} evolve as
\begin{equation}
    \label{FlucEqn1}
    \delta \dot{x}(t)= \nabla F\bigl(x_{p}(t)\bigr)\cdot\delta x(t).
\end{equation}
$\nabla F$ is simply the Jacobian $J$ of the transformation $x\rightarrow F(x)$. 

We may also express this evolution in terms of a linear time evolution operator $U(t)$ that maps an arbitrary initial fluctuation $\delta x(0)$ to $\delta x(t)$. $U$ is thus a time dependent square matrix of dimension $n$. The monodromy matrix $\mathcal{U}$ is then simply defined as $\mathcal{U}\equiv U(T_{p})$. In other words, the monodromy matrix tells us what happens to an infinitesimal fluctuation as it cycles the periodic orbit once. 

The eigenvalues of $\mathcal{U}$ yield information on the stability of the periodic orbits \cite{Seydel1987}. Since $\mathcal{U}$ is a real-valued matrix, its eigenvalues must come in complex conjugate pairs. 
A periodic orbit is unstable iff at least one of its eigenvalues lies \textit{strictly} outside the unit circle $\vert z\vert=1$. For Hamiltonian systems, the symplectic structure of the function $F$ can be used to show that the eigenvalues of the corresponding $\mathcal{U}$ come in reciprocal pairs: $\frac{1}{\lambda}$ is an eigenvalue if $\lambda$ is. In addition, Hamiltonian systems \textit{always} have at \textit{least} two unit eigenvalues \cite{Seydel1987}. The corresponding eigenvectors are either directed along the periodic trajectory or connect the periodic trajectory to one of infinitesimally higher/lower energy. 
To summarize, the following properties are inherent to $\mathcal{U}$'s arising from Hamiltonian systems:
\begin{enumerate}
\item At least two eigenvalues are unity.
\item  If $\lambda$ is an eigenvalue, then so are $\frac{1}{\lambda}$ and $\overline{\lambda}$. 
\end{enumerate}

Eigenvalues of $\mathcal{U}$ are usually computed numerically,  since most periodic orbits can only be found numerically to begin with. Analytic results may become available only when we have explicit expressions for the time evolution of the orbit in question. In our case,  it turns out that the symmetry and the analytically tractable time evolution of the $A_{3}$ and $A_{4}$ orbits simplify monodromy computations enormously and some analytic statements \textit{can} be made. 

It is not possible to obtain exact expressions for the time evolution of any of the remaining NLNMs or the geometric orbits. Nevertheless,  the symmetries of the latter and their persistence over a large range of energy endows them with unexpected stability properties which we explore numerically.  In the subsequent subsections, we will thus extensively analyze the stability of the $A_{3}$, $A_{4}$ and $\Pi$ orbits.
\subsection[A4 Orbits]{$A_{4}$ Orbits}
The $A_{4}$ orbits are the simplest to analyze, since their harmonic nature leads to a straightforward time dependence.  With our chosen conventions,  we will work exclusively with $A_{4}$ orbits that have $a_{2}$ and $a_{3}$ frozen to 0,  and $a_{1}$ varying sinusoidally with unit angular frequency.  To find $\mathcal{U}$, we must first set up the equations governing infinitesimal fluctuations about such $A_{4}$ orbits.  An arbitrary fluctuation about a generic trajectory may be quantified by a six-dimensional phase space vector $\delta x(t)\equiv\Bigl(\delta a_{1}(t),\delta p_{a_{1}}(t),\delta a _{2}(t),\delta p_{a_{2}}(t),\delta a_{3}(t),\delta p_{a_{3}}(t)\Bigr)$.  The fluctuation equations \eqref{FlucEqn1} and the functional form of the $A_{4}$ orbits derived in section \ref{A4Defn} then yield
\begin{align}
    \label{FlucEqnsA4}
    \delta \dot{a}_{1}(t)&=\delta p_{a_{1}}(t),\quad\delta \dot{a}_{2}(t)=\delta p_{a_{2}}(t),\quad\delta \dot{a}_{3}(t)=\delta p_{a_{3}}(t),\quad \\
    \delta \dot{p}_{a_{1}}(t)&=-\delta a_{1}(t),\quad\\
    \delta \dot{p}_{a_{2}}(t)&=-\delta a_{2}(t)(1+A^{2}\cos^{2}t)+3A\cos t \ \delta a_{3}(t),\quad \\
    \delta \dot{p}_{a_{3}}(t)&=-\delta a_{3}(t)(1+A^{2}\cos^{2}t)+3A\cos t \ \delta a_{2}(t).
    \label{FlucEqnsA4E}
\end{align}
A complete decoupling can be achieved by the canonical rotation $a_{\pm}\equiv\frac{a_{2}\pm a_{3}}{\sqrt{2}}$. The fluctuation equations then read
\begin{align}
    \label{FlucEqnsA4V2}
    \delta \dot{a}_{1}&(t)=\delta p_{a_{1}}(t),\quad\delta \dot{a}_{+}(t)=\delta p_{a_{+}}(t),\quad\delta \dot{a}_{-}(t)=\delta p_{a_{-}}(t),\quad \\
    \delta \dot{p}_{a_{1}}&(t)=-\delta a_{1}(t),\quad \\
    \delta \dot{p}_{a_{+}}&(t)=-\Bigl(1+A^{2}\cos^{2} t -3A\cos t\Bigr)\delta a_{+}(t),\quad \\
    \delta \dot{p}_{a_{-}}&(t)=-\Bigl(1+A^{2}\cos^{2} t+3A\cos t\Bigr)\delta a_{-}(t).
    \label{FlucEqnsA4V2E}
\end{align}

The geometry of the $A_{4}$ orbits thus naturally induces a separation of perturbations into `longitudinal' modes ($\delta a_{\pm},\delta p_{a_{\pm}}= 0$) and `transverse' modes ($\delta a_{1},\delta p_{a_{1}}=0$). The  fluctuation equations {\eqref{FlucEqnsA4V2}-\eqref{FlucEqnsA4V2E}}  pick out the unique basis in which the two transverse modes decouple from one another. The (almost) identical forms of the equations governing the evolution of $\delta a_{+}$ and $\delta a_{-}$  simply confirm that there is no discernible structural difference between the two transverse modes. 

We must now attempt to make sense of the fluctuation equations {\eqref{FlucEqnsA4V2}-\eqref{FlucEqnsA4V2E}}.  In principle,  we could do this by using these equations to obtain formal expressions for $\mathcal{U}$ and then numerically solve for its eigenvalues.  As it turns out,  the symmetries of the $A_{4}$ orbits heavily simplify the calculations,  so that a full computation of $\mathcal{U}$ is not necessary.

It is useful to view the generic fluctuation equations \eqref{FlucEqn1} as a single \textit{matrix} equation $\delta \dot{x}(t)=J(t)\delta x(t)$. This has the formal solution
 \begin{equation}
    \label{FlucEqnSoln}
    \delta x(t)=T\{e^{\int_{0}^{t}{J(s)}ds}\}\delta x(0),
\end{equation}
where $T$, the time ordering operator, accounts for the non-commutativity of $J$'s evaluated at different times. Since the $A_{4}$ orbits are $2\pi$ periodic, the monodromy matrix $\mathcal{U}$ is simply $T\{e^{\int_{0}^{2\pi}{J(s)}ds}\}$. 

We can obtain explicit expressions for $J$ by reading off its matrix elements from the fluctuation equations {\eqref{FlucEqnsA4V2}-\eqref{FlucEqnsA4V2E}}. The $J$ matrix splits as a direct sum $J(t)=J_{1}(t)\oplus J_{+}(t)\oplus J_{-}(t)$ in the $\{a_{1},p_{a_{1}},a_{+},p_{a_{+}},a_{-},p_{a_{-}}\}$ basis, where 
\begin{align}
\label{JformA4}
J_{1}(t)&=\begin{pmatrix}
0 & 1\\
-1 & 0\\
\end{pmatrix},\\
J_{+}(t)&=\begin{pmatrix}
0 & 1\\
-(1+A^{2}\cos^{2} t-3A\cos t) & 0\\
\end{pmatrix},\\
J_{-}(t)&=\begin{pmatrix}
0 & 1\\
-(1+A^{2}\cos^{2} t+3A\cos t) & 0\\
\end{pmatrix}
\end{align}
Since the matrices $J_{1},J_{+},J_{-}$ lie on different blocks of $J$,  $\mathcal{U}$ also splits as  $\mathcal{U}=\mathcal{U}_{1}\oplus \mathcal{U}_{+}\oplus \mathcal{U}_{-}$, where $\mathcal{U}_{1/+/-}=T\{e^{\int_{0}^{2\pi}J_{1/+/-}(t) dt}\}$ . Since $J_{1}$ is just $i$ times the Pauli matrix $\sigma_{2}$, we obtain $\mathcal{U}_{1}=\mathbb{I}_{2}$.

In fact,  we could have arrived at this result without any calculation whatsoever.  Since the spin-0 sector is a Hamiltonian system,  the $A_{4}$ monodromy matrix must have two eigenvectors of unit eigenvalue,  one describing time-translations along a single $A_{4}$ orbit, and the other connecting the $A_{4}$ orbit in question to one with infinitesimally higher/lower energy.  It is not hard to see that the required eigenvectors are precisely the longitudinal modes: longitudinal fluctuations with $\delta p_{a_{1}}=0$ clearly just shift one's position along a given orbit, while longitudinal fluctuations with $\delta a_{1}=0$ simply changes the momentum slightly. This alters the energy of the trajectory while retaining its identity as an $A_{4}$ orbit.

Thus the non-trivial features of the stability of the $A_{4}$ orbits reside in the $\mathcal{U}_{\pm}$ matrices. We can further simplify using the symmetry between $\delta a_{+}$ and $\delta a_{-}$. Since $J_{-}(t+\pi)=J_{+}(t)$ and the integral of a periodic function over a single period is independent of the lower limit of integration, we have
\begin{equation}
\label{4Dto2DComp}
 \mathcal{U}_{+}=T\{e^{\int_{0}^{2\pi}J_{+}(t) dt}\}=T\{e^{\int_{\pi}^{3\pi}J_{+}(t) dt}\}=T\{e^{\int_{\pi}^{3\pi}J_{-}(t+\pi) dt}\}=T\{e^{\int_{0}^{2\pi}J_{-}(t) dt}\}=\mathcal{U}_{-}.
\end{equation}
The last equality makes use of the substitution $t\rightarrow t+\pi$. Thus, while the blocks $J_{+}(t)$ and $J_{-}(t)$ differ in form, their time ordered integrals are exactly the same.  As a result, we may confine our attention to either one of the transverse modes. 

There exists a final simplification. Since $\mathcal{U}_{+}=\mathcal{U}_{-}$ and eigenvalues of $\mathcal{U}$ must come in conjugate pairs and reciprocal pairs, we can constrain its spectrum to be of the form $\{1,1,\mu,\lambda,\mu,\lambda\}$
, where $\mu$ and $\lambda$, the eigenvalues of $\mathcal{U}_{+}$ (or $\mathcal{U}_{-})$, must satisfy either of the two following conditions:
\begin{enumerate}
    \item \textbf{$\mu$ and $\lambda$ are real:} In this case, we have $\mu=\frac{1}{\lambda}$. Barring the trivial cases  $\mu=\lambda=\pm1$, either $\mu$ or $\lambda$ will lie outside the unit circle, leading to an unstable orbit. So $\vert\lambda+\mu\vert=\vert\lambda+\frac{1}{\lambda}\vert>2$. 
    \item \textbf{$\mu$ and $\lambda$ are complex conjugates:}  Now we have $\lambda=\frac{1}{\mu}=\overline{\mu}$, so that $\vert\mu\vert=\vert\lambda\vert=1$. Barring the trivial cases  $\mu=\lambda=\pm1$, $\mu$ and  $\lambda$ are thus complex conjugates lying \textit{on} the unit circle, resulting in a stable orbit. In this case, we may represent the pair $\{\mu,\lambda\}$ as $\{e^{i\theta},e^{-i\theta}\}$ for some $\theta$ in $(0,2\pi)$ so that $\vert\mu+\lambda\vert=2\vert\cos\theta\vert<2$.\end{enumerate}
 Thus, we see that the (in)stability of any $A_{4}$ orbit is beautifully captured by \textit{a single number}: $\gamma\equiv\mu+\lambda$. Explicitly, the orbit is stable (unstable) depending on whether $\vert\gamma\vert<2 \,(>2)$ with transitions occurring when $\vert\gamma\vert=2$. Note that in terms of $\mathcal{U}$, we have $\gamma = \frac{\text{Tr}\,\mathcal{U}-2}{2}$
  
The stability of a periodic orbit is thus captured by \textit{trace} of $\mathcal{U}$, rather than its full spectrum. This is a standard feature of four-dimensional Hamiltonian systems \cite{Brack2001}. No such simplifications exist for higher dimensional systems. We again emphasise that it is the special symmetries of the $A_{4}$ orbits (and more generically the Hamiltonian of the spin-0 sector) \eqref{Hamiltonian5} that have produced this extreme simplification.

\begin{figure}[H]
     \centering
         \includegraphics[width=0.4\textwidth]{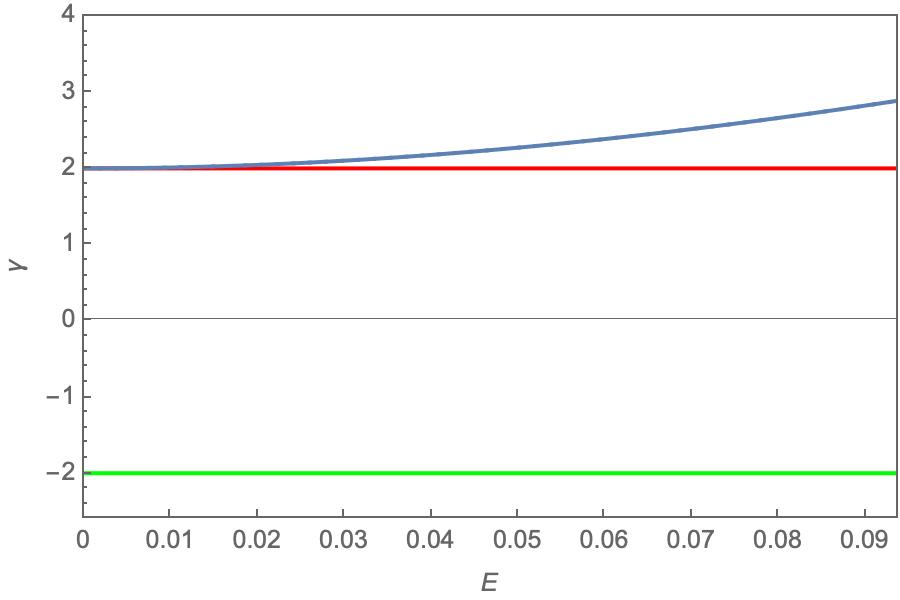}
     \caption{$\gamma$ vs $E$ for subcritical $A_{4}$ orbits}
     \label{A4Sub}
     \hfill
     \end{figure}

Having substantially simplified our computations, we now turn to numerics.  We compute $\gamma$ as a function of energy  in the range $E\in(0,500)$.  Figure~\ref{A4Sub} depicts  $\gamma$ as a function of energy $E$ in the region $0<E<E_{c}$. The key takeaway is that $\gamma$ never dips below 2, so that subcritical $A_{4}$ orbits are unstable without exception. Additionally, the increase of $\gamma$ with $E$ suggests an increase in the `amount of instability'. This notion is indeed true, and can be precisely quantified by chaos theory measures, such as Lyapunov exponents, which we will analyze in section \ref{LE}.

\begin{figure}[H]
     \centering
     \begin{subfigure}[b]{0.4\textwidth}
         \centering
         \includegraphics[width=\textwidth]{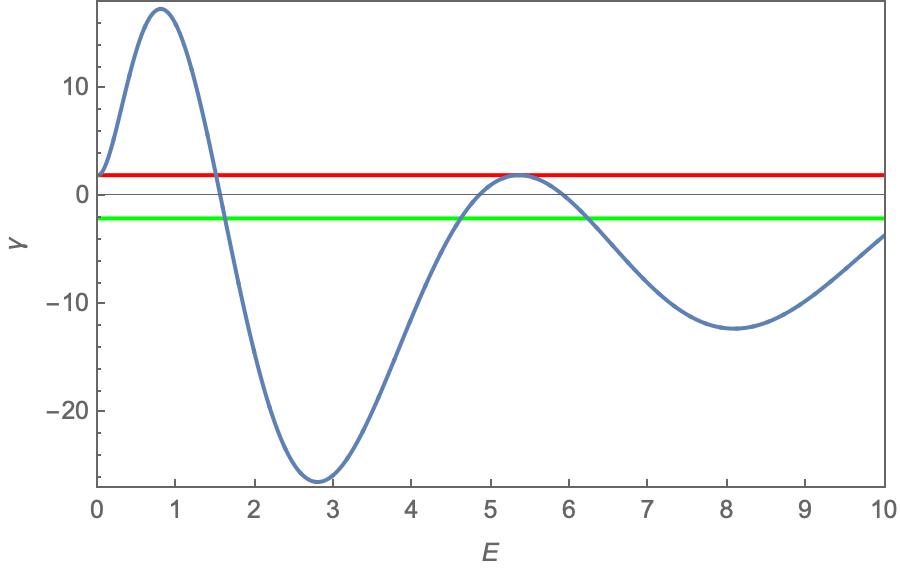}
     \end{subfigure}
     \hfill
     \begin{subfigure}[b]{0.4\textwidth}
         \centering
         \includegraphics[width=\textwidth]{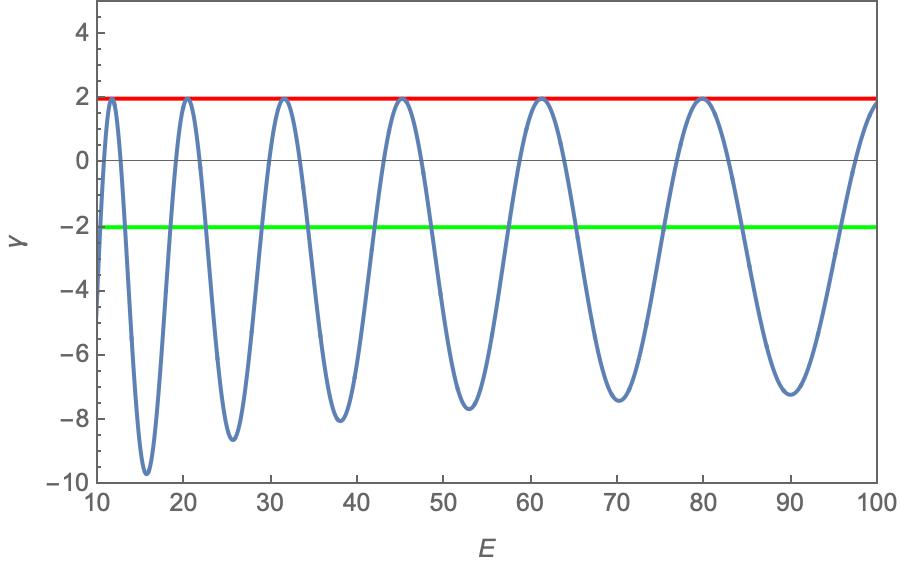} 
     \end{subfigure}
      \hfill
       \begin{subfigure}[b]{0.4\textwidth}
         \centering
         \includegraphics[width=\textwidth]{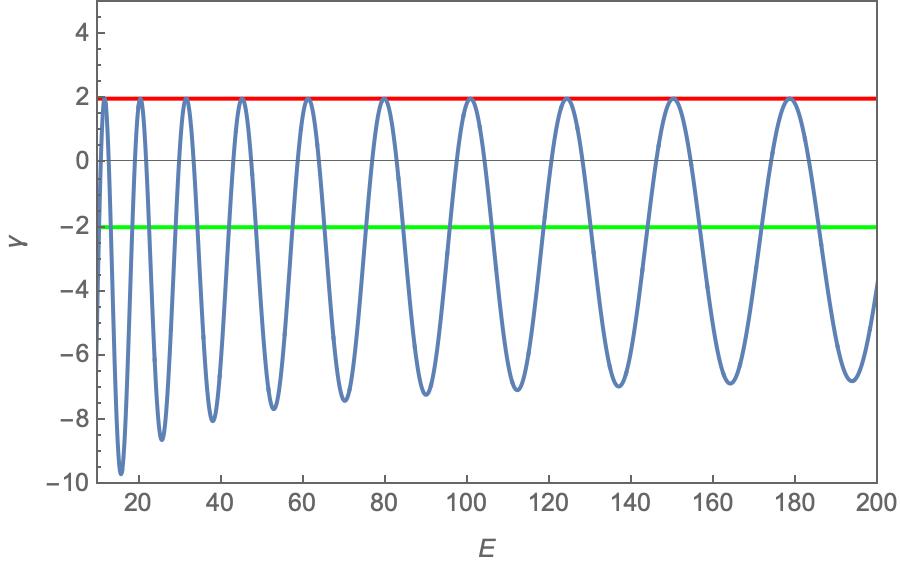}
     \end{subfigure}
         \hfill
     \begin{subfigure}[b]{0.4\textwidth}
         \centering
         \includegraphics[width=\textwidth]{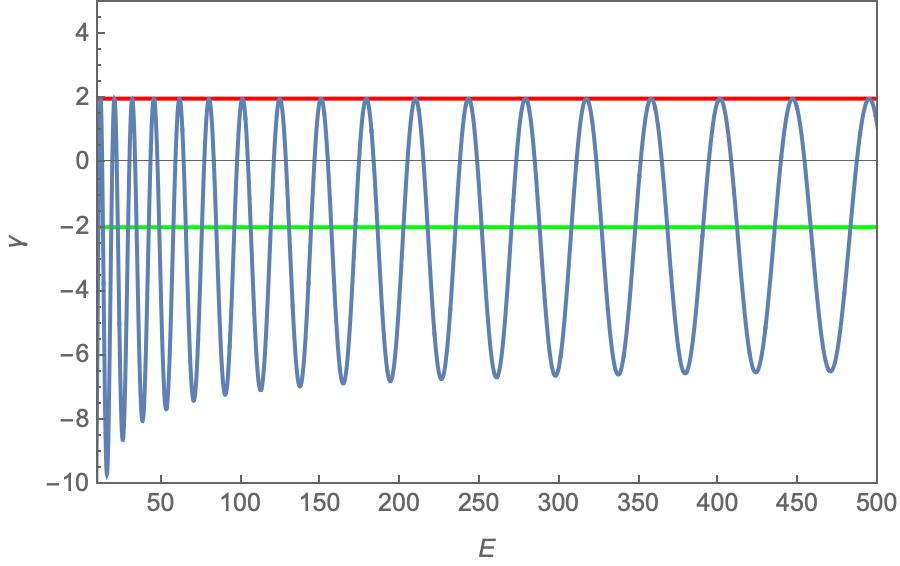}  
     \end{subfigure}
        \caption{$\gamma$ vs $E$ for supercritical $A_4$ orbits}
     \label{A4Sup}
\end{figure}

Our results for subcritical $A_{4}$ orbits are not surprising, as one would expect heightened instabilities with increasing energies. The supercritical regime displays a much more surprising behaviour,  as is clear from Figure~\ref{A4Sup}. From these plots,  we see that the stability of supercritical $A_{4}$ orbits is characterized by \textit{oscillations between stability and instability} with a monotonically decreasing frequency.  These transitions seem to repeat ad infinitum.  Curiously, stability plots of a very similar nature have been observed in literature,  albeit in the seemingly unrelated context of solitonic solutions of the non-linear Schr\"{o}dinger equation \cite{https://doi.org/10.48550/arxiv.2005.12708}. The connections between such themes and our gauge matrix model need to be better understood.

\begin{figure}[H]
     \centering
     \begin{subfigure}[b]{0.4\textwidth}
         \centering
         \includegraphics[width=\textwidth]{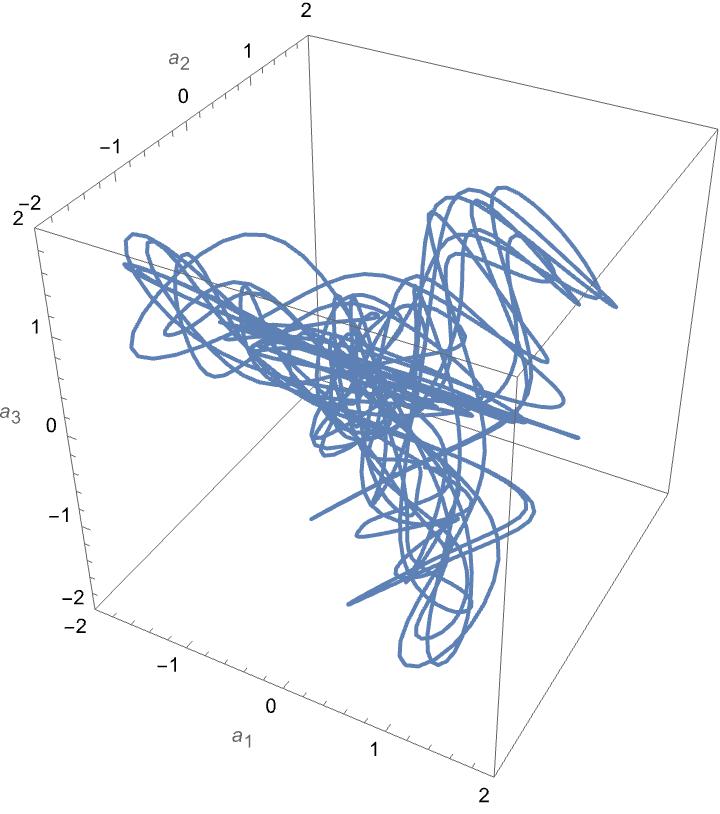}
             \caption{$E=1.499$}
     \end{subfigure}
     \hfill
         \begin{subfigure}[b]{0.4\textwidth}
         \centering
         \includegraphics[width=\textwidth]{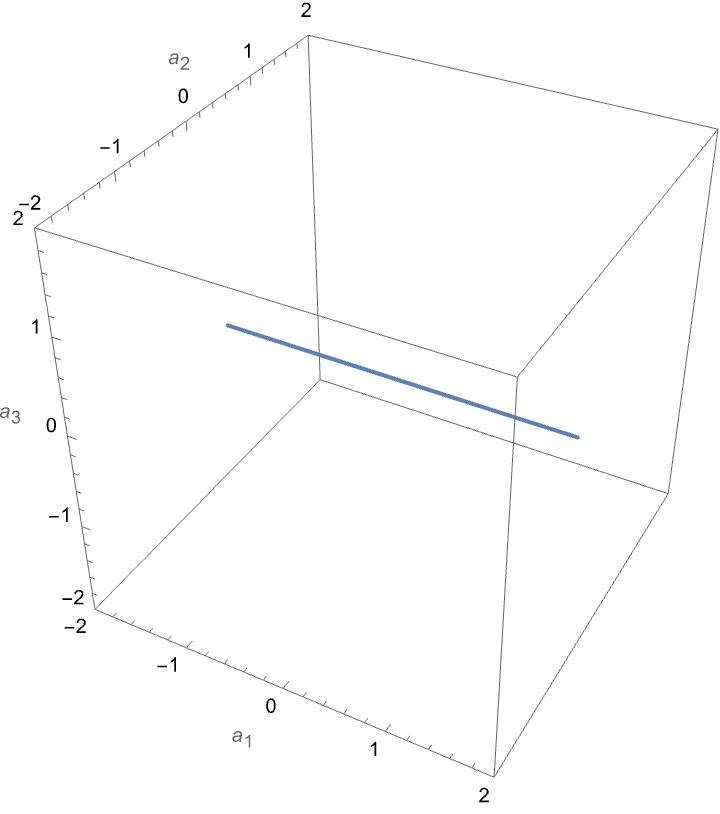}
         \caption{$E=1.501$}
     \end{subfigure}
        \caption{Unstable to stable flip for $A_4$}
     \label{PPA4_1}
\end{figure}

\begin{figure}[H]
     \centering
     \begin{subfigure}[b]{0.4\textwidth}
         \centering
         \includegraphics[width=\textwidth]{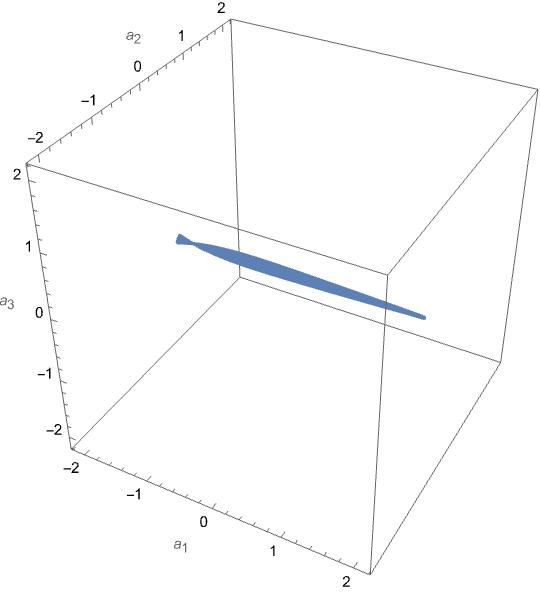}
             \caption{$E=1.628$}
     \end{subfigure}
     \hfill
         \begin{subfigure}[b]{0.4\textwidth}
         \centering
         \includegraphics[width=\textwidth]{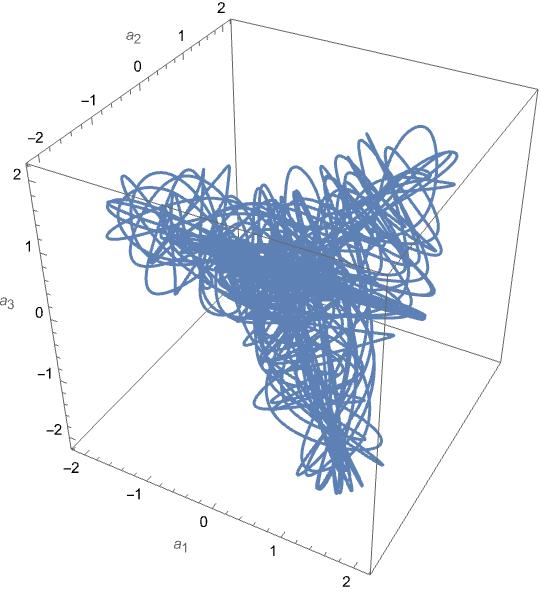}
         \caption{$E=1.629$}
     \end{subfigure}
        \caption{Stable to unstable flip for $A_4$}
     \label{PPA4_2}
\end{figure}

Before seeking analytic explanations for these transitions, it must be noted that the $\gamma$-$E$ plots are just \textit{one} of many signatures of these stability flips. Indeed, we shall encounter more signatures as we proceed with our analysis. One particular signature, however, is worthy of immediate attention. Since the $a_{i}$'s are after all the fundamental observables of our theory, it is natural to look for the imprints of these stability flips on their time evolution. Parametric plots of trajectories in configuration space provide a beautiful way to illustrate these effects.  We construct configuration space parametric plots at energies marginally above and marginally below a transition energy, with initial conditions deviating very slightly from the initial conditions required for relevant $A_{4}$ orbits. The results are displayed for the first transition point $E=\frac{3}{2}$ (we will derive this value later) in Figure~\ref{PPA4_1}.  We see that energies \textit{marginally} above $E=\frac{3}{2}$  yield perfectly regular trajectories barely distinguishable from their $A_{4}$ parent orbits, while energies \textit{marginally} below the transition point yield chaotic trajectories which rapidly fill a sizeable fraction of the available configuration space. Figure \ref{PPA4_2} shows an analogous flip in stability in the opposite direction (stable below the transition energy, unstable above it). Note, in this case, that the transition point is approximate.

We shall now use the fluctuation equations \eqref{FlucEqnsA4} to better understand the stability and establish some quantitative results. It is useful to eliminate the $\delta p$'s from the fluctuations equations and regard them as second order in $\delta a$'s. The equivalence of $a_{+}$ and $a_{-}$ fluctuations means that we may restrict our studies to just one of these modes. Without loss of generality, we choose to work with the $a_{+}$ modes,  whose fluctuation equation reads
\begin{equation}
    \label{Hillb2}
    \delta \ddot{a}_{+}(t)+\bigl(1-3A\cos t+A^2\cos^{2}t\bigr)\delta a_{+}(t)=0
\end{equation}
Rescaling $t=2s$, we obtain 
\begin{equation}
    \label{WHb3}
      \delta \ddot{a}_{+}(s)+\bigl(\eta+2\alpha\cos 2s+2\beta\cos 4s \bigr)\delta a_{+}(s)=0,
\end{equation}
 with $\alpha=-6A,\beta=A^{2}$ and $\eta=4+2A^{2}$,  which is the standard form of the Whittaker-Hill (WH) equation (see \cite{1964} for example).  
 
This WH equation has exactly the form of a Schrodinger equation with a periodic potential, typically encountered in Bloch theory of solids. Hence we expect to see a band structure with bands and band gaps corresponding to stability snd instability.
 
Floquet theory tells us that \textit{any} solution to the WH equation can be expressed in the form $e^{px}g(x)$ for some complex number $p$ and a periodic function $g(x)$. This is usually about as far as we can go, as closed form expressions are generally not available.  However, we can make progress towards finding the locations of transition points, as this requires a study of only the \textit{periodic} solutions to the WH equation. This is because $\vert\gamma\vert= 2$ corresponds to  $\mathcal{U}_{\pm}$ being $\pm\mathbb{I}_{2}$ which in turn leads to periodic behaviour of the fluctuations. 

The WH equation is usually solved by an expansion into a sine or cosine series followed by solving recursion relations that emerge between the Fourier coefficients. As such an approach no doubt reminds the reader of the more common Frobenius methods, it is natural to question whether we can carry over techniques from power series expansions to our case. In particular, since Frobenius type problems often have parameter choices that lead to \textit{finite} termination of the recursion series, we may naturally wonder whether such truncations are possible for the WH equation too. This is unfortunately not the case as the pertinent recurrence relations involve \textit{five} coefficients at a time. However, a remarkable transformation, $\delta z(s)\equiv \delta a_{+}(s)e^{\sqrt{\beta}\cos2s}$, of our WH equation yields the differential equation \cite{1964,magnus2013hill}
\begin{equation}
\label{IE}
    \delta \ddot{z}(s)+4\sqrt{\beta}\sin2s\,\delta \dot{z}(s)+[\eta+2\beta+\bigl(2\alpha+4\sqrt{\beta}\bigr)\cos 2s]\,\delta z(s)=0,
\end{equation}
 the \textit{Ince equation}, which can be solved by \textit{three} term recursions. If $\beta=\frac{\alpha^{2}}{4(p+1)^{2}}$ for some $p\in\mathbb{Z}^{+}$, $\eta$ can be chosen in order to make the recursion relation eventually terminate. In such situations, the Ince equation possesses finite series solutions, known as Ince polynomials, which can then be recast, via the $\delta z \rightarrow\delta a_{+}$ transform, to closed periodic solutions (though not \textit{polynomial solutions}) of the WH equation.
In our case, the coefficients $\alpha, \beta$ and $\eta$ are \textit{additionally} constrained to be related to one another via the amplitude $A$. It turns out that $\alpha$ and $\beta$ indeed satisfy the necessary relations for finite solutions, with $p=2$. However, the restrictions on $\eta$ only grant us finite solutions for \textit{one} value: $A=\sqrt{3}$. This corresponds to a stability flip at $E=\frac{A^{2}}{2}=\frac{3}{2}$. 
 The corresponding Ince polynomial can be worked out to be $1+\frac{2}{\sqrt{3}}\cos2s$. Reverting to the WH equation, we obtain 
 \begin{equation}
     \label{WHSoln3/2}
     \delta a_{+}(s)=\left(1+\frac{2}{\sqrt{3}}\cos2s\right)e^{-\sqrt{3}\cos2s}.
 \end{equation}
 A second, linearly independent periodic solution for this equation for the WH equation can be obtained using the well-known variation of parameters method. Suitably applied to our case, this method tells us that if $w_{1}(s)$ is a solution to the WH equation at $E=\frac{3}{2}$, then so is $w_{1}(s)\int_{0}^{s}{(1/w_{1}(t)^{2})dt}$. We may thus write a second independent solution to the WH equation at $E=\frac{3}{2}$ in quadrature form as 
 \begin{equation}
     \label{WHSoln3/22}
    \delta a_{+}(s)=\left(1+\frac{2}{\sqrt{3}}\cos2s\right)e^{-\sqrt{3}\cos2s}\int_{0}^{s}{\frac{e^{2\sqrt{3}\cos2s}}{\left(1+\frac{2}{\sqrt{3}}\cos2s\right)^2}ds}  
 \end{equation}
 
 This integral unfortunately cannot be evaluated in terms of elementary functions, but we nevertheless have a passable inventory of the solutions of the WH equation at this energy. The numerically obtained $\gamma-E$ plots confirm that $E=\frac{3}{2}$ is indeed a transition point.
%

While we cannot evaluate the precise locations of any other transition points, it is possible to ascertain their asymptotic behaviour. We see that at large enough energies, the coefficient of $\cos2s$ in \eqref{WHb3} dies out far more rapidly (as a function of $A$) than either of the other two coefficients. So we can derive asymptotic expressions for transition points by neglecting this term in the large $A$ limit. Reverting back to $t=2s$, we then see that the WH equation reduces to the far simpler Mathieu equation
\begin{equation}
    \label{MEqn1}
    \delta \ddot{a}_{+}(t)+\left(1+\frac{A^{2}}{2}+\frac{A^{2}}{2}\cos2t\right)a_{+}(t)=0.
\end{equation}
Our problem now simplifies to studying the periodic solutions of the \textit{Mathieu} equation (see \cite{1964} for example). While still non-trivial, this is at least a well documented problem with at least a few known simple analytical results. In general, the Mathieu equation in its standard form
\begin{equation}
    \label{MEqn2}
    \delta \ddot{y}(x)+\bigl(a-2q\cos2x\bigr)y(x)=0
\end{equation}
has periodic solutions only for  special set of parameter values $(a,q)$. These sets are described by two \textit{Mathieu characteristic functions}, a pair of functions defined from $\mathbb{Z}\times\mathbb{R}$ to $\mathbb{R}$ that take in a pair $(n,q)$ and yield a unique value for $a$  that in turn gives the $n\textsuperscript{th}$ (odd/even) Mathieu function as a periodic solution to the Mathieu equation with parameter set $(a,q)$. Given that $(a,q)=(1+\frac{A^{2}}{2},-\frac{A^{2}}{4})$ for us, we see that the locations of the $n\textsuperscript{th}$ family of transition points are asymptotically given by solutions to the equations
\begin{equation}
    \label{TPLoc1}
    \xi_{1}(n,-\frac{A^{2}}{4})=1+\frac{A^{2}}{2}
\end{equation}
and 
\begin{equation}
    \label{TPLoc2}
    \xi_{2}(n,-\frac{A^{2}}{4})=1+\frac{A^{2}}{2}
\end{equation}
where $\xi_{1,2}$ are the Mathieu characteristic functions of the first/second kind. 
Recasting the above equations in terms of the energy $E$, we obtain
\begin{equation}
    \label{TPLoc3}
    \xi_{1}(n,-\frac{E}{2})=1+E
\end{equation}
and 
\begin{equation}
    \label{TPLoc4}
    \xi_{2}(n,-\frac{E}{2})=1+E
\end{equation}

Transition points computed in this manner can be compared with numerically obtained results (see Appendix~\ref{A4Results}), and we observe excellent agreement between the two sets of values.

\subsection[A3]{$A_3$ Orbits}
We consider the $A_{3}$ orbits specified by initial conditions of the form $(a_{0},a_{0},a_{0},p_{a_{0}},p_{a_{0}},p_{a_{0}})$. The fluctuation equations for $A_{3}$ orbits are
\begin{align}
    \label{FlucEqnsA3}
    \delta \dot{a}_{1}(t)& =\delta p_{a_{1}}(t), \quad\delta \dot{a}_{2}(t) =\delta p_{a_{2}}(t),\quad\delta \dot{a}_{3}(t) =\delta p_{a_{3}}(t),  \\
    \delta \dot{p}_{a_{1}}(t) &=-\delta a_{1}(t)(1+2a(t)^{2}) +(3a(t)-2a(t)^{2})(\delta a_{2}(t) +\delta a_{3}(t)),    \\
    \delta \dot{p}_{a_{2}}(t) &=-\delta a_{2}(t)(1+2a(t)^{2}) +(3a(t)-2a(t)^{2})(\delta a_{3} (t)+\delta a_{1}(t)), \\
    \delta \dot{p}_{a_{3}}(t) &=-\delta a_{3}(t)(1+2a^{2}) +(3a(t)-2a(t)^{2})(\delta a_{1}(t) +\delta a_{2}(t)),
        \label{FlucEqnsA3F}
\end{align}
which decouple in the canonical basis $\{b_{1},b_{2},b_{3}\}=\{\frac{a_{1}+a_{2}+a_{3}}{\sqrt{3}},\frac{a_{1}-a_{2}}{\sqrt{2}},\frac{a_{2}-a_{3}}{\sqrt{2}}\}$ into three pairs of independent equations:
\begin{align}
    \label{FlucEqnsA3DeC}
    \delta \dot{b}_{1}(t) &=\delta p_{b_1}(t),  \quad \delta \dot{b}_{2}(t) =\delta p_{b_2} (t), \quad\delta \dot{b}_{3}(t) =\delta p_{b_3}(t),  \\
    \delta \dot{p}_{b_1}(t) &=-(1+6a(t)^{2}-6a(t))\delta b_{1}(t), \\
    \delta \dot{p}_{b_2}(t) &=-(1+3a(t))\delta b_2(t), \\
    \delta \dot{p}_{b_3}(t) &= -(1+3a(t))\delta b_3(t).
        \label{FlucEqnsA3DeC2}
\end{align}
We thus obtain once more a block decomposition of the full monodromy matrix $\mathcal{U}$ into three blocks $\mathcal{U}_{1/2/3}$. Denoting the matrices corresponding to the $b$ variables by $J_{1/2/3}$, we see that $J_{1}$ (and consequently $\mathcal{U}_{1}$) describes the evolution of fluctuations \textit{along} the orbit. We therefore do not expect any non trivial results from this sector. The $J_{2}$ and $J_{3}$ matrices are \textit{manifestly} equal. As a result, the full simplification of the previous subsection carries through for the $A_3$ orbits as well. We need only study $\gamma$, the trace of the $\mathcal{U}_{2} \,(=\mathcal{U}_{3})$ matrix.  

From a physical standpoint, we thus expect the same trends as were observed for the $A_4$ orbits. Since our perturbations once again take the form of time-independent Schrodinger equation characterized by a periodic potential (in this case an elliptic integral), we anticipate alternating \textit{bands} of stability and instability. Indeed, we find that we have oscillations between stability and instability, with the separation between adjacent transition points varying geometrically as we approach the critical energy $E_{c}$ from either side.

\begin{figure}[H]
     \centering
     \begin{subfigure}[b]{0.4\textwidth}
         \centering
         \includegraphics[width=\textwidth]{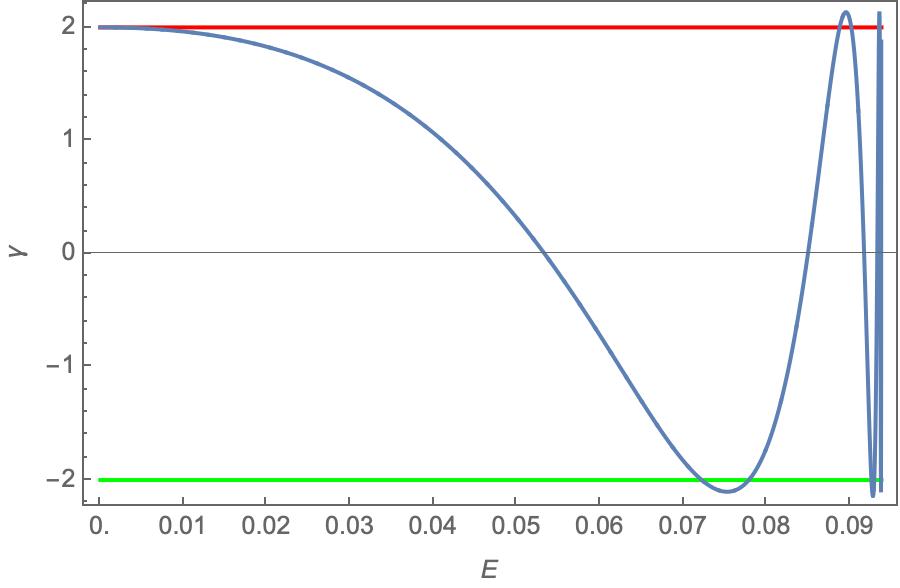}
     
     \end{subfigure}
     \hfill
     \begin{subfigure}[b]{0.4\textwidth}
         \centering
         \includegraphics[width=\textwidth]{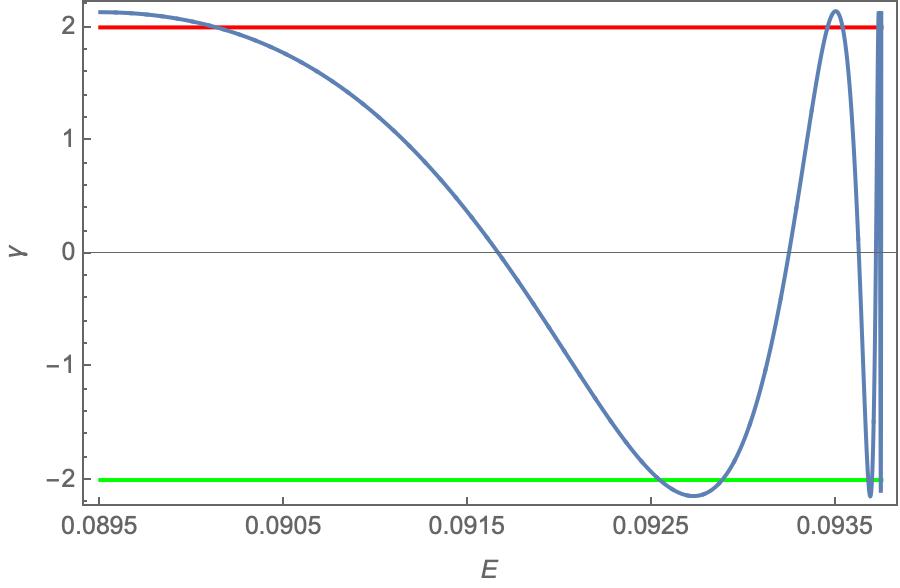}
  
     \end{subfigure}
     \hfill
     \begin{subfigure}[b]{0.4\textwidth}
         \centering
         \includegraphics[width=\textwidth]{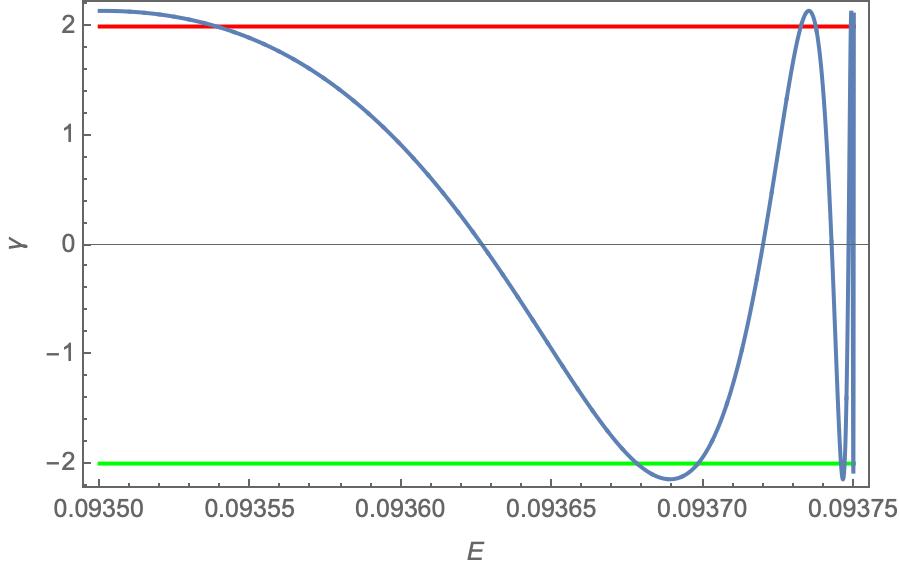}
      
     \end{subfigure}
      \hfill
       \begin{subfigure}[b]{0.4\textwidth}
         \centering
         \includegraphics[width=\textwidth]{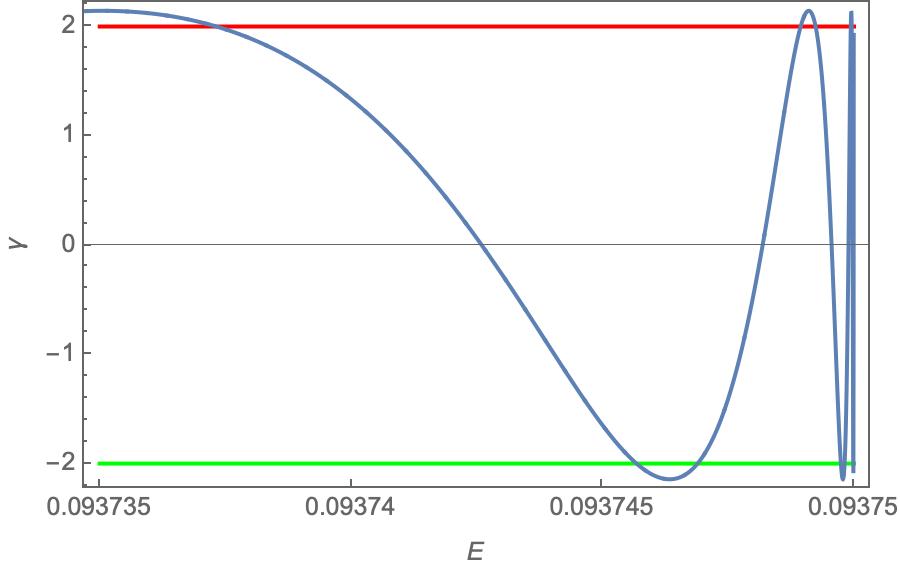}
     \end{subfigure}
        \caption{$\gamma$ vs $E$ for subcritical $A_3$ orbits}
     \label{A3SubCrit}
\end{figure}

\begin{figure}[H]
     \centering
     \begin{subfigure}[b]{0.4\textwidth}
         \centering
         \includegraphics[width=\textwidth]{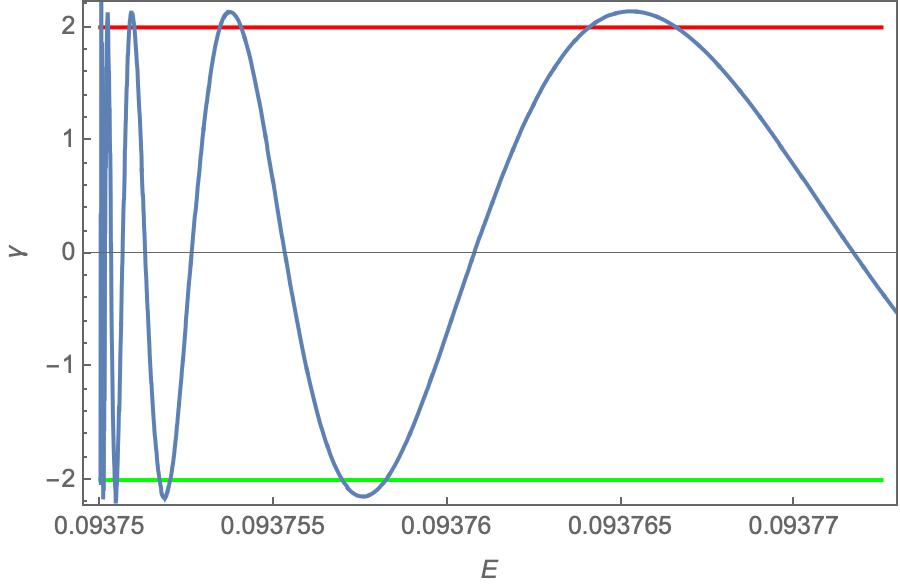}
     \end{subfigure}
     \hfill
     \begin{subfigure}[b]{0.4\textwidth}
         \centering
         \includegraphics[width=\textwidth]{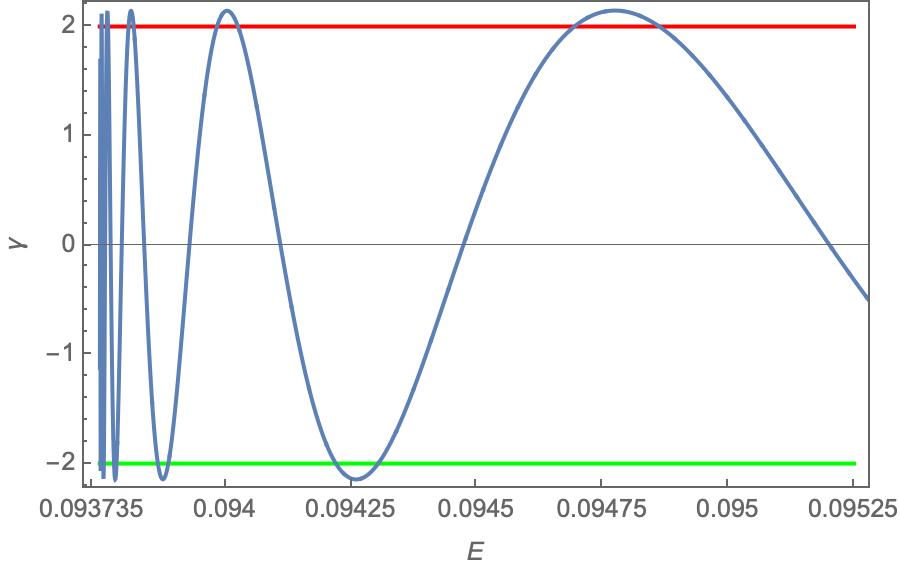}
     \end{subfigure}
     \hfill
     \begin{subfigure}[b]{0.4\textwidth}
         \centering
         \includegraphics[width=\textwidth]{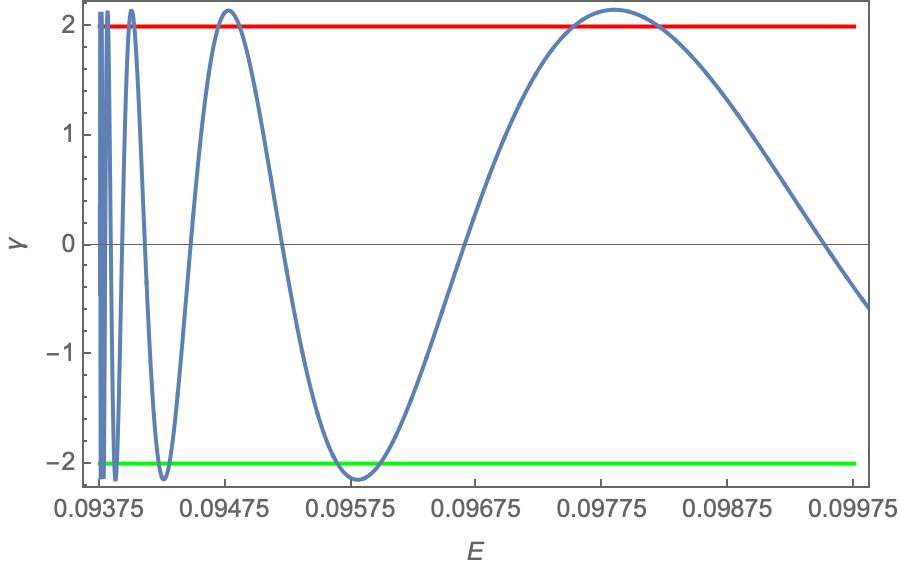}
     \end{subfigure}
      \hfill
       \begin{subfigure}[b]{0.4\textwidth}
         \centering
         \includegraphics[width=\textwidth]{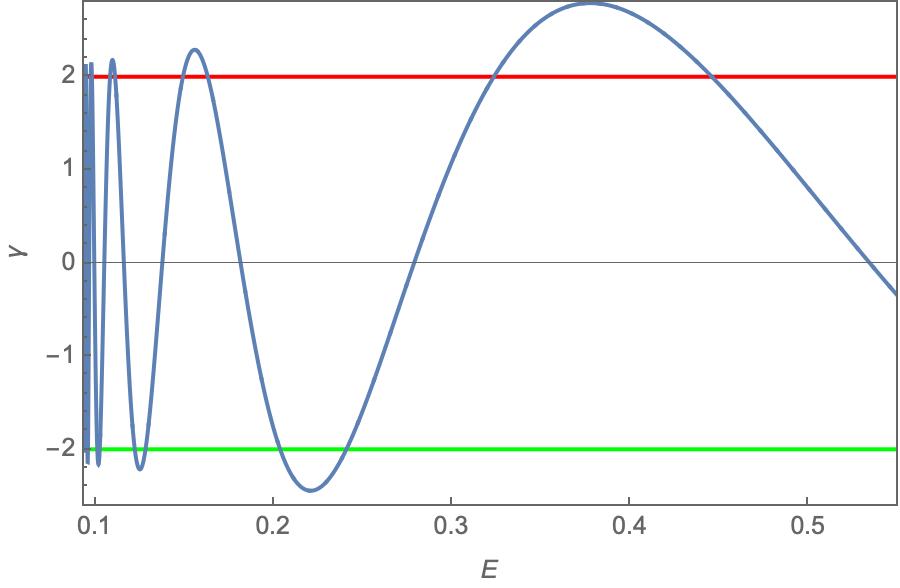}
     \end{subfigure}
        \caption{$\gamma$ vs $E$ for supercritical $A_3$ orbits}
       \label{A3SupCrit}
\end{figure}

Specifically, we will show that
\begin{enumerate}
\item For subcritical energies, the quantities $1-\frac{E^{1}_{n}}{E_{c}}$ and $1-\frac{E^{2}_{n}}{E_{c}}$  where $E^{1}_{n}(E^{2}_{n})$ are the energies corresponding to the $n\textsuperscript{th}$ transition of $\gamma$ from $2^{+}$ to $2^{-}$($2^{-}$ to $2^{+}$) form a geometric series, with common ratio $\delta_{1}=e^{\frac{\pi}{2\sqrt{5}}}$ as we approach $E_{c}$ from below.
\item For supercritical energies, the quantities $1-\frac{E^{1}_{n}}{E_{c}}$ and $1-\frac{E^{2}_{n}}{E_{c}}$  where $E^{1}_{n}(E^{2}_{n})$ are the energies corresponding to the $n\textsuperscript{th}$ transition of $\gamma$ from $2^{+}$ to $2^{-}$($2^{-}$ to $2^{+}$) form a geometric series, with common ratio $\delta_{2}=e^{\frac{\pi}{\sqrt{5}}}$ as we approach $E_{c}$ from above.
\end{enumerate}
The plots of $\gamma$ vs $E$ thus exhibit a self-similar structure as shown in Figures~\ref{A3SubCrit} and \ref{A3SupCrit}. Analogous phenomena, studied in \cite{Brack2001}, were described as `Feigenbaum-like'. While such self-similar structures and `Feigenbaum like' oscillations have been previously observed for Hamiltonian systems \cite{Brack2001}, the spin-0 sector is, to our knowledge unique, as it contains not one, but \textit{two} independent self-similar cascades, one for the supercritical and one for subcritical regimes. Furthermore, these ratios are \textit{distinct} (albeit simply related).


We now present a rigorous derivation of the preceding results, following extensively the methods adopted in \cite{Brack2001}.  Eliminating the $p$'s from the fluctuation equations {\eqref{FlucEqnsA3DeC}-\eqref{FlucEqnsA3DeC2}}, we are left with a single non-trivial second order fluctuation equation
\begin{align}
\label{FE1}
\delta\ddot{a}(t)+\bigl(1+3a(t)\bigr)\delta a(t)=0.
\end{align}
The $a$ appearing in the above equation describes the periodic time evolution along the $A_{3}$ orbits, and can be explicitly expressed in terms of elliptic integrals, with time period
 \begin{equation}
\label{TPA3}
T_{p}(E) = 
\left\{
    \begin{array}{lr}
     \left(\frac{6}{E}\right)^{1/4}F\left(\text{sec}^{-1}\left(\frac{1+\eta}{\sqrt{\eta^{2}-1}}\right),\sqrt{\frac{1+\eta}{2}}\right), & \text{if } E<E_{c}\\
    2\left(\frac{6}{E}\right)^{1/4}K\left(\sqrt{\frac{1+\eta}{2}}\right), & \text{if } E>E_{c}
    \end{array}
\right\}.
\end{equation}
Here $F$ and $K$ are the incomplete and complete elliptic integrals of the first kind respectively, and $\eta\equiv\sqrt{E_{c}/E}$. These time periods diverge at $E_{c}$, i.e. $\eta=1$. The near critical behaviour of $\gamma$ depends on the nature of the divergence of $T_{p}(E)$. This is best brought out using the integral representation of $F$:

\begin{equation}
F(\alpha,k)=\int_{0}^{\sin(\alpha)}{\frac{dx}{\sqrt{(1-x^{2})(1-k^{2}x^{2})}}}.
\end{equation}
We are interested in the singular behaviour of the integral as $\alpha\rightarrow\pi/2$ and $k\rightarrow 1$. The latter portion of the denominator splits as $(1-kx)(1+kx)$ so that when $k \sim 1$, the divergence of the integral stems solely from the $1-kx$ and the additional $1-x$ term from the first term under the square root. We thus immediately see that A) The $1+kx$ term can simply be replaced by $1+x$ as it contributes nothing to the divergence and B) since $1-kx\sim 1-x$ when $k\sim 1$ and since the product of these terms is nested under a square root, we see that we should naively \textit{expect} the integral to diverge as $\log(1-\sin(\alpha))$. \\
Replacing the non singular $1+kx$ term by $1+x$ leads to the analytically tractable integral
\begin{equation}
F(\alpha,k)\sim\int_{0}^{\sin(\alpha)}\frac{dx}{(1+x)\sqrt{(1-x)(1-kx)}}.
\end{equation}
The series of substitutions $x\rightarrow y\equiv\frac{1}{1+x}, y\rightarrow z\equiv 2y-1$ and use of standard integrals then also us to evaluate this integral as
\begin{equation}
    F(\alpha,k)\sim\Biggl(\frac{1}{\sqrt{2(1+k)}}\ln[z+\frac{1-k}{2(1+k)}+\sqrt{\biggl(z+\frac{1-k}{2(1+k)}\biggl)^2-\frac{1}{4}\biggl(\frac{1-k}{1+k}\biggl)^{2}}] \Biggl)\bigg|_{\frac{1-\sin(\alpha)}{1+\sin(\alpha)}}^{1}.
\end{equation}
It is easy to see that only the lower limit contributes to the divergence, so that we may further write
\begin{equation}
\label{DivFormFinal}
    F(\alpha,k)\sim -\Biggl(\frac{1}{\sqrt{2(1+k)}}\ln(z+\frac{1-k}{2(1+k)}+\sqrt{\biggl(z+\frac{1-k}{2(1+k)}\biggl)^2-\frac{1}{4}\biggl(\frac{1-k}{1+k}\biggl)^{2}} \Biggl)\bigg|_{\frac{1-\sin(\alpha)}{1+\sin(\alpha)}}.
\end{equation}
We now apply the above to \eqref{TPA3}:
\begin{enumerate}
    \item \textbf{Supercritical}: Setting $\alpha$ to $\frac{\pi}{2}$ and $k$ to $\sqrt{\frac{1+\eta}{2}}$, we obtain
        \begin{equation}
        \label{DivSup}
        T_{p}(E)\sim 2\sqrt{2}\ln\frac{1}{1-\eta}\sim 2\sqrt{2}\ln\frac{1}{1-\eta^{2}}=2\sqrt{2}\ln\frac{1}{1-\frac{E_{c}}{E}}.
    \end{equation}
    as $\eta\rightarrow 1^{-}$.
    \item \textbf{Subcritical}: Setting $\alpha$ to $\text{sec}^{-1}\left(\frac{1+\eta}{\sqrt{\eta^{2}-1}}\right)$ and $k$ to $\sqrt{\frac{1+\eta}{2}}$, we obtain      
      \begin{equation}
        \label{DivSub}
        T_{p}(E)\sim \sqrt{2}\ln\frac{1}{\eta-1}\sim\sqrt{2}\ln\frac{1}{\eta^{2}-1}=\sqrt{2}\ln\frac{1}{\frac{E_{c}}{E}-1}
    \end{equation}
as $\eta\rightarrow 1^{+}$.
\end{enumerate}
 We next study the variation of $\gamma$ with the time period. Following \cite{Brack2001}, we see that $\gamma$ can be expressed as a trigonometric Fourier series in $T_{p}$, with the leading Fourier coefficient yielding the only non-trivial contribution in the limit of $\eta\rightarrow 1$. We thus have
\begin{equation}
    \label{TracetoTP}
    \gamma(T_{p})\sim 2\cos\omega T_{p}
\end{equation}
where $\omega$ can be worked out as follows: since in the limit of $\eta\rightarrow 1$, $a$ spends an increasingly large amount of time near the saddle point $\frac{1}{2}$, we may estimate the asymptotic period by simply replacing $a$ by $\frac{1}{2}$ in  {\eqref{FlucEqnsA3}-\eqref{FlucEqnsA3F}}. Then the dynamical equations \eqref{FE1} simply reduce to those of an oscillator with period $\Delta T=2\sqrt{\frac{2}{5}}{\pi}$. Thus, we have $\omega=\sqrt{\frac{5}{2}}$. 

We can now compute the geometric ratios for the $A_{3}$ oscillations. From \eqref{TracetoTP}, we see that the transition points are evenly spaced in intervals of $\Delta T$ when viewed as functions of the time period $T_{p}$. The logarithmic dependence of $T_{p}$ with $\vert 1-\eta\vert$, captured by \eqref{DivSup} and \eqref{DivSub} tells us that the locations of the transition points, as measured by the quantity $1-\eta$, must asymptotically form a geometric series. It is also easily seen from \eqref{DivSup} and \eqref{DivSub} that the relevant common ratios are $e^{\frac{\Delta T}{2\sqrt{2}}}=e^{\frac{\pi}{\sqrt{5}}}$ for the supercritical oscillations and $e^{\frac{\Delta T}{\sqrt{2}}}=e^{\frac{\pi}{2\sqrt{5}}}$ for the subcritical oscillations.

\begin{figure}[H]
     \centering
     \begin{subfigure}[b]{0.4\textwidth}
         \centering
         \includegraphics[width=\textwidth]{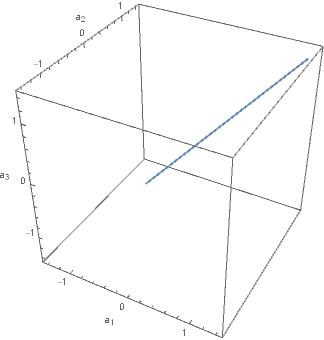}
             \caption{$E=0.323$}
     \end{subfigure}
     \hfill
         \begin{subfigure}[b]{0.4\textwidth}
         \centering
         \includegraphics[width=\textwidth]{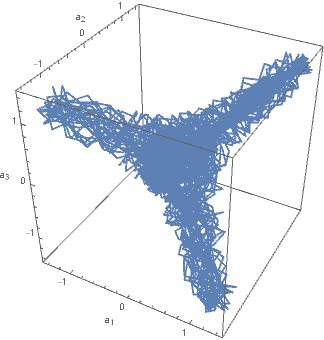}
         \caption{$E=0.324$}
     \end{subfigure}
        \caption{Stable to unstable flip for $A_3$}
     \label{PPA3_1}
\end{figure}

\begin{figure}[H]
     \centering
     \begin{subfigure}[b]{0.4\textwidth}
         \centering
         \includegraphics[width=\textwidth]{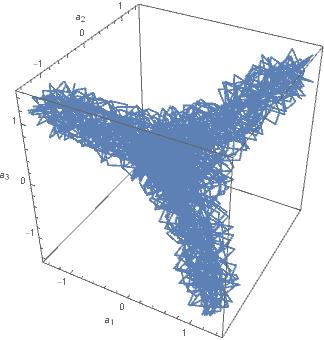}
             \caption{$E=0.445$}
     \end{subfigure}
     \hfill
         \begin{subfigure}[b]{0.4\textwidth}
         \centering
         \includegraphics[width=\textwidth]{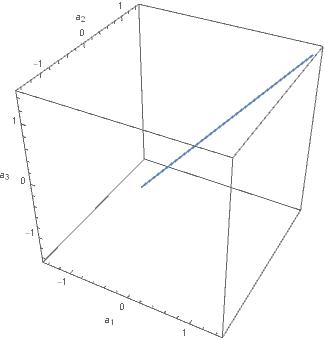}
         \caption{$E=0.446$}
     \end{subfigure}
        \caption{Unstable to stable flip for $A_3$}
     \label{PPA3_2}
\end{figure}

Figures \ref{PPA3_1} and \ref{PPA3_2} demonstrate the flip from stability to instability and vice versa as the energy is varied across a transition point.

\subsection[Pi1 Orbits]{$\Pi_{1}$ Orbits}
The wealth of results obtained for $A_{3}$ and $A_{4}$ ultimately traces back to the high symmetry of these orbits.  These symmetries are enough to completely decouple the fluctuation equations, which eventually lead to a significant simplification.  Since the geometric orbits do \textit{not} originate from Weinstein's theorem,  they are less symmetric and the fluctuation equations remain partially coupled. As a result, a single spectral invariant (like $\gamma$) is not enough to capture the stability properties of the geometric orbits.  Nevertheless,  as the geometric orbits reside on RDS phase spaces,  a partial decoupling of the fluctuations can indeed be accomplished,  with two modes spanning fluctuations confined to the relevant RDS subspace, and the third mode generating fluctuations orthogonal to this subspace.  Consequently,  some simplifications can be made before reverting to numerics.

The equations describing fluctuations about $\Pi_{1}$ orbits are given by 
\begin{align}
    \label{Pi1Perturbations}
    \delta \dot{a}_{1}(t) =&\,\,\,\delta p_{a_1}(t), \quad\delta \dot{a}_{2}(t) =\delta p_{a_2}(t), \quad\delta \dot{a}_{3}(t) =\delta p_{a_3}(t), \\
    \delta \dot{p}_{a_1}(t) =&-\delta a_1(t) [1+2a(t)^{2}]+ 3a(t)[\delta a_3(t) +\delta a_2(t)]-2a_1(t) a(t)[\delta a_2(t)+\delta a_3(t)], \\
    \delta \dot{p}_{a_2}(t) =&-\delta a_2(t)[1+a(t)^{2}+a_1(t)^{2}]+ 3a(t)\delta a_1(t) +3a_1(t)\delta a_3(t) \nonumber \\
    &-2a(t)[a(t)\delta a_3(t)+a_1(t)\delta a_1(t)], \\
    \delta \dot{p}_{a_3}(t) =&-\delta a_3(t) [1+a_1(t)^{2}+a^{2}(t)]+ 3a_1(t)\delta a_2(t) +3a(t)\delta a_1(t) \nonumber \\
    &-2a(t)[a_1(t)\delta a_1(t)+a(t)\delta a_2(t)],
     \label{Pi1Perturbations1}
\end{align}
where $a(t),a_{1}(t)$ and their momenta describe time evolution along the unperturbed $\Pi_{1}$ orbit.  Using the canonical rotation, $a_{\pm}=\frac{a_{2}\pm a_{3}}{\sqrt{2}}$, as before,  we may restate these equations as
\begin{align}
     \label{Pi1Perturbations2}
    \delta \dot{a}_{1}(t) &=\delta p_{a_1}(t),\quad\delta \dot{a}_{+}(t) =\delta p_{a_+}(t), \quad\delta \dot{a}_{-}(t) =\delta p_{a_-}(t), \\
    \delta \dot{p}_{a_1}(t) &=-\delta a_1(t) (1+2a(t)^{2})+ 3a(t)(\delta a_3(t) +\delta a_2(t))-2a_1(t) a(t)(\delta a_2(t)+\delta a_3(t)),  \\
    \delta \dot{p}_{a_+}(t) &=-\delta a_{+}(t)(1+a(t)^{2}+a_1(t)^{2})+ 3\sqrt{2}a(t)\delta a_1(t) +3a_1(t)\delta a_{+}(t) \nonumber \\
    &\quad-2\sqrt{2}a(t)(a_1(t)\delta a_1(t))-2a(t)^{2}\delta a_{+}(t),  \\
    \delta \dot{p}_{a_-}(t) &=-\delta a_{-}(t) (1+a(t)^{2}+a_1(t)^{2})  -3a_1(t)\delta a_{-}(t)+2a(t)^{2}\delta a_{-}(t),
     \label{Pi1Perturbations3}
\end{align}
Since we had earlier restricted ourselves to a  concrete instance of an RDS (see section \ref{RMMDefn}) by fixing $a_{2}$ to $a_{3}$, and $p_{a_{2}}=p_{a_{3}}$,  it is evident that fluctuations with $\delta a_{-}=\delta p_{a_{-}}=0$ yield trajectories that deviate from the $\Pi_{1}$ orbit,  but are confined to the phase space of the \textit{RDS}.  On the other hand,  fluctuations with $\delta a_{-}\neq 0$  destroy the equality of $a_{2}$ and $a_{3}$. Such fluctuations lead to trajectories that are not confined to the RDS, but span the full six-dimensional phase space of the spin-0 sector.  In short,  an arbitrary fluctuation can be split into an `orthogonal' mode perpendicular to the RDS phase space,  and a pair of coupled `tangential' modes living in the RDS phase space. 
The independence of the orthogonal modes from the tangential modes results in the factorization of the monodromy matrix into a 2+4 block diagonal form.  Unlike with the NLNMs, no further simplifications can be made at this point and numerical evaluations are the only way forward.  

As before, the fluctuation equations {\eqref{Pi1Perturbations2}-\eqref{Pi1Perturbations3}} retain the form of a Schr\"{o}dinger equation,  albeit with a two-component `wavefunction' unlike the previous two instances. 

\begin{figure}[H]
     \centering
     \begin{subfigure}[b]{0.4\textwidth}
         \centering
         \includegraphics[width=\textwidth]{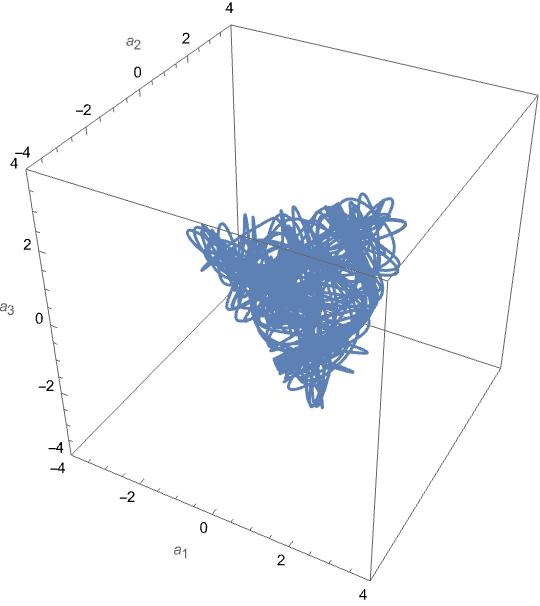}
             \caption{$E=2.475$}
     \end{subfigure}
     \hfill
         \begin{subfigure}[b]{0.4\textwidth}
         \centering
         \includegraphics[width=\textwidth]{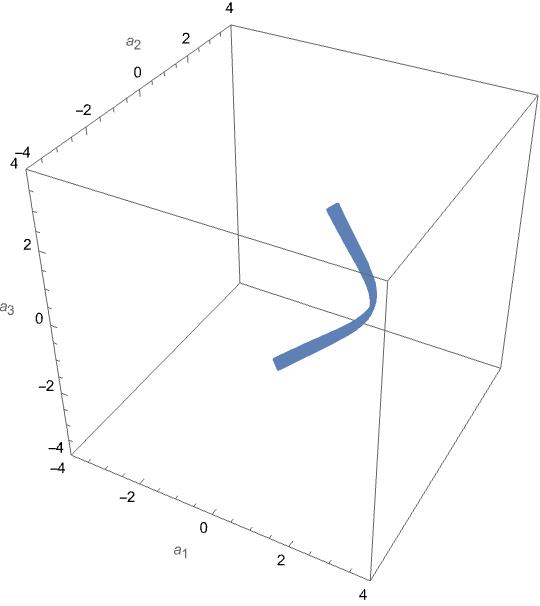}
         \caption{$E=2.476$}
     \end{subfigure}
        \caption{Unstable to stable flip for $\Pi_1$ orbits}
     \label{PPPi1_1}
\end{figure}

\begin{figure}[H]
     \centering
     \begin{subfigure}[b]{0.4\textwidth}
         \centering
         \includegraphics[width=\textwidth]{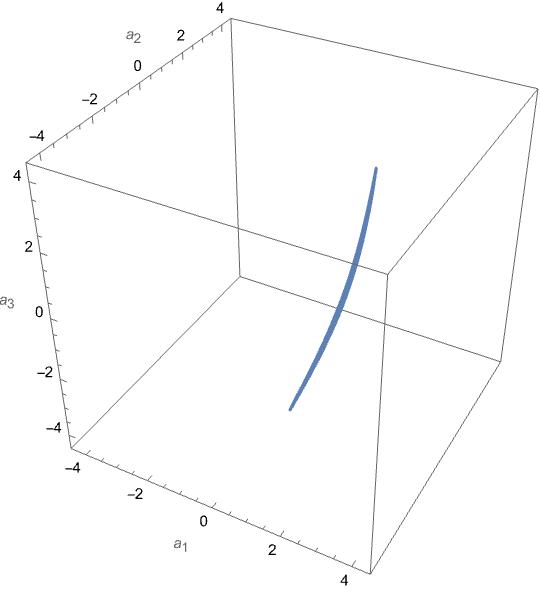}
             \caption{$E=9.786$}
     \end{subfigure}
     \hfill
         \begin{subfigure}[b]{0.4\textwidth}
         \centering
         \includegraphics[width=\textwidth]{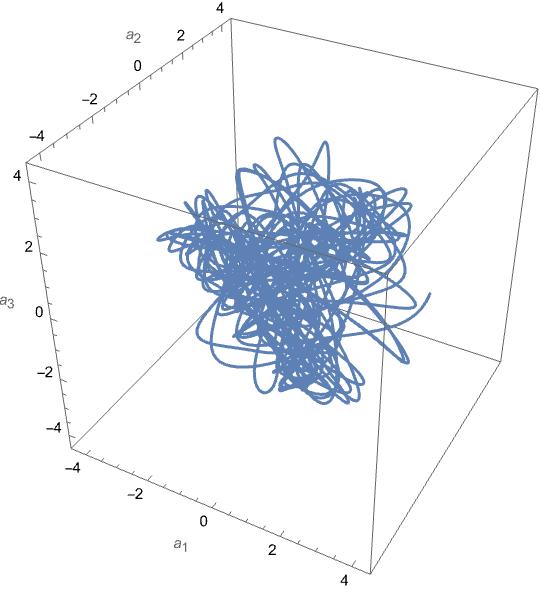}
         \caption{$E=9.787$}
     \end{subfigure}
        \caption{Stable to unstable flip for $\Pi_1$ orbits}
     \label{PPPi1_2}
\end{figure}

Numerics once again reveal the presence of bands: there exist energy bands displaying regular behaviour, with chaos ensuing outside these bands. However it turns out that the bands are finite in number, as opposed to the cases of $A_4$ and $A_3$.  Specifically,  we find that $\Pi_{1}$ orbits are always unstable for subcritical energies and undergo just \textit{four} stability flips,  with two `bands' of stability from $2\lesssim E\lesssim 3$ and $5\lesssim E\lesssim 10$.   Note that since $ \text{Tr}\,\mathcal{U}$ does not directly correlate with stability as it did for $A_{3/4}$,  stability can only be ascertained by looking at the full spectrum of the monodromy matrix.  The requisite numerics is not very illuminating, so we do not present the full calculations here. Graphical evidence for these flips (demonstrated in Figures \ref{PPPi1_1} and \ref{PPPi1_2}) comes from the Lyapunov exponent plots, which we will display in section \ref{LE}.

\subsection[Pi2 Orbits]{$\Pi_{2}$ Orbits}
$\Pi_{2}$ orbits are investigated using the same methodology as $\Pi_{1}$ orbits.  We shall not go over our procedures again,  and will simply state the results of our numerics.  We find that subcritical $\Pi_{2}$ orbits are always stable under generic fluctuations.  Supercritically, we find two stable but small bands, the first near $E\sim3$ and the second near $E\sim4$. These results will be corroborated by plots of Lyapunov exponents in section \ref{LE}.

\section{Progression to Chaos}\label{CT}
In the previous section, we analyzed the stability of several sets of orbits by drawing information from their monodromy matrices. Here, we will pursue another traditional tool to study chaos, utilising the standard techniques of Poincar\'e sections and Lyapunov exponents.  In so doing,  we will come across numerous novel and peculiar features which,  using our prior monodromy analysis, will correlate beautifully to the periodic orbits and ultimately the symmetries of the spin-0 sector. 
\subsection{Poincar\'e Sections}
As defined in \cite{goldstein:mechanics}, a Poincar\'e section for an $N$-dimensional Hamiltonian system is a $2N-2$ dimensional slice through a $2N-1$ dimensional constant energy hypersurface.  Poincar\'e sections are thus most effective for four dimensional Hamiltonian systems,  and are in general not useful for higher-dimensional systems. 

 However,  we find that a simple variation of the usual construction can serve as an excellent visual aid. Specifically,  we locate points on a given trajectory where a particular
coordinate/momentum is zero. We then project the collection of such points onto a hyperplane spanned by
three of the five remaining coordinates/momenta.  With this construct (which we continue to refer to as a Poincar\'e section), the usual rules for distinguishing regular trajectories from chaotic ones no longer hold.  In particular,  regular orbits could yield (our version of) Poincar\'e sections that are a collection of randomly scattered points.  
This is not a matter
of concern for us since our current aim is to study only chaotic trajectories, having carried out an extensive study of regular solutions earlier.  We will work exclusively with trajectories that monodromy computations certify as unstable.
We consider small fluctuations about periodic orbits and construct Poincar\'e sections at various energies by projecting points on these trajectories having  $p_{a_{1}}=0$  onto the hyperplane spanned by the $a_{i}$'s. 

\begin{figure}[H]
     \centering
     \begin{subfigure}[b]{0.4\textwidth}
         \centering
         \includegraphics[width=\textwidth]{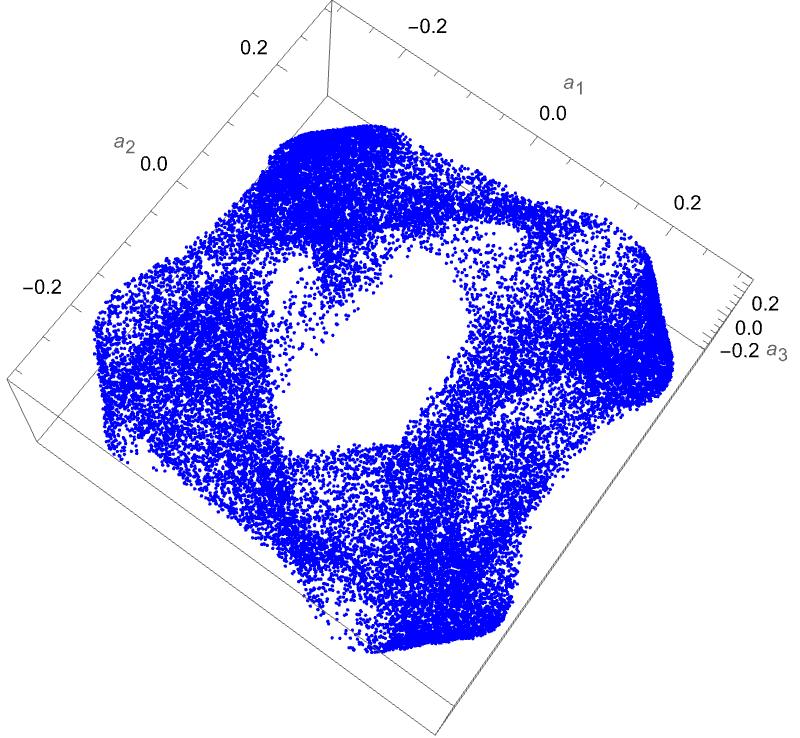}
            \caption{$A_{4}$ section at $E=0.05$}
     \end{subfigure}
     \hfill
     \begin{subfigure}[b]{0.4\textwidth}
         \centering
         \includegraphics[width=\textwidth]{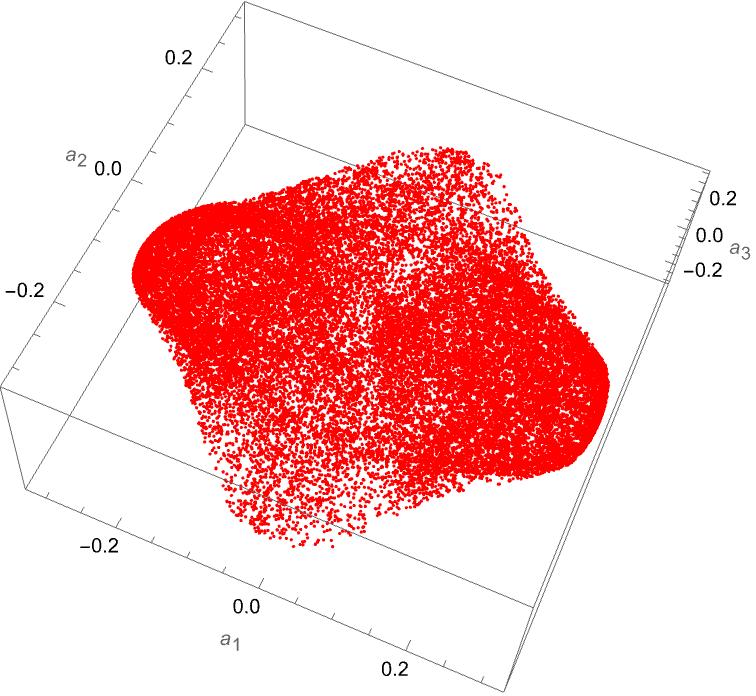}
                \caption{$\Pi_{1}$ section at $E=0.05$}
     \end{subfigure}
          \begin{subfigure}[b]{0.4\textwidth}
         \centering
         \includegraphics[width=\textwidth]{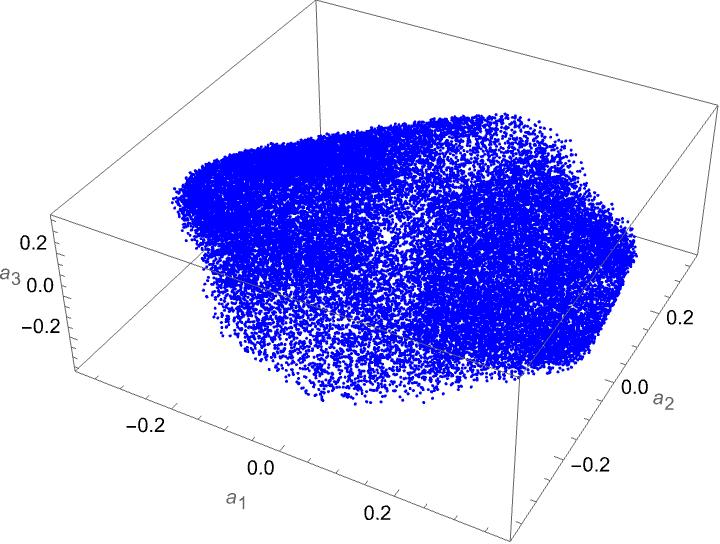}
            \caption{$A_{4}$ section at $E=0.09$}
     \end{subfigure}
     \hfill\
     \begin{subfigure}[b]{0.4\textwidth}
         \centering
         \includegraphics[width=\textwidth]{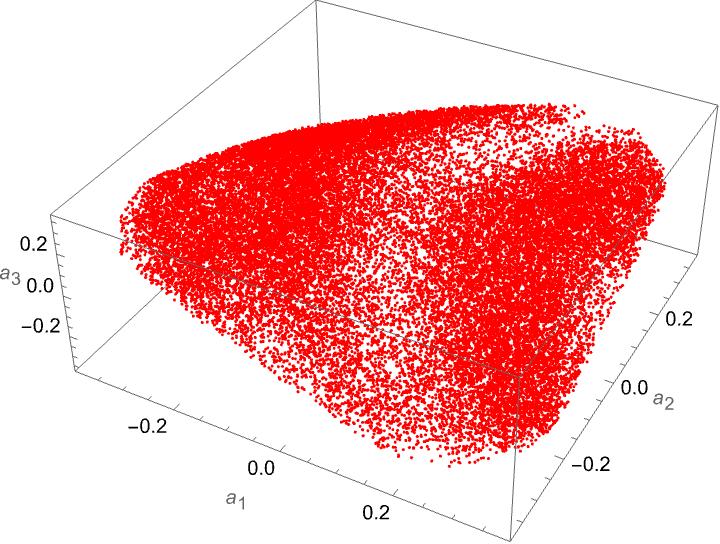}
            \caption{$\Pi_{1}$ section at $E=0.09$}
     \end{subfigure} 
     \hfill
           \begin{subfigure}[b]{0.4\textwidth}
         \centering
         \includegraphics[width=\textwidth]{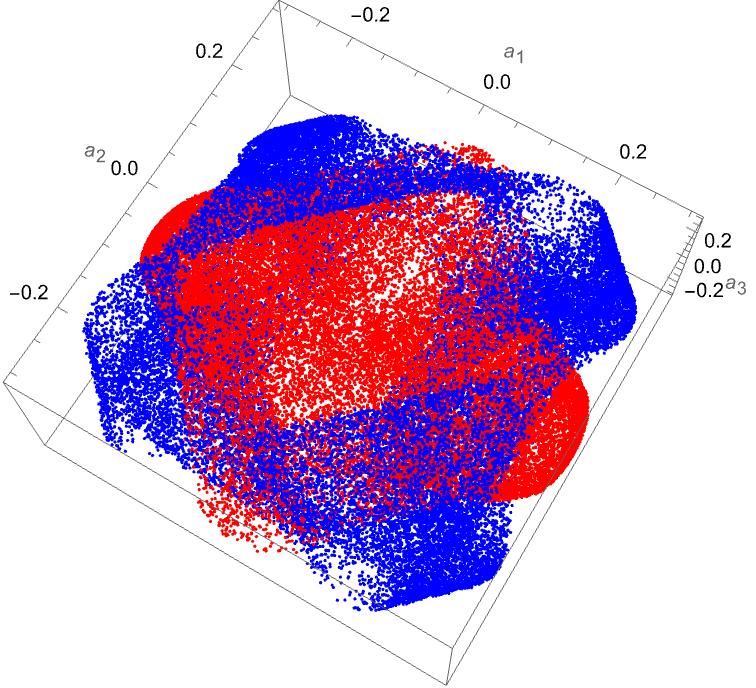}
         \caption{$A_{4}$ and $\Pi_{1}$ sections at $E=0.05$}
     \end{subfigure}
     \hfill
  \begin{subfigure}[b]{0.4\textwidth}
         \centering
         \includegraphics[width=\textwidth]{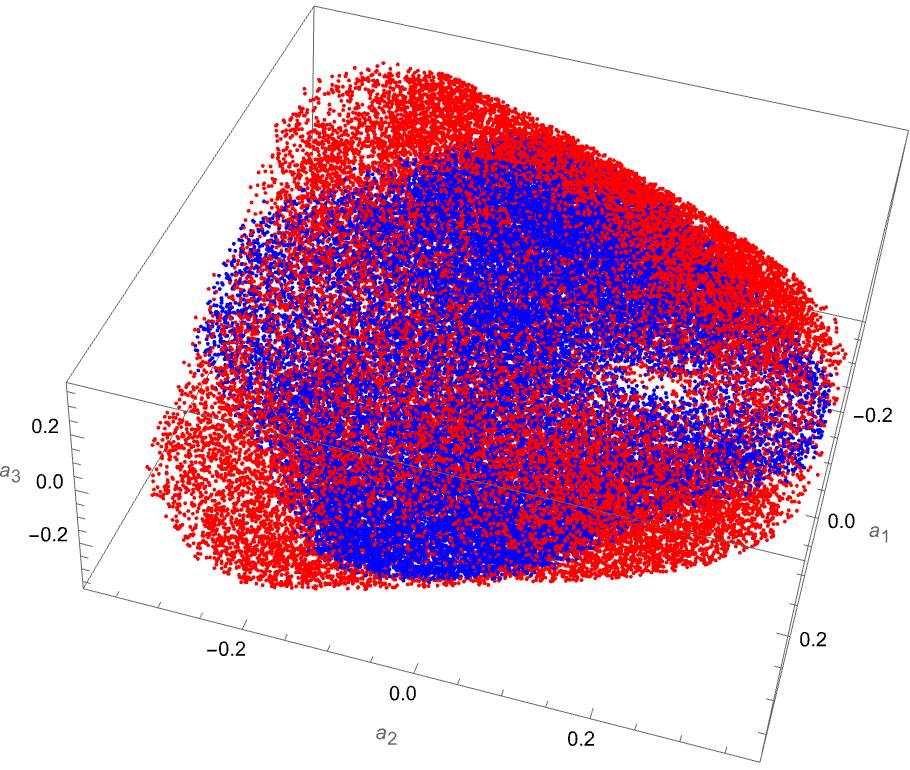}
         \caption{$A_{4}$ and $\Pi_{1}$ sections at $E=0.09$}
     \end{subfigure}

              \caption{Poincar\'e sections at subcritical energies}
              \label{PSSub3D}
              \end{figure}

 Since the $A_{4}$ and $\Pi_{1}$ orbits are always unstable at subcritical energies,  we expect the Poincar\'e sections to be sets of randomly scattered points. While we do find that the sections are indeed scattered and locally random, there are large scale patterns. These patterns depend solely on the `parent orbit'.  Poincar\'e sections for subcritical $A_{4}$ and $\Pi_{1}$ orbits are shown in Figure \ref{PSSub3D}. 
     We thus conclude that we have a set of co-existing chaotic basins,  one for each family of unstable orbits! 
     
     To illustrate a second peculiar feature of the dynamics, we recall that in addition to the chaotic dynamics of the full spin-0 sector,  unstable trajectories confined to RDS phase spaces may well display chaotic dynamics of their own.  This leads to chaotic basins embedded in a four-dimensional subset nested within the full six-dimensional chaotic dynamics! This extraordinary feature of the dynamics is the combined result of the large dimensionality and the tetrahedral symmetry.  This `nested' chaos, as part of a genuine four-dimensional system, can be analyzed using Poincar\'e sections in the usual sense.  
     
     We thus generate Poincar\'e sections for chaotic trajectories of the RDS,  both to study the nested chaos and to look for similarities to the full six-dimensional dynamics.  In particular, since most of the interesting orbits of the full spin-0 sector have RDS analogs,  we would naturally expect a similar substructure of  multiple chaotic basins,  one for each class of orbit.  This substructure is indeed replicated in the RDSs, as evidenced by Figure \ref{PSSub}. 

\begin{figure}[H]
     \centering
 \begin{subfigure}[b]{0.4\textwidth}
         \centering
         \includegraphics[width=\textwidth]{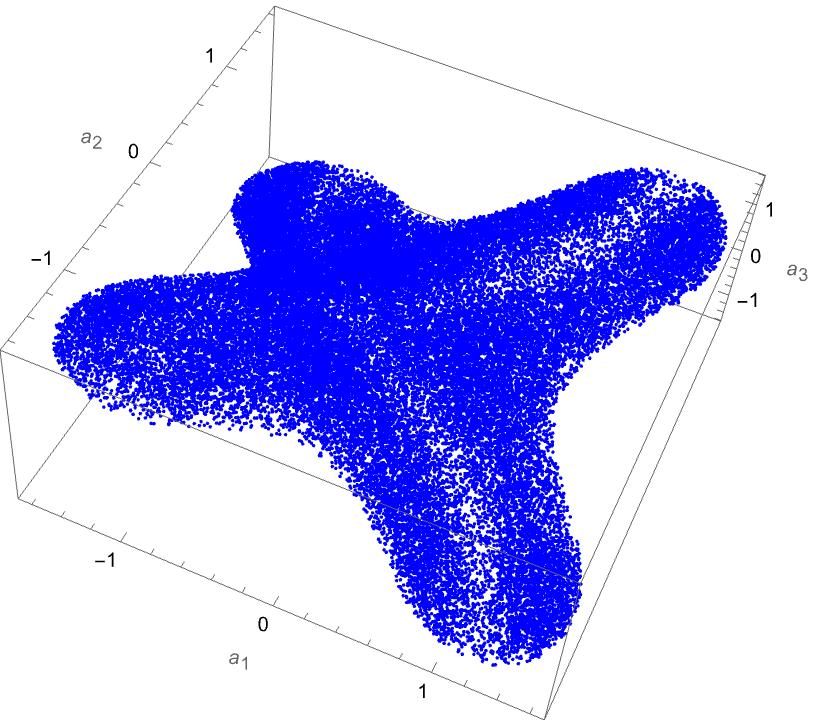}
            \caption{$A_{4}$ section at $E=0.5$}
     \end{subfigure}
     \hfill
          \begin{subfigure}[b]{0.4\textwidth}
         \centering
         \includegraphics[width=\textwidth]{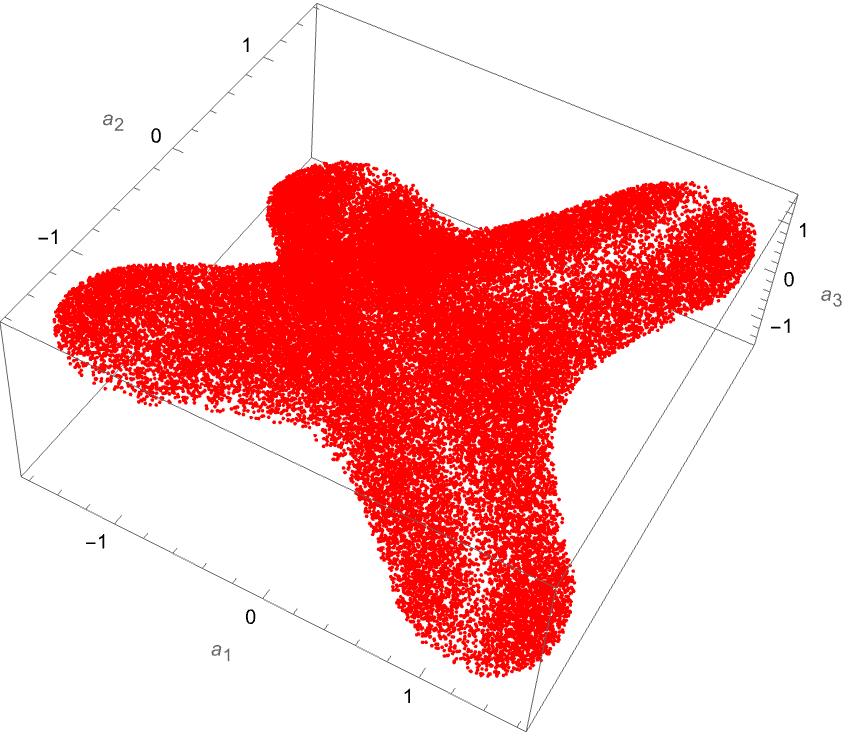}
         \caption{$\Pi_{1}$ section at $E=0.5$}
     \end{subfigure}
     \hfill

              \caption{Poincar\'e sections at supercritical energies}
              \label{PSSup3D}
              \end{figure}

\begin{figure}[H]
     \centering
     \begin{subfigure}[b]{0.4\textwidth}
         \centering
         \includegraphics[width=\textwidth]{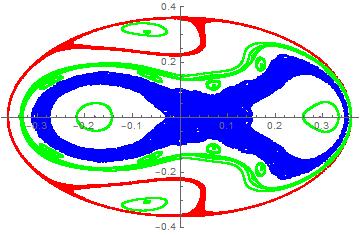}
            \caption{E=0.065}
     \end{subfigure}
     \hfill
     \begin{subfigure}[b]{0.4\textwidth}
         \centering
         \includegraphics[width=\textwidth]{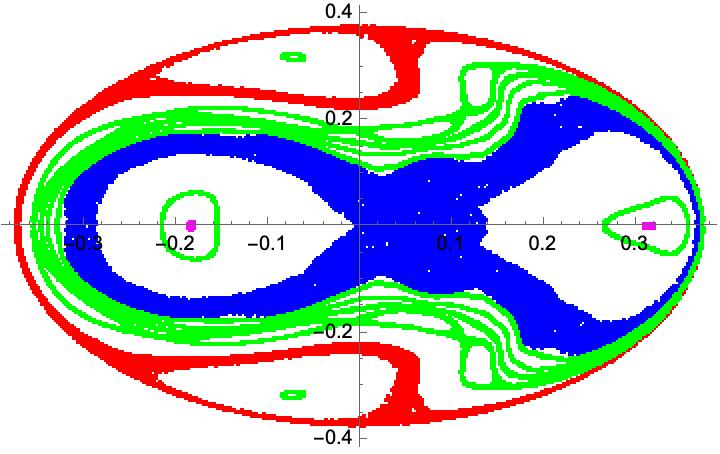}
            \caption{E=0.070}
     \end{subfigure}
          \begin{subfigure}[b]{0.4\textwidth}
         \centering
         \includegraphics[width=\textwidth]{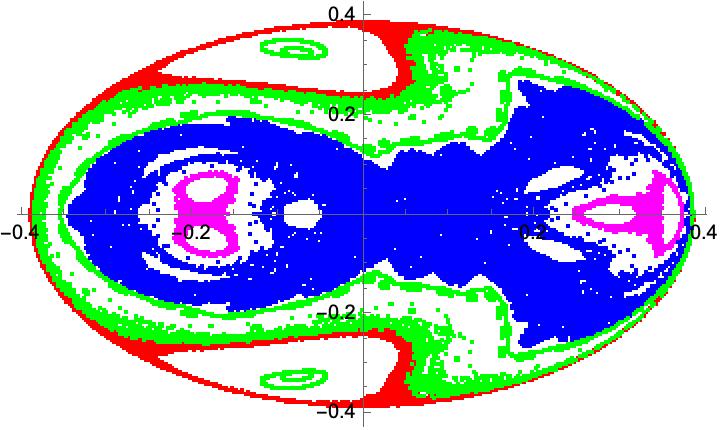}
            \caption{E=0.075}
     \end{subfigure}
     \hfill
     \begin{subfigure}[b]{0.4\textwidth}
         \centering
         \includegraphics[width=\textwidth]{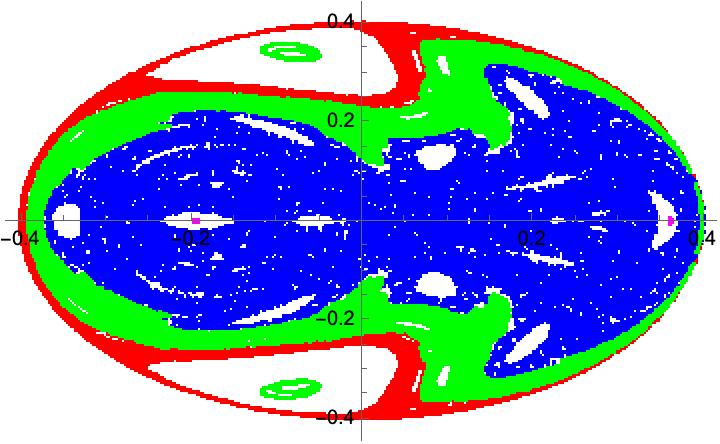}
            \caption{E=0.08}
     \end{subfigure} 
     \hfill
     \begin{subfigure}[b]{0.4\textwidth}
         \centering
         \includegraphics[width=\textwidth]{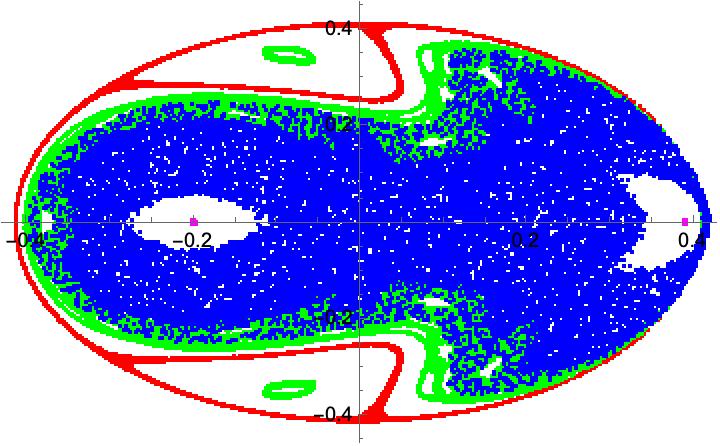}
            \caption{E=0.085}
     \end{subfigure}
     \hfill
          \begin{subfigure}[b]{0.4\textwidth}
         \centering
         \includegraphics[width=\textwidth]{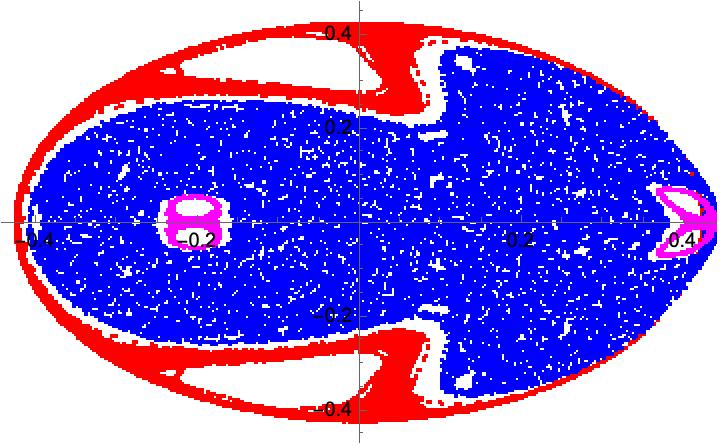}
         \caption{E=0.09}
     \end{subfigure}
     \hfill
              \caption{Poincare Sections at Subcritical Energies}
              \label{PSSub}
              \end{figure}

\begin{figure}[H]
     \centering
     \begin{subfigure}[b]{0.3\textwidth}
         \centering
         \includegraphics[width=\textwidth]{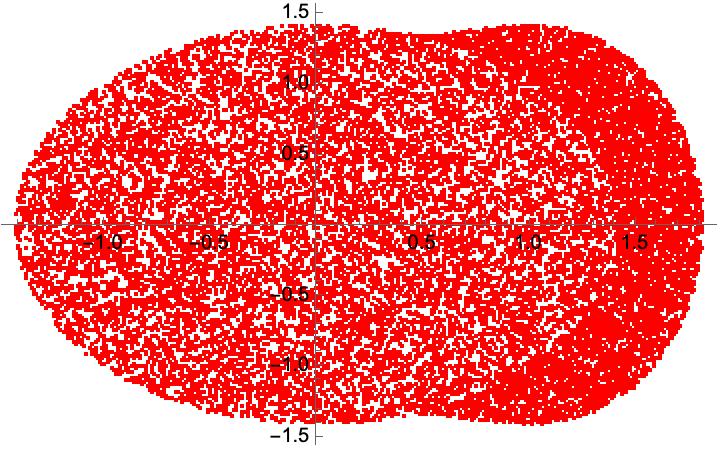}
      \caption{$\Pi_{1}$ sector}
     \end{subfigure}
   \hfill
     \begin{subfigure}[b]{0.3\textwidth}
         \centering
         \includegraphics[width=\textwidth]{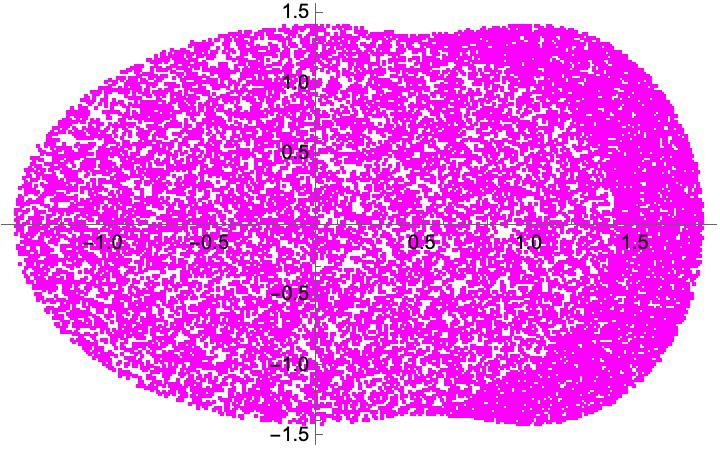}
          \caption{$A_3$ sector}
     \end{subfigure}
      \hfill
             \begin{subfigure}[b]{0.3\textwidth}
         \centering
         \includegraphics[width=\textwidth]{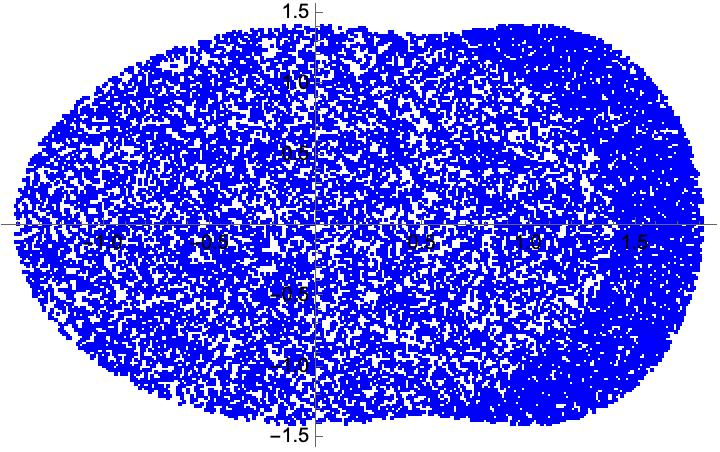}
         \caption{$A_4$ sector}
     \end{subfigure}
     \hfill
     \caption{Poincare Sections at Supercritical Energies ($E$=1)}
     \label{PSSup}
\end{figure}

Next, we see that the supercritical regime appears to comprise of just a single chaotic basin (Figure \ref{PSSup3D}).  The mechanism responsible for separating chaotic subsectors in the subcritical regions is no longer operative,  so that fluctuations about unstable periodic orbits rapidly grow and eventually cover the entire available phase space,  losing memory of their initial conditions.  Analogous results hold for the supercritical regimes of the RDSs, as is seen from the Poincar\'e sections of Figure \ref{PSSup}.

We now turn to Lyapunov exponents which will provide additional confirmation for our already established results while also motivating the `thermodynamic' perspective we will encounter in section \ref{TD}.

\subsection{Lyapunov Exponents} \label{LE}

Recall that the maximal Lyapunov exponent (LE) at a phase point $P$ is defined as
\begin{equation}
\lambda_{P}\equiv \lim_{T\rightarrow\infty}\lim_{\vert\vert\delta x(0)\vert\vert\rightarrow 0}\frac{1}{T}\log\left(\frac{\vert\vert\delta x(T)\vert\vert}{\vert\vert\delta x(0)\vert\vert}\right),
\end{equation}
where $\delta x$ is a small fluctuation about a given trajectory $x(t)$ starting at $P$.

\begin{figure}[H]
     \centering
     \begin{subfigure}[b]{0.45\textwidth}
         \centering
         \includegraphics[width=\textwidth]{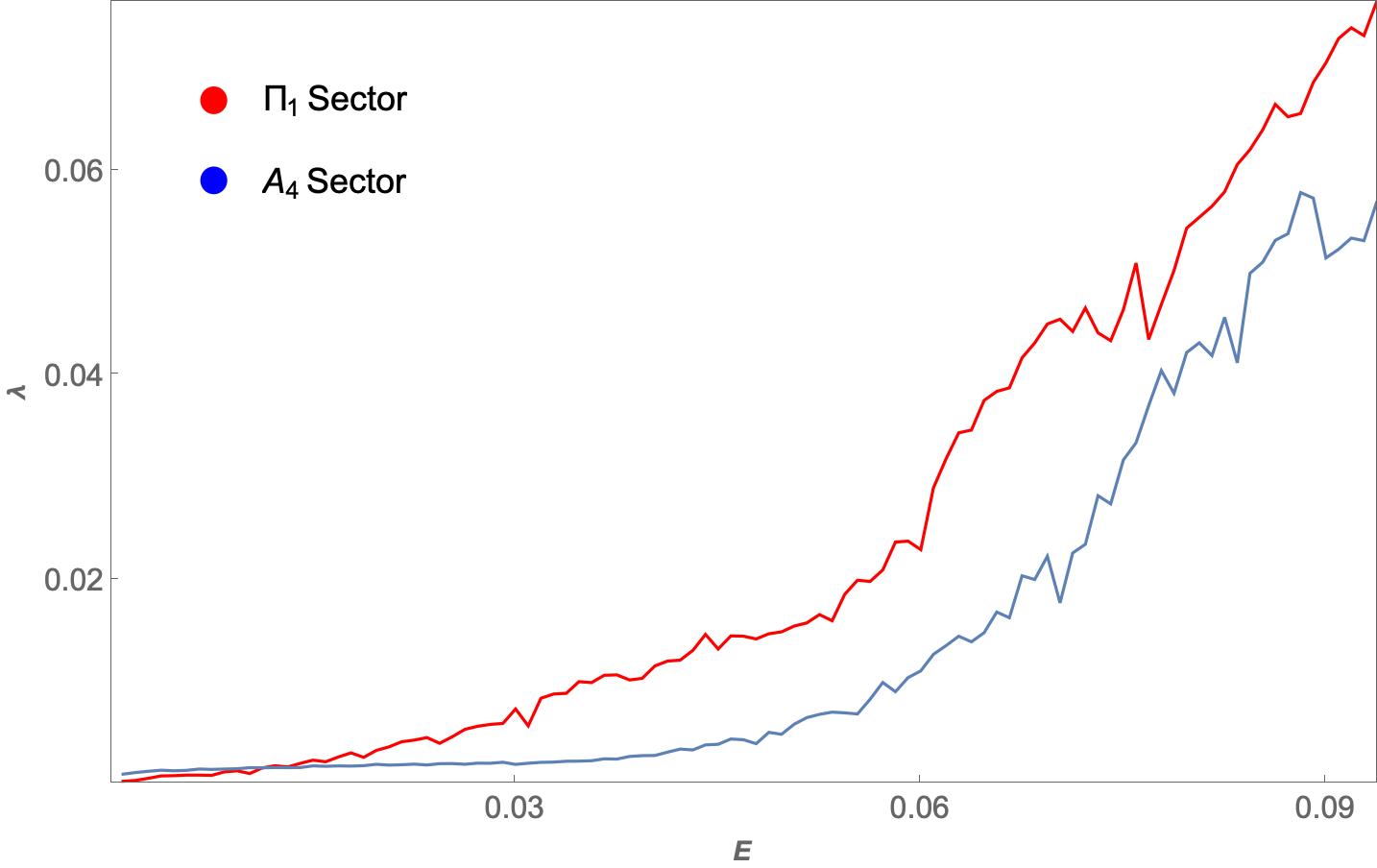}
     \end{subfigure}
     \hfill
         \caption{LEs for subcritical $A_{4}$ and $\Pi_{1}$ Orbits}
         \label{a4pi1}
     \hfill

\end{figure}

%
We compute the LEs for the $A_{4}$ and $\Pi_{1}$ basins separately by considering arbitrary fluctuations about these orbits.  The results for the subcritical zone are shown in Figure \ref{a4pi1}.  Consistent with our interpretation as co-existing chaotic basins,  we see that the exponents of the $\Pi_{1}$  and $A_{4}$ orbits differ from one another.  The stability of all other periodic orbits at nearly all subcritical energies means that we may confine our analysis to these two families of orbits.

\begin{figure}[H]
     \centering
     \begin{subfigure}[b]{0.45\textwidth}
         \centering
         \includegraphics[width=\textwidth]{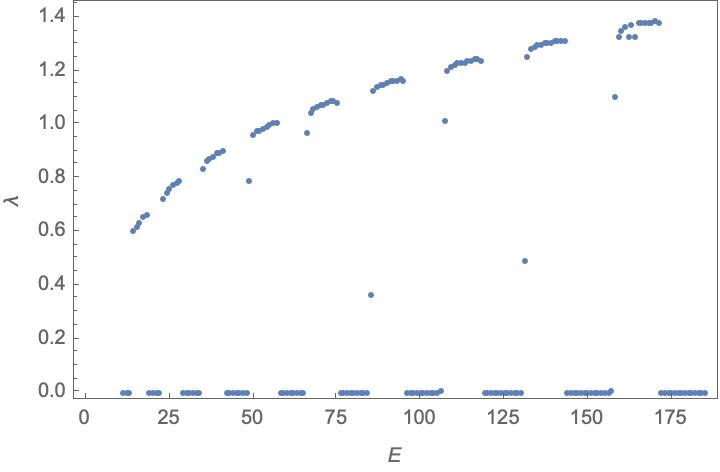}
     \end{subfigure}
     \hfill
     \begin{subfigure}[b]{0.45\textwidth}
         \centering
         \includegraphics[width=\textwidth]{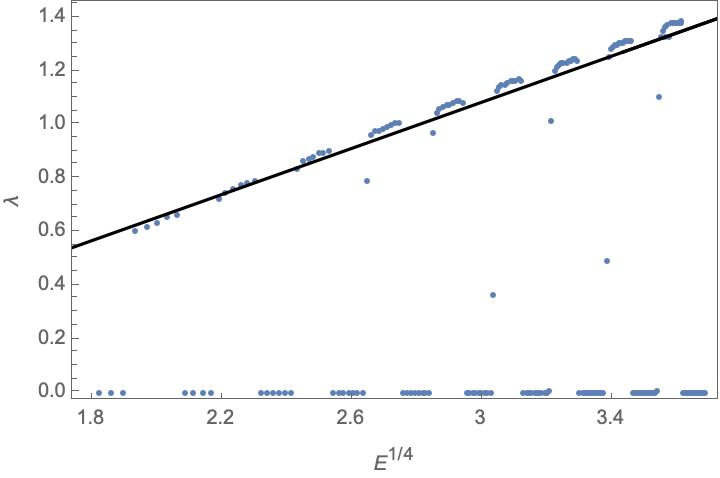}
     \end{subfigure}
         \caption{LEs and Fits for $A_4$ orbits}
         \label{A4s}
     \hfill

\end{figure}

\begin{figure}[H]
     \centering
     \begin{subfigure}[b]{0.45\textwidth}
         \centering
         \includegraphics[width=\textwidth]{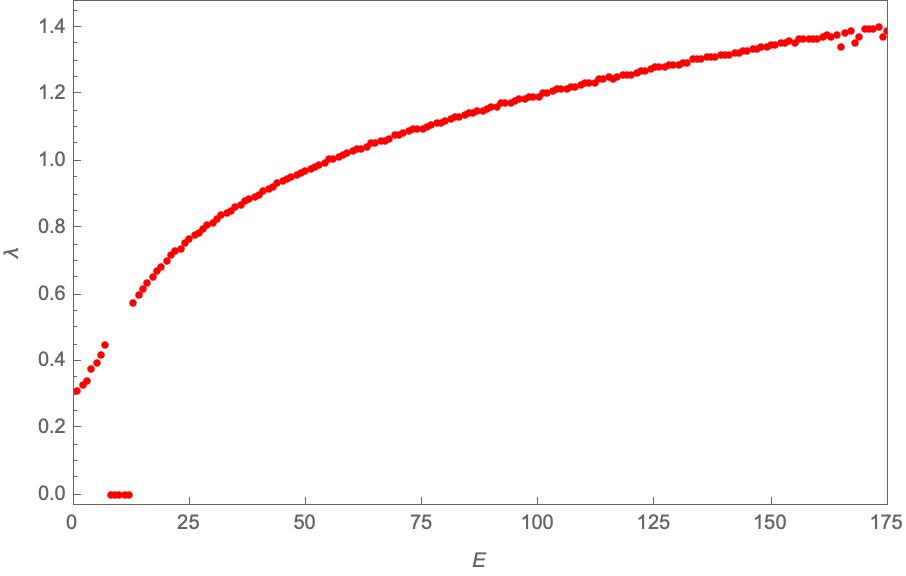}
     \end{subfigure}
     \hfill
     \begin{subfigure}[b]{0.45\textwidth}
         \centering
         \includegraphics[width=\textwidth]{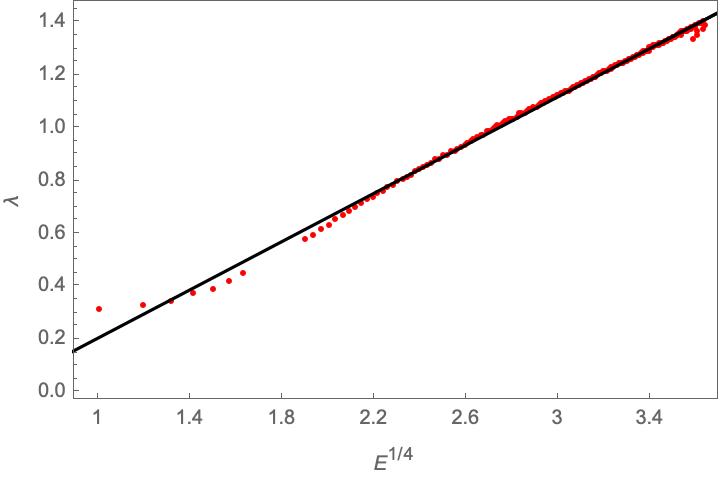}
     \end{subfigure}
         \caption{LEs and Fits for $A_3$ orbits}
         \label{A3s}
     \hfill

\end{figure}

\begin{figure}[H]
     \centering
     \begin{subfigure}[b]{0.45\textwidth}
         \centering
         \includegraphics[width=\textwidth]{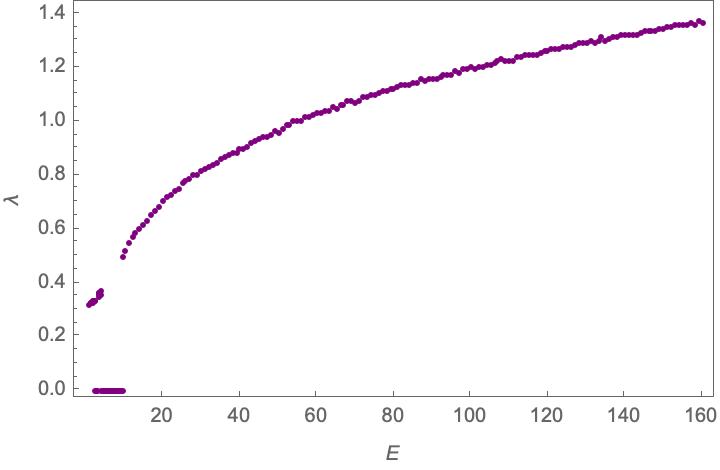}
     \end{subfigure}
     \hfill
     \begin{subfigure}[b]{0.45\textwidth}
         \centering
         \includegraphics[width=\textwidth]{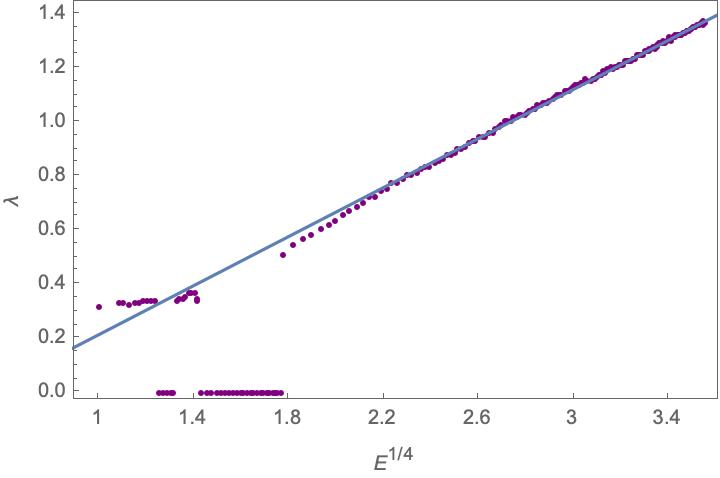}
     \end{subfigure}
         \caption{LEs and Fits for $\Pi_{1}$ orbits}
         \label{pi1s}
     \hfill
     \end{figure}

\begin{figure}[H]
     \centering
     \begin{subfigure}[b]{0.45\textwidth}
         \centering
         \includegraphics[width=\textwidth]{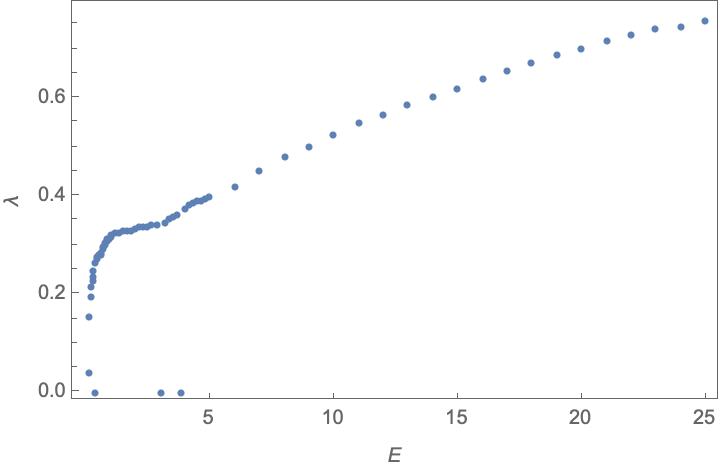}
     \end{subfigure}
      \caption{LEs for $\Pi_{2}$ orbits}
      \label{pi2s}
     \hfill
\end{figure}

Since the $\Pi_{2}$ and $A_{3}$ orbits can destabilize for $E>E_{c}$, the supercritical analysis must include these orbits as well.  The LEs for the basins corresponding to each of these orbits have been shown in Figures~{\ref{A4s}-\ref{pi2s}}.  There are three features of interest here.
\begin{enumerate}
\item The exponents for each basin frequently alternate between regions of steady concave growth and regions where the exponent is identically zero.  A comparision with the monodromy matrix computations shows that the regions of zero exponents precisely correspond to the stable bands of the relevant periodic orbits.  
\item The non-zero portions of each of the four curves fit nicely onto one another.  Chaotic trajectories are thus characterized by a \textit{single} Lyapunov exponent at large enough supercritical energies.  Barring stability-instability transitions of the periodic orbits,  this agrees with our earlier assertion of a \textit{single} chaotic basin at sufficiently high supercritical energies.
\item The non-zero sectors of the exponent plots are neatly captured by a $E^{\frac{1}{4}}$ fit. The exponents thus steadily develop an algebraic dependence on $E$,  at least to the leading order. The same scaling has been observed in \cite{Hashimoto2016}.
\end{enumerate}


\section{On Thermalization in the Matrix Model}\label{TD}
For chaotic systems that are also ergodic, the Birchoff-Khinchin theorem holds: for almost any dynamical observable, the time-average is equal to the ensemble average. Berdichevsky has suggested that for such ergodic systems, the laws of equilibrium statistical mechanics may be adapted, making the systems amenable to thermodynamic discussion.

The Hamiltonian (\ref{Hamiltonian5}) describes a small system: the phase space is only six-dimensional. 
Nevertheless, as we have demonstrated in the previous sections, the system becomes chaotic as the energy $E$ (or more accurately, $g^2 E$) increases. Thermodynamics of small systems is a subject of active research \cite{Hilbert2014}, the starting point of this discussion being the formula for entropy first given by Gibbs \cite{Gibbs2010} (see also \cite{Khinchin2014-aj}) for a microcanonical ensemble. If $\Gamma (E)$ is the volume of the region $H \leq E$, then the Gibbs entropy $S_{\Gamma}$ is
\begin{align}
S_{\Gamma}=\ln \Gamma+\textrm{const.} \label{entropy}
\end{align}

The two other definitions of entropy
\begin{align}
S&=\ln \frac{\partial\Gamma}{\partial E}\delta E +\textrm{const}. \\
S&=\ln \frac{\partial \Gamma}{\partial E}+\textrm{const.}
\end{align}
agree with (\ref{entropy}) in the limit when the number of degrees of freedom $N$ becomes large \cite{Huang1987-mx}. However, only (\ref{entropy}) satisfies the equipartition theorem for small systems \cite{Hilbert2014}.

Given the expression for entropy, one can define a temperature $T$ as
\begin{align}
\frac{1}{T}=\frac{\partial S}{\partial E}. \label{ergodicity1}
\end{align}

For \textit{ergodic} systems, there exists another definition of temperature coming from the equipartition theorem:
\begin{align}
T=\left\langle p_i \frac{\partial H}{\partial p_i} \right\rangle=\left\langle p_i^2 \right\rangle, \textrm{ } i=1,2, \cdots N .\label{ergodicity2}
\end{align}
Here $\langle \cdot \rangle$ denotes temporal average over a time interval $\tau$ as $\tau \to \infty$. 

In this section, we will explore in some detail issues related to thermalization in our model, and its relation to earlier discussions of stability and ergodicity. As in the earlier discussion, we will use group theory to guide us in this exploration. 


%
\subsection{Equipartition Theorem and Ergodicity}
\begin{figure}[H]
\centering
\begin{subfigure}[c]{0.49\textwidth}
\includegraphics[scale=0.66]{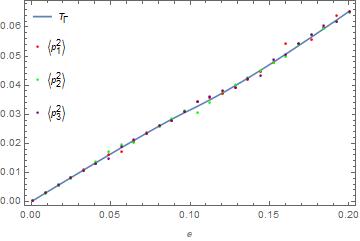}
\caption{}
\end{subfigure}
%
\begin{subfigure}[c]{0.49\textwidth}
\includegraphics[scale=0.64]{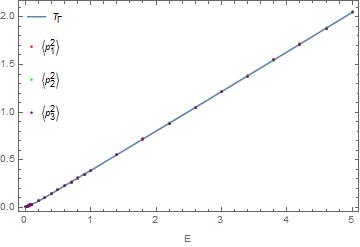}
\caption{}
\end{subfigure}
\caption{Temperatures vs energy for random initial conditions}
\label{Randomplots}
\end{figure}

We can use the equipartition theorem and ergodicity to decide if our system has thermalized.
Our procedure is as follows:
\begin{enumerate}
\item Choose an energy $E$ and compute the phase space volume $\Gamma(E)$, the volume of the region $H \leq E$. 
\item Use (\ref{entropy}) to compute the Gibbs entropy $S_{\Gamma}$, and the Gibbs temperature $T_{\Gamma}=\left(\frac{\partial S_{\Gamma}}{\partial E}\right)^{-1}$.
\item Generate 10 random sets of initial conditions corresponding to the energy $E$.
\item Calculate the temporal averages $ \langle p_1^2 \rangle $, $ \langle p_2^2 \rangle$, $ \langle p_3^2 \rangle $ and $\langle p_i p_i \rangle /3$ for each of these initial conditions, and their means and standard deviations.
\item Compare $T_{\Gamma}$ with the above temporal averages (see Figure \ref{Randomplots}). The figures do not include the error bars because they are negligible compared to the mean values. Also, excluding the error bars provides clarity.
\end{enumerate}

The extent of agreement between $ \langle p_1^2 \rangle $, $ \langle p_2^2  \rangle $, $ \langle p_3^2 \rangle $ and $T_{\Gamma}$ tells us the extent of `thermalization' in the system. As Figure \ref{Randomplots} shows, this agreement is excellent. It is surprising to see that thermodynamic ideas like temperature and equipartition come together as an equality even in a system as small as ours.

\subsection{Ergodicity breaking}
The analysis of chaos presented in the previous sections was quite nuanced because we were able to study that using group theoretic and geometric arguments. Now we analyze the same from a thermodynamical and statistical point of view and find their imprints here as well.


The procedure we use to study ergodicity of various orbits is similar to that described in the previous subsection, the only difference being that in step 3, we generate random initial conditions \textit{belonging to a specific orbit}.

Before looking at the results, it is worth mentioning that if an orbit is stable, it is not ergodic. If it is unstable, it may or may not be ergodic. For $A_4$ and $A_3$ orbits, we have knowledge about stability from our previous monodromy matrix considerations, and this information had better agree with the thermodynamic considerations that follow. Remarkably, we find that they do.

There is a similar connection between sensitivity to initial conditions and ergodicity. If the Lyapunov exponent corresponding to some orbit is zero (within limits of numerical accuracy), we `almost' always expect it to be non-ergodic. We say `almost' because an orbit may have a negligible Lyapunov exponent and still be ergodic.
\subsubsection[Ergodicity of A4 orbits]{Ergodicity of $A_4$ orbits}
$A_4$ orbits, by our monodromy matrix results, are unstable upto energy $E=\frac{3}{2}$, after which there is an alternation of stable bands and band gaps, with their lengths increasing with energy.

\begin{figure}[H]
\centering
\begin{subfigure}[c]{0.49\textwidth}
\includegraphics[scale=0.67]{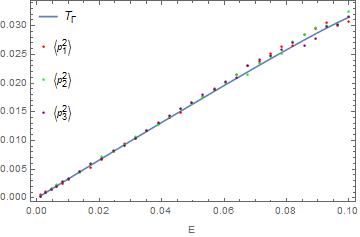}
\caption{}
\end{subfigure}
%
\begin{subfigure}[c]{0.49\textwidth}
\includegraphics[scale=0.63]{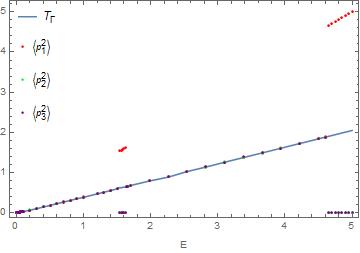}
\caption{}
\end{subfigure}
\caption{Temperatures vs energy for $A_4$ orbits}
\label{A4plots}
\end{figure}

The two plots in Fig. \ref{A4plots} clearly agree with these results: there is ergodicity till $E=\frac{3}{2}$, after which ergodicity is broken because the orbit is stable. The next stable (non-ergodic) region appears for energies $4.6 \lesssim E \lesssim 5$. This is also in agreement with curves of Lyapunov exponents vs energy: ergodicity is broken whenever the Lyapunov exponent vanishes (Figure \ref{A4s}).
\subsubsection[Ergodicity of A3 orbits]{Ergodicity of $A_3$ orbits}
$A_3$ orbits exhibit a self-similar structure where the bands keep getting narrower as one approaches the critical energy $E_c$. This structure is apparent in the ergodicity plots below as well: ergodicity is absent (i.e. the orbit is stable) till $ E \simeq 0.07$. But note that despite the presence of an unstable band near $E \simeq 0.075$, ergodicity is still broken. Stability implies ergodicity breaking, but instability does not necessarily imply ergodicity restoration.

Further bands are also visible and in agreement with monodromy matrix results. Again, we mention that ergodicity is broken whenever the Lyapunov exponent vanishes, so these plots agree with Lyapunov exponent considerations as well.

\begin{figure}[H]
\centering
\begin{subfigure}[c]{0.49\textwidth}
\includegraphics[scale=0.65]{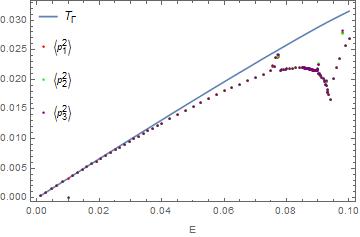}
\caption{}
\end{subfigure}
\begin{subfigure}[c]{0.49\textwidth}
\includegraphics[scale=0.65]{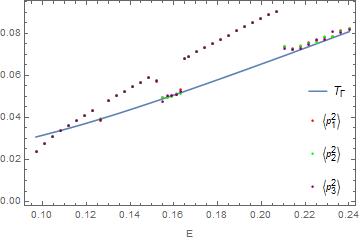}
\caption{}
\end{subfigure}
\begin{subfigure}[c]{0.49\textwidth}
\includegraphics[scale=0.65]{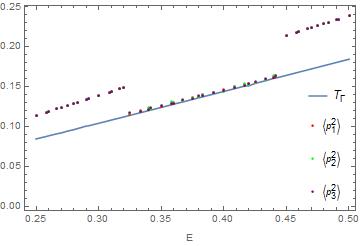}
\caption{}
\label{fig:A3plot2}
\end{subfigure}
\begin{subfigure}[c]{0.49\textwidth}
\includegraphics[scale=0.65]{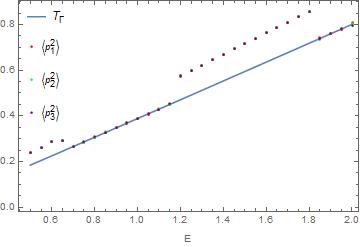}
\caption{}
\end{subfigure}
\begin{subfigure}[c]{0.49\textwidth}
\includegraphics[scale=0.65]{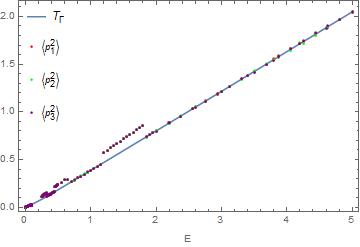}
\caption{}
\end{subfigure}
\caption{Temperatures vs energy for $A_3$ orbits}
\label{A3plots}
\end{figure}
\subsubsection[Ergodicity of A2 orbits]{Ergodicity of $A_2$ orbits}
For $A_2$ and the rest of the orbits remaining, we do not have the monodromy matrix tool at our disposal, so it is not possible to study the correlation between stability and ergodicity. We can, however, study ergodicity and its breaking.

Our plots in Fig. \ref{A2plots} extend to regions of energy high enough so that the periodic orbits do not exist at all. This is possible to do because despite the orbits losing periodicity, we still have initial conditions from Section \ref{others}. We do this in order to  investigate eventual fate of periodic orbits.

The plots in Fig. {\ref{A2plots} do not seem to possess a neat band structure as in the case with $A_4$ and $A_3$. However, we have the following conclusions for $A_2$:
\begin{enumerate}
\item Ergodicity is clearly broken for small energies $0.001 \lesssim E \lesssim 0.01$.
\item Ergodicity is restored afterwards except for an energy region in the range $0.05 \lesssim E \lesssim 0.08$. Ergodicity breakage is clearly visible.
\item Ergodicity is restored for energies $E \gtrsim 0.1$.
\end{enumerate}

\begin{figure}[H]
\centering
\begin{subfigure}[c]{0.49\textwidth}
\includegraphics[scale=0.65]{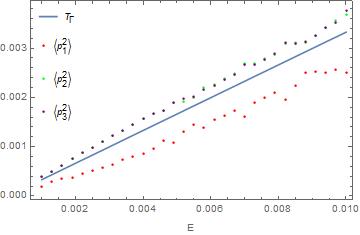}
\caption{}
\end{subfigure}
\begin{subfigure}[c]{0.49\textwidth}
\includegraphics[scale=0.65]{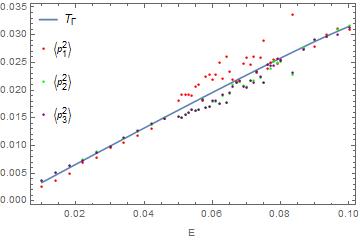}
\caption{}
\end{subfigure}
\begin{subfigure}[c]{0.49\textwidth}
\includegraphics[scale=0.65]{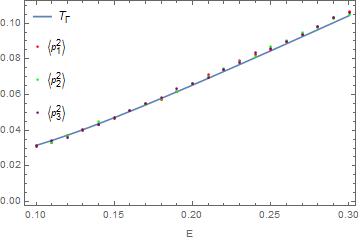}
\caption{}
\end{subfigure}
\begin{subfigure}[c]{0.49\textwidth}
\includegraphics[scale=0.65]{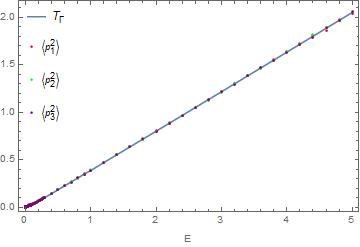}
\caption{}
\end{subfigure}
\caption{Temperatures vs energy for $A_2$ orbits}
\label{A2plots}
\end{figure}
\newpage
\subsubsection[Ergodicity of B4 orbits]{Ergodicity of $B_4$ orbits}
Fig. {\ref{B4plots} shows that $B_4$ orbits are non-ergodic till around energy $E \simeq 0.16$, above which they are ergodic.
\begin{figure}[H]
\centering
\begin{subfigure}[c]{0.49\textwidth}
\includegraphics[scale=0.67]{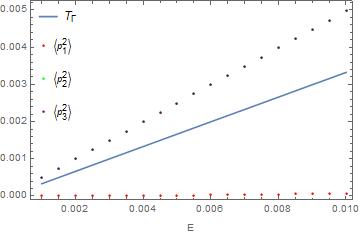}
\caption{}
\end{subfigure}
\begin{subfigure}[c]{0.49\textwidth}
\includegraphics[scale=0.65]{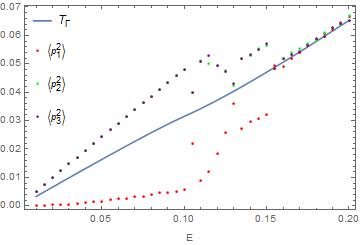}
\caption{}
\end{subfigure}
\begin{subfigure}[c]{0.49\textwidth}
\includegraphics[scale=0.65]{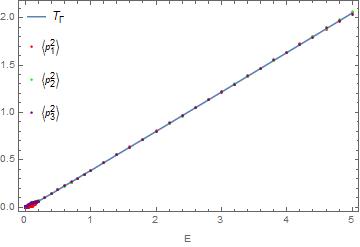}
\caption{}
\end{subfigure}
\caption{Temperatures vs energy for $B_4$ orbits}
\label{B4plots}
\end{figure}
\newpage
\subsubsection[Ergodicity of B3 orbits]{Ergodicity of $B_3$ orbits}
$B_3$ orbits are ergodic in the energy range considered ($0 \lesssim E \simeq 5)$.
\begin{figure}[H]
\centering
\begin{subfigure}[c]{0.49\textwidth}
\includegraphics[scale=0.65]{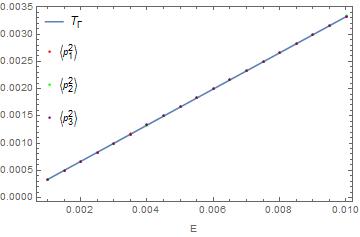}
\caption{}
\end{subfigure}
\begin{subfigure}[c]{0.49\textwidth}
\includegraphics[scale=0.65]{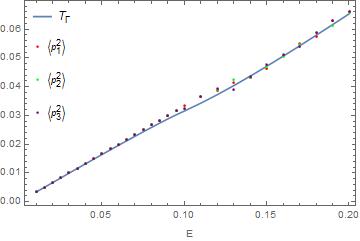}
\caption{}
\end{subfigure}
\begin{subfigure}[c]{0.49\textwidth}
\includegraphics[scale=0.65]{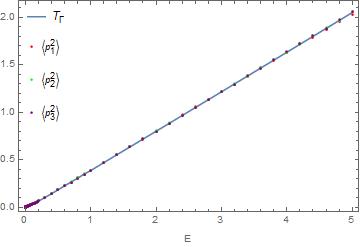}
\caption{}
\end{subfigure}
\caption{Temperatures vs energy for $B_3$ orbits}
\label{B3plots}
\end{figure}

\newpage
\subsubsection{Ergodicity of non-critical NLNMs}
Non-critical NLNMs (Non Linear Normal Modes) remain non-ergodic till energy $E \simeq 0.11$ and become ergodic above this energy (Fig. \ref{NCESplots}).
\begin{figure}[H]
\centering
\begin{subfigure}[c]{0.49\textwidth}
\includegraphics[scale=0.65]{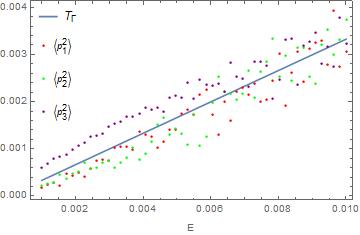}
\caption{}
\end{subfigure}
\begin{subfigure}[c]{0.49\textwidth}
\includegraphics[scale=0.65]{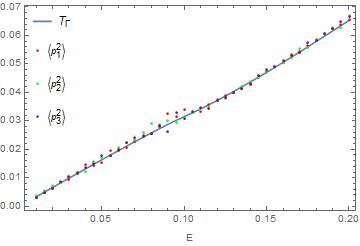}
\caption{}
\end{subfigure}
\begin{subfigure}[c]{0.49\textwidth}
\includegraphics[scale=0.65]{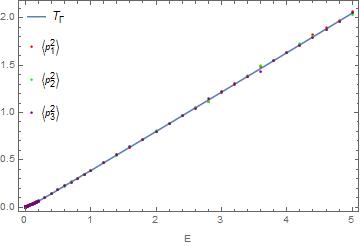}
\caption{}
\end{subfigure}
\caption{Temperatures vs energy for non-critical orbits}
\label{NCESplots}
\end{figure}
\subsubsection[Ergodicity of Pi1 orbits]{Ergodicity of $\Pi_1$ orbits}
$\Pi_1$ orbits are important because they complement the chaotic basin formed by $A_4$ orbits at energies $E \lesssim E_c$. Figure \ref{Pi1plots} shows that at these energies, they do indeed have the same temperature as the corresponding $A_4$ orbits. However, ergodicity is broken below energy $E \simeq 0.06$, despite $\Pi_1$ orbits being unstable for all subcritical energies, as shown by the monodromy matrix plot (not presented here) as well as Lyapunov exponent considerations. Again, we see that stability implies ergodicity breaking but instability does not necessarily imply ergodicity restoration.
\begin{figure}[H]
\centering
\begin{subfigure}[c]{0.49\textwidth}
\includegraphics[scale=0.65]{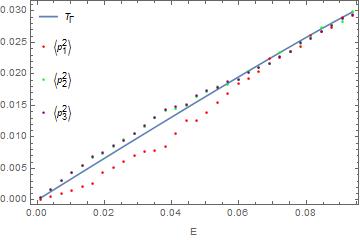}
\end{subfigure}
\caption{Temperatures vs energy for $\Pi_1$ orbits}
\label{Pi1plots}
\end{figure}
%
\subsubsection[Ergodicity of Pi2 orbits]{Ergodicity of $\Pi_2$ orbits}
$\Pi_2$ orbits are found to be non-ergodic for all subcritical energies and ergodic for all supercritical energies considered.
\begin{figure}[H]
\centering
\begin{subfigure}[c]{0.49\textwidth}
\includegraphics[scale=0.65]{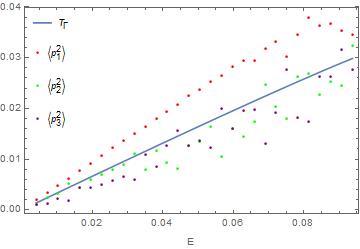}
\caption{}
\end{subfigure}
\begin{subfigure}[c]{0.49\textwidth}
\includegraphics[scale=0.65]{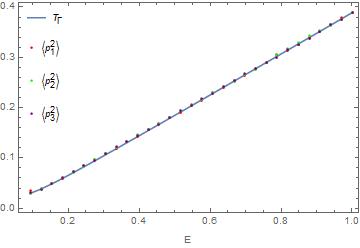}
\caption{}
\end{subfigure}
\begin{subfigure}[c]{0.49\textwidth}
\includegraphics[scale=0.65]{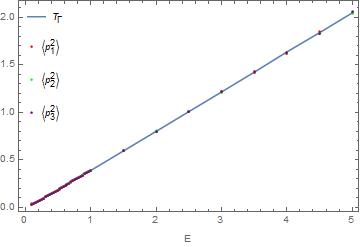}
\caption{}
\end{subfigure}
\caption{Temperatures vs energy for $\Pi_2$ orbits}
\label{Pi2plots}
\end{figure}
%

\subsection{Other Ergodic Averages}
The equipartition theorem is, more generally,
\begin{align}
\left\langle x_i \frac{\partial H}{\partial x_j} \right\rangle=T \delta_{ij},\ i,j=1,2,...,2N
\label{equipartition_general}
\end{align}
where $x_i$ are {\it any} phase space coordinates. We can compute these averages  in addition to the $p_i^2$, to confirm ergodicity. Computations show that the quantities $\langle a_i \frac{\partial H}{\partial a_j} \rangle$ and $\langle p_{a_i} \frac{\partial H}{\partial p_{a_j}} \rangle$, agree exactly in the ergodic regime.

For example, the following plot depicts $\langle p_{a_1} p_{a_2} \rangle$ and $\langle a_1 \frac{\partial H}{\partial a_1} \rangle$ for $A_3$ orbits, for $0.25 \leq E \leq 0.5$. In the region $0.325 \lesssim E \lesssim 0.45$ the system can clearly be seen to be ergodic and, remarkably, agrees with Fig. \ref{fig:A3plot2}. Outside this region, ergodicity is broken.
\begin{figure}[H]
\centering
\begin{subfigure}[c]{0.49\textwidth}
\includegraphics[scale=0.65]{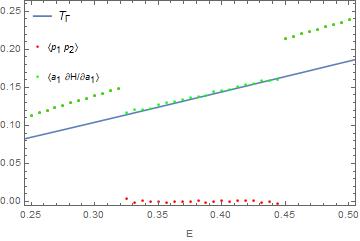}
\end{subfigure}
\caption{Time averages $\left\langle a_1 \frac{\partial H}{\partial a_1} \right\rangle$ and $\langle p_{1} p_{2} \rangle$ for $A_3$ orbits. $T_{\Gamma}$ is also plotted for comparison.}
\end{figure}

Similar plots of time averages for other orbits and energies also confirm the expected connection between ergodicity and the equipartition theorem.

It is striking that the system obeys the general version of the equipartition theorem (Eq. \ref{equipartition_general}).

\section{Classical Phases of the Matrix Model}
\label{sec8}

We have investigated the dynamical behaviour of a large number of subsectors of our model in different regimes using a variety of techniques.
We note that the unusual diversity in dynamics - subsectors, nested dynamics and ergodicity breaking - is highly reminiscent of an underlying phase structure and associated phase transitions. 
In fact, ordered and chaotic regimes have indeed been identified as distinct classical phases, particularly in the context of matrix models  \cite{Berenstein2017}, \cite{Hashimoto2016}.
 Additionally, the exotic dynamics uncovered here hints at an uncommonly rich phase structure. 
 There is an even more suggestive reason to believe that a phase study is the way to go, which we shall outline later on. In this section, we will just press forward with this viewpoint and outline the phase structure of the matrix model.

Phases are usually identified by regions in an appropriate phase diagram, labelled by a set of independent variables. Taking the quintessential example of ice-water-steam phase diagram, pressure, volume and temperature serve as the distinguishing parameters. The most obvious parameter that we could utilise for the matrix model is, of course, the energy. Although slightly unusual in a more `physical’ sense (where temperature is the natural choice), energy is a natural variable to use in the more abstract context of non-linear systems. Alternatively, our stand simply reflects the `microcanonical’ nature of our setup, as outlined in section \ref{TD}. 

Generally, energy is sufficient to capture the phase structure, with low energies yielding regular behaviour and chaos taking over later on. As we have seen however, the matrix model may display several distinct types of dynamics even at a given energy. An exact characterization using just the energy is therefore incomplete. Furthermore, there is no precise list of variables which, together with the energy, \textit{do} completely characterize the phase structure. Our previous analysis tells us that the symmetries of the Hamiltonian are the key players, but that is about as far as we can go. Nevertheless, the absence of such a list does not prevent us from \textit{enumerating} the numerous existing phases, following the generic methodology of identifying ordered and chaotic regimes as distinct classical phases.

With this viewpoint, we see that the multiple chaotic subsectors in the subcritical range (and their regular counterparts) have a
natural interpretation as \textit{co-existing} classical phases. The disjoint Poincar\'e sections of Figure \ref{PSSub3D} neatly illustrate the `chaotic $A_{4}$ phase’ and the `chaotic $\Pi_{1}$’ coexisting with the `$\Pi_{2}$ phase’ and the `$A_{3}$ phase’ (not shown in the figure).
The phenomenon of co-existing phases is a fairly well-known
one, with water-steam-ice \cite{IcePhases} serving as a well documented example.
As phases are typically distinguished by differing expectations of certain interesting observables,
it is natural to list out such observables for our model as well. Since the Poincar\'e sections of the
$A_{4}$ sector are more concentrated near the edges of the allowed configuration space, while those of the
$\Pi_{1}$ sector group near the centre, it is reasonable to expect that the squares of the $a_{i}$
’s (the second
moments, so to speak) serve as distinguishing observables. Indeed computing the time average of these
observables for $\Pi_{1}$ based trajectories and $A_{4}$ based trajectories of equal energy yield noticeably different
results. A more sophisticated distinguishing observable is, of course, the Lyapunov exponent. The computations of
section \ref{LE} indeed corroborate this view, with the exponents of the $A_{4}$ sector being marginally lower than their $\Pi_{1}$
counterparts. Additionally, as seen from the monodromy plots (see Figure 12), the $A_{3}$ orbits describe
chaotic bands of their own at suitable subcritical energies, implying that we can have three
chaotic phases intermixing with one another at certain $E < E_{c}$.

Next, translating the phenomenon of nested chaos to our phase centred viewpoint implies the existence of yet another collection of phases, this time dimensionally distinct from our earlier sets. Since the RDSs inherit nearly all of the peculiarities of the full dynamics, the structure of this lower dimensional collection of phases is just as intricate as the full 6D phase structure. Indeed, one can draw correspondences between the RDS phases and those of the full model. The notion of `lower-dimensional’ phases in a physical system is rather unusual, though not unheard of, with edge states in topological physics serving as a good example. It is therefore interesting to see such themes emerge naturally in the context of a gauge matrix model.


Much like the subcritical regime, nested phases are also a feature of the supercritical regime, with supercritical nested phases appropriately inheriting the phase structure of their parent 6D phases. It is interesting to note that the notion of symmetry breaking persists in this model despite the symmetries of the RDS only encompassing a small subgroup of the full tetrahedral group. 

Fascinating as this game of coexistence and mergers is, it involves only the chaotic phases of the model. The transitions between ordered and chaotic phases are no less interesting. We have already encountered numerous signatures of these transitions, via monodromy plots, Lyapunov exponents and thermodynamics. These analyses neatly corroborate one another and clearly indicate alternations between ordered and chaotic regimes, and thus, between ordered and chaotic \textit{phases}. Specifically, the phase structure involves an  \textit{alternation} between individual regular phases and the `global’ chaotic phase. These alternations happen at energies that are specific to the parent orbit in question. 

As regards the (breaking of) the symmetries of the system, we thus see that each the symmetry of the parent orbits is after all \textit{not} completely lost at high energies, but is retained solely by the \textit{ordered} phases, insofar as they exist at high energies.  As we have seen, symmetries bifurcate the dynamics into a host of basins, one each for the $A_{3},A_{4}, \Pi_{1}$ and the $\Pi_{2}$ orbits. The transitions for the first two of this set continue ad infinitum, implying that these symmetry classes persist at arbitrarily high energies. In contrast, the $\Pi_{1}$ and $\Pi_{2}$ orbits  cease to alternate in stability at high enough energies, so that any memory of these symmetry classes is erased at suitably high energies. As before, the nested dynamics presents the same systematics, despite its reduced symmetries. Curiously enough, we will see later that this notion of finite versus infinite alternations has some ties to the quantum dynamics of the model.
\section{Quantum Connections}
\label{sec9}
While the previous sections have firmly established the $SU(2)$ QCD matrix model as a classical non-linear system of great interest, its primary usage as a tool, is in a quantum setting. From a pure gauge theory point of view, what then do we learn about the quantum theory from perusing its classical aspects? Given that we know of certain features of the quantum theory \cite{Balachandran2015}, it 
is thus worth investigating how the `memory' of these quantum features is retained in the classical limit. On the flip side, one might also be interested in using the above classical analysis to search for more elusive quantum features. 

Some quantum aspects of the $SU(2)$ matrix model coupled to massless quarks have already been studied in the 
`Born-Oppenheimer' limit of the theory: in this limit, the quarks are the fast degrees of freedom, and the gauge field the slow mode. The quarks are quantized in the background of the classical gauge field, and the gauge field is then quantized. The quarks produce an emergent Berry connection (a vector potential) as well as a scalar potential on the gauge configuration space. The gauge field is then quantized taking these additional emergent potentials into account.

Inclusion of the quark leads to an unexpected benefit even for investigations of the pure gauge theory: it provides for 
a much more refined understanding of the gauge configuration space. 
One can show that in terms of 
\begin{eqnarray}
x= \text{Tr}(M^T M),\quad y = \det{M}, \quad z = \frac{1}{16} \Big( 2 \text{Tr}(M^T M M^T M) - [\text{Tr}(M^T M)]^2 \Big)
\end{eqnarray}
the function $F(M) = F(x,y,z)$ obeys the inequality
\begin{equation}
F(M) = \frac{1}{2}\left(2 x^4 z+x^3 y^2-64 x^2 z^2-144 x y^2 z-54 y^4+512 z^3 \right) \geq 0.
\end{equation}

With
\begin{eqnarray}
{\mathbf g}_3\equiv \frac{\det M}{\left(\frac{1}{3}\textrm{Tr}(M^TM)\right)^{3/2}}, \quad\quad 
{\mathbf g}_4 \equiv \frac{1}{16}\left[\frac{2\textrm{Tr}(M^TM)^2}{\left(\frac{1}{3}\textrm{Tr}(M^TM)\right)^2}-9\right],
\end{eqnarray}
the condition $F \geq 0$ becomes
\begin{eqnarray}
\Delta =\frac{1}{2}\Big(27{\mathbf g}_3^2-54{\mathbf g}_3^4+162{\mathbf g}_4-432{\mathbf g}_3^2{\mathbf g}_4-576{\mathbf g}_4^2+512{\mathbf g}_4^3\Big) \geq 0.
\label{identity}
\end{eqnarray}
In other words, $F \geq 0$ (or equivalently $\Delta \geq 0$) gives us the set of all gauge-invariant spin-zero gauge field configurations. This parametrization of the gauge configuration space explicitly brings out the fact that it has corners ($A, B$ and $C$) and edges ($AB, BC$ and $AC$).

We can plot the region bounded by the above inequality:
\begin{figure}[hbtp]
\centering
\includegraphics[height=10cm,width=14cm]{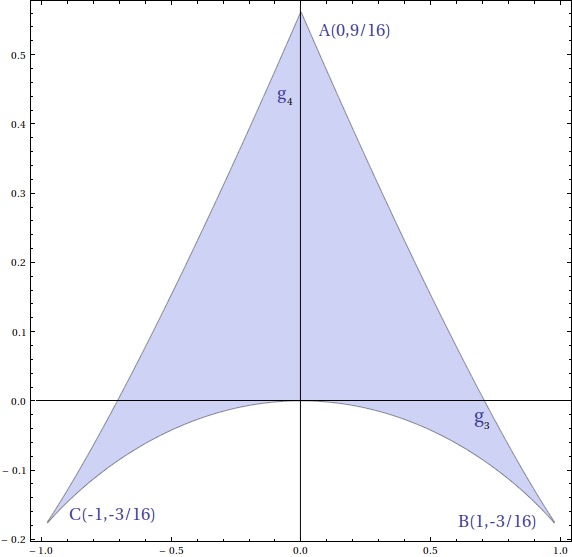}
\caption{(Scaled) Configuration space of $SU(2)$ gauge matrix model}
\label{fig2}
\end{figure}

The $\mathbf{g}_{3}-\mathbf{g}_{4}$ plot is an `arrowhead' curve consisting of configuration space points satisfying $\Delta \geq 0$.

In terms of the coordinates $(R, A, S)$, the functions $g_3$ and $g_4$ take a rather simple form
\begin{equation}
    \label{g3Defn}
    g_3(a_{1},a_{2},a_{3})\equiv \frac{a_1 a_2 a_3}{\left(\frac{a_1^2 + a_2^2 + a_3^2}{3}\right)^{\frac{3}{2}}}
\end{equation}
and 
\begin{equation}
    \label{g4Defn}
    g_4(a_{1},a_{2},a_{3})\equiv \frac{9 (a_1-a_2+a_3)(a_2-a_3+a_1)(a_3-a_1+a_2)(a_1+a_2+a_3)}{
 16 (a_1^2 + a_2^2 + a_3^2)^2}.
\end{equation}

It was argued in \cite{Pandey2017} that quarks `condense' at these corners and edges, leading to quantum phases. These phases, obtained via superselection sectors can be distinguished using two scale invariant configuration space functions $g_{3}$ and $g_{4}$ defined as above.

The figures below provides a graphical depiction of the quantum phases.  The quantum phases are distinguished by their relative positions on the $g_{3}-g_{4}$ plot, with the interior of the arrowhead depicting a `bulk phase' while the sides of the arrowhead model `edge' phases. The three tips of the arrowhead also represent distinct phases, with the phases corresponding to the two lower tips of the arrowhead related to one another by a parity transform. 

While the $g_{3}-g_{4}$ plot and relevant machinery concerned were developed in a purely quantum setting, it turns out to be very useful for discussing aspects of classical dynamics as well. Specifically, we may associate each classical trajectory with a given trajectory traversing the boundary and interior of the arrowhead. Identifications between classical and quantum phases can then be made by comparing \textit{classically} generated $g_{3}-g_{4}$ plots with the pictorial hierarchy of quantum phases mentioned in the above paragraph. For instance, a general chaotic trajectory unsurprisingly covers the bulk of the $g_{3}-g_{4}$ plot and thus is evidently in loose correspondence with the `bulk' quantum phase. On the other hand, the ``2 equal $a$'s" trajectories that make up the 4-dimensional RDS are, from the definition of the $\Delta$ function, confined to lie on the edges of the arrowhead and thus are in loose correspondence with the edge phases of the model. That the correspondence is not exact is obvious as, for instance, generic trajectories may have, at some points of times, two equal $a$s thereby landing themselves on the edges of the $g_{3}-g_{4}$ rather than the bulk. Additionally, as we have mentioned, the arrowhead comprises numerous disconnected edge phases in addition to three `point phases', while generic trajectories in the 4-dimensional RDS span the entire arrowhead, so that they mix the edge phases and cross over the point phases at least partly. The 4D restrained $\Pi_{1}$ orbits for instance cover only the right half (or only the left half, in case of parity reversal) of the $g_{3}-g_{4}$ plots although even they encompass four quantum phases. While far from perfect, such correspondences are about as much as we may expect from a preliminary analysis and nevertheless have \textit{some} semblance to a deeper correspondence, telling us that we are after all on the right track.

There is also a reasonably clear correspondence between the $A_3$ and $A_4$ orbits (or more precisely the phases they map to) and the ``point phases" of the quantum model. Indeed, $g_{3}-g_{4}$  plots of the \textit{exact} $A_3$ and $A_4$ trajectories are \textit{perfectly} confined to the top tip ($A_4$) and lower right/left tips ($A_3$) of the arrowhead. Chaotic dynamics about these orbits is associated with space filling $g_{3} g_{4}$ plots while band gaps are only associated with minor spillovers from the tips of the arrowhead. The $A_4$ and the $A_3$ orbits are, at least at first glance, the apparent classical remnants of the quantum point phases. Interestingly enough, these are the only two periodic orbits whose phases underwent an infinite cascade of flips. On conjecture at least, this cascade has \textit{something} to do with quantum properties of the matrix model.
\begin{figure}[H]
     \centering
     \begin{subfigure}[b]{0.45\textwidth}
         \centering
         \includegraphics[width=\textwidth]{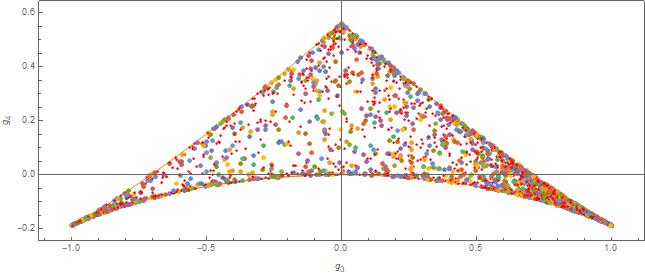}
         \caption{$E=0.324$}
         \label{A3g3g41}
     \end{subfigure}
     \hfill
     \begin{subfigure}[b]{0.45\textwidth}
         \centering
         \includegraphics[width=\textwidth]{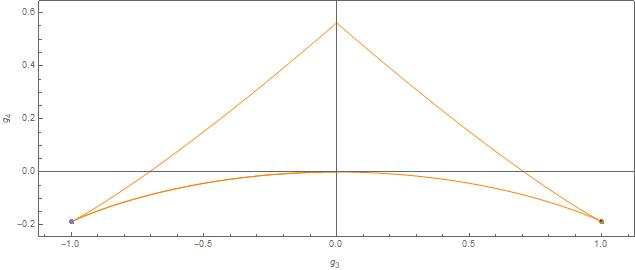}
         \caption{$E=0.323$}
         \label{A3g3g42}
     \end{subfigure}
         \caption{$g_3-g_4$ plots for the $A_3$ orbits}
     \hfill

\end{figure}
\begin{figure}[H]
     \centering
     \begin{subfigure}[b]{0.45\textwidth}
         \centering
         \includegraphics[width=\textwidth]{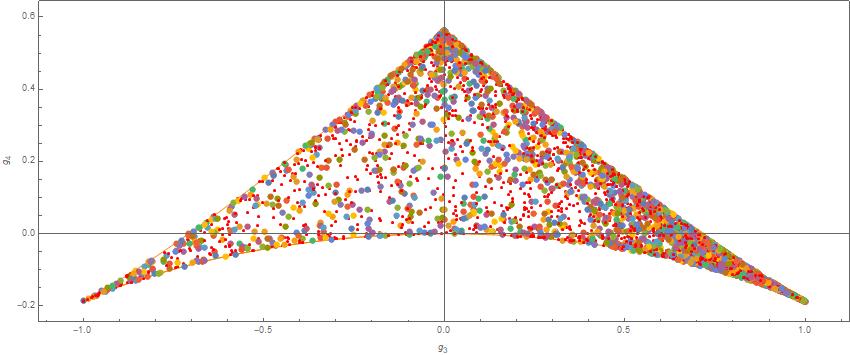}
         \caption{$E=1.499$}
         \label{A4g3g41}
     \end{subfigure}
     \hfill
     \begin{subfigure}[b]{0.45\textwidth}
         \centering
         \includegraphics[width=\textwidth]{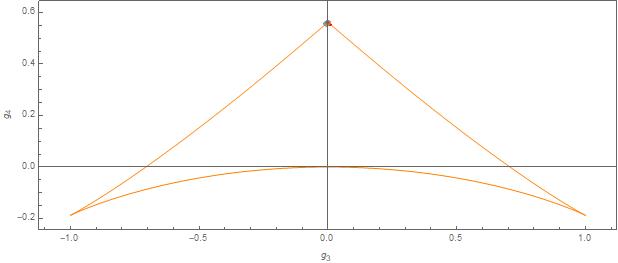}
         \caption{$E=1.501$}
         \label{A4g3g42}
     \end{subfigure}
         \caption{$g_3-g_4$ plots for the $A_4$ orbits}
     \hfill

\end{figure}
\section{Conclusions}
\label{sec10}
In this article, we pursued a detailed study of the classical dynamics of the spin-0 sector of an $SU(2)$ gauge-matrix model. The presence of an unexpected tetrahedral symmetry greatly enriched the resulting dynamics, endowing the system with several distinctive features such as co-existing chaotic basins, ergodicity breaking and nested chaos. The tetrahedral symmetry also allowed us to better adapt standard techniques to bring out the salient features of the model. We utilized a three-pronged approach comprising monodromy analysis, chaos-theoretic studies, and  statistical mechanical methods. The last of these motivated a transition from a non-linear dynamical perspective to a thermodynamic one, wherein we identified the regular and chaotic sectors of the model as classical phases. The intricacies of the classical dynamics translated into a rich phase structure consisting of co-existing chaotic phases protected by their respective symmetries at subcritical energies. The underlying protective mechanism seemed to degrade at suitably high supercritical energies, culminating with a merger into a single supercritical chaotic phase. Also observed were quasi-periodic transitions between ordered and chaotic phases and a collection of lower dimensional \textit{nested} phases. Surprisingly, a selection of classical phases bore tantalizing resemblances to \textit{quantum} phases stemming from superselection sectors. This correspondence had benefits for both sides. In one direction, the quantum sector naturally yielded refined tools (i.e. the $g_{3}-g_{4}$ plots) for identifying classical phases. In the other direction, the classical phase structure could potentially give signatures for further investigations of the quantum phase structure of the matrix model.

Broadly speaking, the questions we aim to answer going forward fall into three categories, the first of which involves investigating the classical dynamics of the spin-0 sector in even more depth. From a non-linear dynamical standpoint, several features of the dynamics beg for deeper explorations. For one, we are yet to understand the mechanism behind the localization of the co-existing chaotic sub-sectors for subcritical energies.  It is also unclear why this mechanism ceases to work at sufficiently high energies. Relevant thermodynamic problems include a better enumeration of the properties of the classical phases, via appropriately chosen observables, and a detailed study of the transitions between these phases. In particular, given that ergodicity breaking is a key ingredient for the emergence of the intricate phase structure of the model, it would be interesting to search for connections to color glasses in non-abelian gauge theories \cite{Gelis2010}.

The second class of questions we wish to explore center around the relations between the classical and quantum phases. Our current understanding of the correlations between the classical phases generated by the $A_{3/4}$ orbits and their quantum counterparts is rather heuristic. A more rigorous study of their connections, possibly via the Gutzweiller trace formula, is thus called for.  Another interesting pathway involves searching for quantum analogs of the phases generated by the remaining NLNMs or the geometric orbits.

Lastly, as illuminating as the spin-0 sector is, its study is only the first half of a broader endeavour. After all, a complete study of the classical dynamics of the full matrix model requires including the effects of angular momentum. We plan to add back the rotational degrees of freedom and analyze the resulting dynamics in a future work. A natural follow up would be to probe the connections between the full classical dynamics and the corresponding quantum analog. 

Although our present discussion has centred on the $SU(2)$ matrix model, it seems unlikely that the peculiarities of the dynamics will disappear as we go over to  the $SU(3)$ model. We expect at least some of these features to persist for $SU(3)$ models, with interesting consequences for real-world QCD.
\section*{Acknowledgements}The work of CB was supported by the PMRF programme. VN acknowledges that a substantial portion of the research was carried out before his affiliation with the Cavendish Laboratory.
\newpage
\begin{appendices}

\section{Asymptotic Locations of A4 Stability Transition Points}\label{A4Results}

\begin{table}[H]
\caption {$A_4$ Transition Points: Stable to Unstable} \label{SU} 
\begin{centering}
\begin{tabular}{ |c|c|c|c|c| } 
\hline

N &Mathieu Index	& Type (A/B)&	Analytically Computed&	Numerically Computed\\
&&&Transition Energy&Transition Energy \\ \hline
9 &  10&	B&	117.163&	118.55\\   \hline 
10&	11&	A&	142.451	&143.85\\   \hline 
11&	12&	B&	170.206	&171.65\\   \hline 
12&	13&	A&	200.429	&201.85\\   \hline 
13&	14&	B&	233.12&	234.65\\   \hline 
14&	15&	A&	268.278	&269.75\\   \hline 
15&	16&	B&	305.904&	307.45\\   \hline 
16&	17&	A&	345.998&	347.55\\   \hline 
17&	18&	B&	388.558&	390.15\\   \hline 
18&	19&	A&	433.587&	435.15\\   \hline 
19&	20&	B&	481.083&482.65\\   \hline 
20&	21&	A&	531.046&	532.65\\   \hline 
21&	22&	B&	583.477&	585.15\\   \hline 
22&	23&	A&	638.375	&640.05\\   \hline 
23&	24&	B&	695.741&	697.45\\   \hline 
24&	25&	A&	755.574	&757.25\\   \hline 
25&	26&	B&	817.875&819.55\\   \hline 
26&	27&	A&	882.643	&884.35\\   \hline 
27&	28	&B&	949.878	&951.55\\   \hline 
28&	29&	A&	1019.58	&1021.35\\   \hline 
29&	30	&B&	1091.75&	1093.45\\   \hline 
30&	31&	A&	1166.39	&1168.15\\   \hline 
31&	32&	B&	1243.49&	1245.25\\   \hline 
32&	33&	A&	1323.07	&1324.85\\   \hline 
33&	34&	B&	1405.11&	1406.85\\   \hline 
34&	35&	A&	1489.62	&1491.35\\   \hline 
35&	36&	B&	1576.59&	1578.35\\   \hline 
36&	37&	A&	1666.03	&1667.85\\   \hline 
\end{tabular}
\end{centering}
\end{table}
\begin{table}[H]
\caption {$A_4$ Transition Points: Unstable to Stable} \label{US} 
\begin{centering}
\begin{tabular}{ |c|c|c|c|c| } 

\hline

N &Mathieu Index	& Type (A/B)&	Analytically Computed&	Numerically Computed\\
&&&Transition Energy&Transition Energy \\ \hline
10&	10	&A&	129.4&	130.05\\   \hline 
11&	11&	B&	155.922&	156.55\\   \hline 
12&	12&	A&	184.911&	185.55\\   \hline 
13&	13&	B&	216.368&	217.05\\   \hline 
14&	14&	A&	250.293	&251.05\\   \hline 
15&	15&	B&	286.685&	287.45\\   \hline 
16&	16&	A&	325.544&	326.35\\   \hline 
17&	17&	B&	366.871	&367.65\\   \hline 
18&	18&	A&	410.666&	411.45\\   \hline 
19&	19&	B&	456.928&	457.75\\   \hline 
20&	20&	A&	505.658&	506.45\\   \hline 
21&	21&	B&	556.855&	557.65\\   \hline 
22&	22&	A&	610.519	&611.35\\   \hline 
23&	23&	B&	666.651&	667.55\\   \hline 
24&	24&	A&	725.251&	726.15\\   \hline 
25&	25&	B&	786.318&	787.25\\   \hline 
26&	26&	A&	849.852	&850.75\\   \hline 
27&	27&	B&	915.854&	916.75\\   \hline 
28&	28&	A&	984.323&	985.25\\   \hline 
29&	29&	B&	1055.26&	1056.25\\   \hline 
30&	30&	A&	1128.66&	1129.65\\   \hline 
31&	31&	B&	1204.54	&1205.45\\   \hline 
32&	32&	A&	1282.87&	1283.85\\   \hline 
33&	33&	B&	1363.68&	1364.65\\   \hline 
34&	34&	A&	1446.95&	1447.95\\   \hline 
35&	35&	B&	1532.7&	1533.65\\   \hline 
36&	36&	A&	1620.9&	1621.95\\   \hline 
\end{tabular}
\end{centering}
\end{table}
%
%
%
%
\end{appendices}


\end{document}